\RequirePackage{ifpdf}
\documentclass[11pt]{JHEP3}

\usepackage[normalem]{ulem}
\usepackage{cite}
\usepackage{epsfig}
\usepackage{amsmath}
\usepackage{bm,longtable,enumerate}
\usepackage{graphicx}
\usepackage{graphics}
\usepackage{mathrsfs}
\usepackage{amsthm}
\usepackage{amssymb}
\usepackage{oldgerm}
\usepackage{indentfirst}
\usepackage{setspace}
\usepackage{multicol}
\usepackage{multirow}

\def\beq{\begin{equation}}
\def\eeq{\end{equation}}
\def\beqa{\begin{eqnarray}}
\def\eeqa{\end{eqnarray}}
\usepackage{array} 

\title{Effective holographic models for QCD: glueball spectrum and trace anomaly}

\author{Alfonso Ballon-Bayona,$^{a}$ Henrique 
Boschi-Filho,$^b$ Luis A. H. Mamani,$^c$ Alex S. 
Miranda,$^d$  and Vilson T. Zanchin$^c$\\
$^a$
Instituto de F\'isica Te\'orica, Universidade Estadual Paulista,\\
01140-070 S\~ao Paulo, SP, Brazil.\\
$^b$Instituto de F\'{i}sica, Universidade
Federal do Rio de Janeiro,\\
Caixa Postal 68528, RJ 21941-972, Brazil\\
$^c$Centro de Ci\^encias Naturais e Humanas,
Universidade Federal do ABC,\\
Rua Santa Ad\'elia 166, 09210-170, Santo Andr\'e, SP, Brazil\\
$^d$Laborat\'orio de Astrof\'{\i}sica Te\'orica e Observacional\\ 
Departamento de Ci\^encias Exatas e Tecnol\'ogicas\\
Universidade Estadual de Santa Cruz, 45650-000, Ilh\'eus, BA, Brazil\\
E-mail: aballonb@ift.unesp.br; boschi@if.ufrj.br;
luis.mamani@ufabc.edu.br; asmiranda@uesc.br;  zanchin@ufabc.edu.br}

\abstract{
 We investigate effective holographic models for QCD arising from five 
dimensional Dilaton-Gravity. The models are characterized by a dilaton with a 
mass term in the UV, dual to a CFT deformation by a relevant operator, and 
quadratic in the IR. The UV constraint leads to the explicit breaking of 
conformal symmetry whereas the IR constraint guarantees linear confinement. We 
propose semi-analytic interpolations between the UV and the IR and obtain a 
spectrum for scalar and tensor glueballs consistent with lattice QCD data. We 
use the glueball spectrum as a physical constraint to find the evolution of the 
model parameters as the mass term goes to zero. Finally, we reproduce the 
universal result for the trace anomaly of deformed CFTs and propose a dictionary 
between this result and the QCD trace anomaly. A nontrivial consequence of this 
dictionary is the emergence of a $\beta$ function similar to the two-loop 
perturbative QCD result.} 

\keywords{AdS/CFT, gauge/gravity duality, holographic QCD, glueballs}

\preprint{}

\begin{document}

\section{Introduction}
\label{introd}

Quantum Chromodynamics (QCD) is probably the most striking example of conformal 
symmetry breaking in a Quantum Field Theory. In QCD conformal symmetry is broken 
already in the UV by a negative $\beta$ function (asymptotic freedom) that, at 
the same time, leads to strong coupling and confinement in the IR. 
Despite the huge success of perturbative QCD in describing hard scattering 
processes, basic QCD features in the IR, such as the hadronic spectrum or chiral 
symmetry breaking, require the development of non-perturbative methods. The 
most well established non-perturbative approach consists of Monte-Carlo 
simulations for QCD on a lattice. The so-called lattice QCD is very successful 
in describing static properties such as the hadronic spectrum and thermodynamic 
properties of the quark-gluon plasma. However, real-time dynamics usually demand 
the development of other non-perturbative methods. 

Holographic QCD provides a non-perturbative description of real-time QCD 
dynamics in terms of the dynamics of a five dimensional gravitational (string) 
theory. In the so-called bottom-up approach, it is used the dictionary arising 
from the AdS/CFT correspondence to build a set of five dimensional fields dual 
to the QCD operators responsible for describing the QCD vacuum. In the 
large-$N_c$ limit one focuses on the stress-energy tensor $T^{\mu \nu}$ of the 
gluon field, as well as on the scalar operator ${\rm Tr}\, F^2$ responsible for 
the gluon condensate and the QCD scale
anomaly, i.e. $T^{\mu}_{\;\;\mu} \sim {\rm Tr}\, F^2$ \cite{Shuryak:1988ck}. In
holography the operators $T^{\mu \nu}$ and ${\rm Tr}\,F^2$ couple to a five
dimensional metric and scalar field (the dilaton), respectively. 
The dynamics of the metric and dilaton fields is determined by the equations of 
the Dilaton-Gravity theory, and the dilaton potential contains a negative 
cosmological constant leading to AdS geometry in the UV. Since Lorentz 
invariance is not broken by the QCD vacuum, in holographic QCD one takes a 
conformally flat metric with a warp factor $A$ depending solely on the radial 
coordinate $z$. Then a non-constant dilaton
$\Phi(z)$ leads to a deformation of the AdS spacetime geometry. This is the 
holographic realization of conformal symmetry breaking in large-$N_c$ QCD. 
Assuming that $\Phi(z)$ couples directly to the operator ${\rm Tr} \, F^2$  at 
the boundary one would expect, according to the AdS/CFT dictionary, $\Phi(z)$ to 
behave as $\phi_0 + G\, z^4$ near the boundary. This assumption, however, was 
shown not to be consistent with the glueball spectrum \cite{Csaki:2006ji}, 
because in that case conformal symmetry is spontaneously broken, leading to a 
zero mode in the
spectrum (a Nambu-Goldstone boson).  

There are two possible solutions to this problem. The first one consists of 
making the dilaton potential compatible with asymptotic freedom in the UV, as 
advocated in \cite{Csaki:2006ji}. This scenario was realized in the so-called 
Improved Holographic QCD models \cite{Gursoy:2007cb,Gursoy:2007er}, proposed by 
Gursoy, Kiritsis and Nitti. These authors also found the right physical 
constraint for the dilaton in the IR: a confining background, which corresponds 
to a quadratic dilaton for large $z$ and leads to an approximate linear 
glueball spectrum. Interestingly, that behaviour had already been anticipated in 
the soft-wall model \cite{Karch:2006pv} from the analysis of the meson spectrum. 
The second solution consists of introducing a mass term for the dilaton 
potential in the UV. In this case the near-boundary behaviour of the dilaton 
field becomes $\phi_0 \, z^{\epsilon} +  G\, z^{4-\epsilon}$, with $\epsilon$ 
related to the  dilaton mass $M$ by $M^2 = - \epsilon ( 4 - \epsilon)$. This 
UV asymptotics was proposed by Gubser, Nellore, Pufu and Rocha in
\cite{Gubser:2008ny,Gubser:2008yx}, within the context of finite-temperature 
holographic QCD as a model for describing the equation of state of a 
non-conformal plasma through a five-dimensional black hole. In 
\cite{Gubser:2008ny,Gubser:2008yx} the parameter $\epsilon$ was 
interpreted in terms of the anomalous dimension of the operator ${\rm Tr}\, F^2$.

Inspired by the seminal works \cite{Gursoy:2007cb,Gursoy:2007er} and 
\cite{Gubser:2008ny,Gubser:2008yx}, we investigate in this paper a family of 
holographic QCD backgrounds where the IR is driven by linear confinement and the 
UV, although asymptotically AdS, is deformed by a dilaton field with nonzero 
mass. We interpret the dilaton mass term in the UV as the holographic 
description of a CFT deformation  $\delta {\cal L} = \phi_{0}\,{\cal O}$, where 
${\cal O}$ is a relevant operator and $\phi_0$ is the corresponding coupling. 
This interpretation was advocated in \cite{Gubser:2008ny,Gubser:2008yx} and is 
consistent with previous studies in holographic renormalization 
\cite{Bianchi:2001de,Skenderis:2002wp}. The main motivation for this work is to 
understand how holography realizes conformal symmetry breaking in the UV without 
introducing explicitly the $\beta$ function of large-$N_c$ QCD. Our guide in this investigation 
will be the spectrum of scalar and tensor glueballs, because it will allow us to 
fix the source and the VEV coefficients in the near-boundary 
expansion.\footnote{This approach is similar to the one used in the holographic 
model for chiral symmetry breaking \cite{Erlich:2005qh,DaRold:2005mxj} where the 
quark masses and chiral condensates are fixed by the meson masses and decay 
constants. See also \cite{Abidin:2009aj,Ballon-Bayona:2017bwk}.} As in refs. 
\cite{Gubser:2008ny,Gubser:2008yx}, we assume that the CFT deformation takes 
place at a UV energy scale $E^{*}$. However, although $4-\epsilon$ is indeed 
the conformal dimension of the operator ${\cal O}$ (responsible for the CFT 
deformation),  the relation between ${\cal O}$ and the QCD operator ${\rm Tr}\, 
F^2$ is not direct. Therefore, we will not interpret the parameter $\epsilon$ as 
the anomalous dimension $\epsilon_{{\rm an}}$ of the operator ${\rm Tr}\, F^2$, as was advocated in \cite{Gubser:2008ny,Gubser:2008yx}. We will, however, propose a relation between these two quantities.

We show in this work that for arbitrary values of $\epsilon$, in the range $ 0.001 < 
\epsilon < 0.1$, it is always possible to reproduce the spectrum of scalar and 
tensor glueballs obtained in lattice QCD. This is achieved by using the first 
two scalar glueball masses as a physical criterion to fix the source and VEV 
coefficients, $\phi_0$ and $G$, respectively, for each value of $\epsilon$. We 
find that the evolution of these parameters as functions of $\epsilon$ admits 
simple fits that allow us to predict their behaviour in the $\epsilon \to 0$ 
limit (where the dilaton becomes massless). On the other hand, from the analysis 
of the vacuum energy density $\langle T^{00} \rangle$ and the VEV of ${\cal O}$ 
we calculate the corresponding trace anomaly $\langle T^{\mu}_{\;\; \mu} \rangle 
 = - \epsilon \phi_0 \langle O \rangle$, which is consistent with the general 
result of deformed CFTs \cite{Skenderis:2002wp,Papadimitriou:2016yit}. We will 
suggest a reinterpretation of this result in terms of the QCD trace anomaly, 
which in turn will suggest a dictionary for the parameters $\phi_0$ and 
$\epsilon$. 

Our approach is quite different from the one considered in
\cite{Gursoy:2007cb,Gursoy:2007er}, where the 4-$d$ theory does not have a UV 
cut-off and the dilaton potential is built to reproduce the two-loop perturbative $\beta$ function of 
large-$N_c$ QCD (asymptotic freedom). It is not clear, however, how the 
holographic map between the energy $E$ and the warp factor $A(z)$, proposed in 
\cite{Gursoy:2007cb,Gursoy:2007er}, should be modified in our case in order to 
accommodate the energy cut-off $E^{*}$. In recent works 
\cite{Bourdier:2013axa,Megias:2014iwa}, a massive term for the dilaton in the UV 
was interpreted as the dual of a CFT deformation by a nearly marginal operator 
and the corresponding RG flow was investigated (see also 
\cite{Kiritsis:2016kog}). Our results, however, support the interpretation of 
the CFT deformation $\delta {\cal L} = \phi_{0}\, {\cal O}$ in terms of the 
large-$N_c$ Yang-Mills Lagrangian. In some sense this is a reinterpretation of 
the RG flow in
holographic QCD where now the operator ${\rm Tr}\,F^2$ is interpreted as the source of a nontrivial $\beta$-function. This may  also shed some light in the origin of 
the dilaton potentials considered in holographic QCD models without relying on 
a stringy top-down approach.

This paper is organized as follows. In section \ref{Sec:HQCD} we review the 
holographic QCD models arising from five-dimensional Dilaton-Gravity theory, 
with focus on the linearised equations that lead to the glueball spectrum. Then, 
in section \ref{Sec:AnomConf}, we describe the physical constraints in the UV 
and IR, compatible with explicit conformal symmetry breaking and linear 
confinement, and  present two interpolations where the dilaton admits an 
analytic form.  In section \ref{Sec:Spectrum}  we find a spectrum for scalar and 
tensor glueballs compatible with lattice QCD data and describe how the model
parameters evolve with the conformal dimension $\epsilon$. In section 
\ref{Sec:TraceAnomaly} we calculate the trace anomaly for our model, which 
agrees with the general expectation of deformed CFTs and compare this result 
with the QCD scale anomaly. We finish this paper with our conclusions and two 
appendices. Appendix \ref{App:massless} describes massless scalar modes in 
holographic QCD. In appendix \ref{App:ModelB} we present two holographic QCD 
models where the warp factor admits an analytic form.

\section{Review of holographic QCD}
\label{Sec:HQCD}

In this section we review the holographic QCD backgrounds (HQCD) in a bottom-up 
approach, which makes use of the field/operator correspondence unveiled by the 
AdS/CFT correspondence. In order to describe the large-$N_c$ QCD vacuum the 
focus is on the stress-energy tensor $T^{\mu \nu}$ as well as the Yang-Mills 
Lagrangian operator ${\rm Tr} \, F^2$.  The former couples to a five dimensional 
metric $g_{mn}$ and the latter couples to the scalar dilaton field $\Phi$. Then 
the natural five dimensional framework is the Dilaton-Gravity theory, where 
besides the Einstein-Hilbert action one considers a dilaton kinetic term and a 
dilaton potential. 

First we will exploit the fact that HQCD backgrounds are conformally flat and map the perturbed metric to linearised gravity around the Minkowski spacetime. Then we will briefly review the general features of the models proposed in \cite{Gursoy:2007cb,Gursoy:2007er}, also known as Improved Holographic QCD. We finish the section with a full description of the linearised Dilaton-Gravity equations leading to the glueball spectrum.

\subsection{Ricci tensor in HQCD backgrounds}

Consider a Weyl transformation for a 5-$d$ metric:
\beqa
g_{mn} = e^{2A(x)} \bar g_{mn} \, . \label{Weyltransf}
\eeqa
We take the transformation (\ref{Weyltransf}) as a field redefinition for the
metric, or as an ansatz for the background spacetime\footnote{This is in 
contrast with the Weyl transformations used in
String Theory where the conformal factor is also a field.}.
In holography the 5-$d$ coordinates $x^m$ of the bulk spacetime decompose as 
$(x^{\mu},z)$
where $x^{\mu}$ are the 4-$d$ coordinates associated with the field theory at 
the boundary and $z$ is the bulk radial coordinate.  

The metrics $g_{mn}$ and  $\bar g_{mn}$ include also the fluctuations around
the unperturbed background, and so they admit the expansions
\beq
\begin{aligned}
& \bar{g}_{mn}= \bar g_{mn}^{(0)} + \bar g_{mn}^{(1)} + \dots, \qquad  & 
\qquad
& g_{mn} = g_{mn}^{(0)} + g_{mn}^{(1)}+\dots,  
\end{aligned}
\eeq
where we take $\bar g_{mn}^{(0)}$ as the metric associated with a reference background. Analogously, the Christoffel symbols
$\Gamma^p_{mn}=\frac 12 g^{pq}(\partial_m g_{qn} + \partial_n g_{qm} - \partial_q g_{mn})$ and  the Ricci tensor $R_{mn}\equiv {R^p}_{mpn}= \partial_p \Gamma^p_{nm} - 
\partial_n \Gamma^p_{pm} + \Gamma^p_{pq}  \Gamma^q_{nm} - \Gamma^p_{nq} \Gamma^q_{pm} $ also admit the expansions
\beq
\begin{aligned}
& \bar{\Gamma}_{mn}^{p}= \bar{\Gamma}_{mn}^{p(0)}+\bar{\Gamma}_{mn}^{p(1)} +
\dots, \qquad &  \qquad 
& \Gamma_{mn}^{p}= {\Gamma}_{mn}^{p(0)}+{\Gamma}_{mn}^{p(1)} + \dots,  \cr
& \bar{R}_{mn}=\bar{R}_{mn}^{(0)} + \bar{R}_{mn}^{(1)} + \dots, \qquad & 
\qquad 
& R_{mn}= R_{mn}^{(0)} + R_{mn}^{(1)} + \dots .
\end{aligned}
\eeq
The Ricci tensor transforms under (\ref{Weyltransf}) as 
\beqa
R_{mn} &=& \bar R_{mn} - 3 \left [ \partial_m \partial_n A - \bar 
\Gamma_{mn}^p
\partial_p A \right ] + 3 \partial_m A \partial_n A \cr
&-& \bar{g}_{mn} \bar{g}^{pq} \left [ \partial_p \partial_q A - \bar 
\Gamma_{pq}^r
\partial_r A + 3 \partial_p A \partial_q A \right ] \,. \label{Riccitransf}
\eeqa 

In HQCD we are interested in the case where the reference background is flat,
i.e. 
\beqa
\bar g_{mn}^{(0)} = \eta_{mn}\,, \quad\quad 
 \bar {\Gamma}_{mn}^{p(0)} = 0\,, \quad \quad 
 \bar R_{mn}^{(0)} = 0 \, ,
\eeqa
and we take the warp factor $A$ as a function of the radial coordinate only, 
i.e. $A = A(z)$ so that 4-$d$ Poincar\'e symmetry, associated with the 
coordinates $x^\mu$, is preserved. 
Then at $0$th order the Ricci tensor of the dual metric takes the form 
\beqa
R_{mn}^{(0)} = - 3 \partial_m \partial_n A + 3
\partial_m A \partial_ n A - \eta_{mn} ( A'' + 3 A'^{\, 2}) \, , 
\eeqa
where $'$ means $d/dz$. Projecting out the Ricci tensor we obtain
the components
\beqa
R_{zz}^{(0)} = - 4 A''\, ,  \quad\quad 
R_{z\mu}^{(0)} = 0\, , \quad\quad 
R_{\mu \nu}^{(0)} = - \eta_{\mu \nu} ( A'' + 3 A'^{\,2}) \, , 
\label{RicciIHQCD}
\eeqa
and the $0$th order Ricci scalar $R \equiv g^{mn} R_{mn}$ takes the form 
\beqa
R^{(0)} = - e^{-2A} (8 A''+ 12 A'^{\,2}) \,. \label{RicciscalarIHQCD}
\eeqa
On the other hand, the reference metric at $1$st order is that of linearised
gravity around flat space, i.e. 
\beqa
\bar g_{mn}^{(1)} &=& h_{mn} \, , \cr
\bar \Gamma_{mn}^{p(1)} &=& \partial_{(m} h_{n)}^{\, \, \, p} -
\frac12 \partial^p h_{mn} \, , \cr
\bar R_{mn}^{(1)} &=&  \partial_p \partial_{(m} h_{n)}^{\, \, \, p} - \frac12
\partial_p \partial^p h_{mn} - \frac12 \partial_m \partial_n h_p^{\, \, \, p} \, 
,
\eeqa
where the parentheses around the indices denote symmetrization,
$V_{(mn)}\equiv(V_{mn}+V_{nm})/2$,
and the indices of $h_{mn}$ are raised (lowered) using the Minkowski metric
$\eta^{mn}$ ($\eta_{mn}$).
Expanding both sides of  (\ref{Riccitransf}), we find that the $1$st order 
perturbations of
the Ricci tensor can be written as
\beqa
R_{mn}^{(1)} &=& \bar R_{mn}^{(1)} + 3 A' \, \bar \Gamma_{mn}^{z(1)} \cr
&+& \left ( A'' + 3 A'^{\,2} \right ) \left ( h_{zz} \eta_{mn}  - h_{mn} \right )
+ A' \eta^{pq} \bar \Gamma_{pq}^{z(1)} \eta_{mn} \, . 
\label{Ricci1storder}
\eeqa
Decomposing the tensor of linearised gravity $h_{mn}$ as 
$(h_{zz},h_{z\mu},h_{\mu \nu})$ and defining
$h_{zz} \equiv 2 \phi$ and $h_{z \mu}\equiv \mathcal{A}_\mu$, we can project out the 
Ricci tensor
(\ref{Ricci1storder}) and find the following components:
\beqa
R_{zz}^{(1)} &=&  \left [ \partial_z + A' \right ]  ( \partial_\mu  
\mathcal{A}^{\mu} -
\frac12 h' ) + 4 A' \phi' - \Box \phi \, ; \cr
R_{z \mu}^{(1)} &=& \frac12 \partial_\nu \, {h'}_\mu^{\, \,  \nu} -
\frac12 \partial_\mu h' + \frac12 \partial_\nu \mathcal{F}_{\mu}^{\, \, \nu} +
3 A' \partial_\mu \phi  - (A'' + 3 A'^{\,2} )  \mathcal{A}_{\mu}  \,; \cr
R_{\mu \nu}^{(1)} &=& \left [ \partial_z + 3 A' \right ]
\partial_{(\mu}  \mathcal{A}_{\nu)} - \frac12 \left [ \partial_z^2 + 3 A' 
\partial_z  +
\Box \right ] h_{\mu \nu}  
+ \partial_\rho \partial_{( \mu } h_{\nu )}^{\, \, \rho} 
- \frac12 \partial_\mu \partial_\nu (2 \phi + h) \cr 
&+& \frac{1}{2} A' \left [ 2\phi'- h' + 2 \partial_\rho  
\mathcal{A}^{\rho}\right ]
\eta_{\mu \nu} + (A'' + 3 A'^{\, 2} ) \left [ 2 \phi \,  \eta_{\mu \nu} -
h_{\mu \nu} \right ] \,; 
\label{Ricci1storderv2}
\eeqa
where $\Box\equiv\partial_{\mu}\partial^{\mu}$ is the d'Alembertian operator in 
the boundary spacetime, the scalar $h$ is defined by the trace $h \equiv h^{\mu}_{\, 
\, \mu}$ and $\mathcal{F}_{\mu \nu} \equiv \partial_\mu  \mathcal{A}_{\nu} -
\partial_\nu  \mathcal{A}_{\mu}$ is the field strength associated with the vector
$\mathcal{A}_{\mu}$.

\subsection{HQCD backgrounds from Dilaton-Gravity}
\label{Sec:Dilaton-Gravity}

The goal of holographic QCD is to find the gravity (string) dual of QCD in the 
large-$N_c$ limit. This is motivated by the 't Hooft planar limit 
\cite{tHooft:1973alw} and the AdS/CFT correspondence 
\cite{Maldacena:1997re,Gubser:1998bc,Witten:1998qj}. QCD is, on the one hand, 
well approximated by a CFT in the UV and, on the other hand, confining in the 
IR. These facts suggest that the holographic dual spacetime should be AdS near 
the boundary. Far from the boundary, it should be such that dual probe fields 
(and strings) living in that 5-$d$ background reproduce confinement and the 
4-$d$ hadronic physics. In the pioneer work of 
\cite{Gursoy:2007cb,Gursoy:2007er}, a very general family of HQCD backgrounds 
was proposed and, based on the work of \cite{Kinar:1998vq}, a universal IR 
criterion for confinement was found. Moreover, the requirement of linear Regge 
trajectories led the authors of \cite{Gursoy:2007er} to  conclude that a 
quadratic dilaton field was necessary in the IR. This supports the early work of 
\cite{Karch:2006pv}, where a quadratic dilaton was proposed on the basis of the 
meson spectrum, the so-called soft-wall model.  

In this subsection, we briefly review the HQCD backgrounds proposed in \cite{Gursoy:2007cb,Gursoy:2007er} focusing on the IR physics. In the next subsection, we will describe the scalar and gravitational perturbations that lead to the Schr\"{o}dinger equations associated with the glueball spectrum. 

In the HQCD approach, we start with a 5-$d$ Dilaton-Gravity action of the 
form 
\beqa
S = M_{p}^3 N_c^2 \int d^{5} x \sqrt{-g} \left [ R + {\cal L}_\Phi \right ] \, 
,
\label{DilatonGravityAction}
\eeqa
where $M_p$ is the 5-$d$ Planck scale, $N_c$ is the number of colors, 
and the
dilaton Lagrangian has a kinetic term and a potential,
\beqa
{\cal L}_\Phi = - \frac43 g^{mn} \partial_m \Phi \partial_n \Phi + V(\Phi) 
\,. 
\eeqa
Variating the action (\ref{DilatonGravityAction}) with respect to $\Phi$ and the metric $g_{mn}$, we obtain the Dilaton-Gravity equations 
\beq
R_{mn} - \frac{1}{2} g_{mn}R =\frac{1}{2M_{p}^{3}N_{c}^{2}}\,T_{mn}\,, 
\label{Einsteineqs}
\eeq
\beq
\frac43 \nabla^2 \Phi + \frac12 \frac{d V}{d \Phi} = 0 \,, \label{KGeq}  
\eeq
where $\nabla^2$ is the Laplacian operator\footnote{The Laplacian 
operator $\nabla^{2}$ applied to a  scalar function $f$ is given by
$
\nabla^{2}f=\frac{1}{\sqrt{-g}}\partial_{m}\left(\sqrt{-g}\,g^{mn}\,\partial_{
n}f\right)
$.}
and we have defined an energy-momentum tensor $T_{mn}$ for the 
dilaton field:
\beq
T_{mn}\equiv M_{p}^{3}N_{c}^{2}\left[\frac{8}{3}\partial_{m}\Phi\partial_{n}\Phi+g_
{mn}\,{\cal L}_\Phi\right].
\eeq
It is also convenient to write the Einstein equations (\ref{Einsteineqs}) in the
Ricci form:
\beqa
R_{mn} = \frac43 \partial_m \Phi \partial_n \Phi - \frac13 g_{mn} V\,.
\label{RicciFormEqs} 
\eeqa

The HQCD backgrounds correspond to solutions for the Dilaton-Gravity 
equations
of the form
\beq
ds^2=e^{2A(z)} \left [ dz^2 + \eta_{\mu \nu} dx^\mu dx^\nu \right ],\qquad
\Phi=\Phi(z) \,. 
\eeq
The warp factor $A(z)$ and dilaton $\Phi(z)$ are usually mapped to the energy scale and coupling of the dual 4-$d$ theory. Using (\ref{RicciIHQCD}) and the definition of the scalar Laplacian 
$\nabla^2$, the
Dilaton-Gravity equations (\ref{KGeq}) and (\ref{RicciFormEqs}) take the form 
\beqa
12 A'' + 4 \Phi'^{\, 2} = e^{2A} V \,, \cr  
3 A'' + 9  A'^{\, 2}  = e^{2A} V  \, , \cr
\frac83 \left [ \partial_z + 3 A' \right ] \Phi' = - e^{2A} \frac{d V}{d \Phi}
\,. \label{IHQCDEqs}
\eeqa
The last equation in (\ref{IHQCDEqs}) can be obtained from the first two, which 
in turn  can be rewritten as 
\beqa
A'^{\, 2} - A'' = \frac49 \Phi'^{\, 2} \,, \cr
3 A'^{\, 2} + A'' = \frac{1}{3}e^{2A}V \, .  \label{IHQCDEqsv2}
\eeqa
At this point it is very convenient to define the quantity 
\beqa
\zeta(z) \equiv \exp[-A(z)] \, , \label{zetadef}
\eeqa
so that the first equation in (\ref{IHQCDEqsv2}) takes a linear form in $\zeta$:
\beqa
\zeta'' - \frac49 \Phi'^{\, 2} \zeta  = 0 \,. \label{zetaEq}
\eeqa
The Schr\"{o}dinger form of this equation is useful to understand the AdS 
deformation due to a non-constant dilaton, which is the dual of a conformal 
symmetry breaking. In the case of a constant dilaton, the interesting solution
for the holography is $\zeta = z/\ell $, corresponding to the usual AdS 
spacetime with curvature radius $\ell$. 
In this paper we will describe how the presence of dilaton mass, associated with 
a CFT deformation, leads to an explicit breaking of conformal symmetry and gives 
a reasonable glueball spectrum. Note from equation (\ref{zetaEq}) that $\zeta'' 
\geq 0$ for all values of $z$. AdS asymptotics implies that
$\zeta'(0)=1/\ell$ so we conclude that  $\zeta' \geq 1/\ell\,$.\footnote{This
is equivalent to the statement $\partial_u A \leq - 1/\ell$, obtained in
\cite{Gursoy:2007er}, where $u$ is the domain-wall coordinate related to
$z$ by $dz = \zeta du$.}  

Following ref. \cite{Gursoy:2007er} we write the equations (\ref{IHQCDEqsv2}) as a system of first order differential equations  
\beqa
\zeta \Phi' &=& \frac{dW}{d\Phi}\,, \quad \quad
\zeta' = \frac49 W \, ,  \cr
V &=& - \frac43 \left ( \frac{dW}{d\Phi} \right )^2 + \frac{64}{27} W^2 \, , 
\label{superpotEq}
\eeqa
where $W=W(\Phi)$ is the superpotential associated with the Dilaton-Gravity dynamics. Another useful quantity is the field $X$ defined by the relations 
\beqa
X &\equiv& \frac13 \frac{d \Phi}{d A} = - \frac13 \frac{ \zeta \Phi'}{\zeta'} =
- \frac34 \frac{d \log W}{d \Phi} \,. 
\label{xEq}
\eeqa
This field can be interpreted as a bulk beta function $X \sim \beta_\Phi$ describing the evolution of the dilaton $\Phi$ with the warp factor $A$. In the next subsection we will describe how this field appears in the Schr\"{o}dinger equation associated with scalar glueballs. In \cite{Gursoy:2007cb,Gursoy:2007er} the authors proposed a dictionary that maps the bulk field $X$ to the $\beta$ function of the 4-$d$ dual theory. In our work
the CFT deformation in the UV implies the existence of a cut-off $E^{*}$ in the energy scale of the 4-$d$ theory. This feature suggests a departure from the map proposed in\cite{Gursoy:2007cb,Gursoy:2007er} so we take $X$ as a pure bulk field. In particular, we will see that while $X$ goes to zero as we approach the boundary the dual $\beta$ function is still finite. 

We finish this subsection describing the confining constraint found in \cite{Gursoy:2007er}. The discussion takes place in the string frame, where the metric is given by 
\beqa
ds^2 &=& e^{2A_s(z)} \left [ dz^2 + \eta_{\mu \nu} dx^\mu dx^\nu \right ] \, ,
\label{MetricSF}
\eeqa
and the string-frame warp factor is related to the Einstein-frame warp factor by 
\beqa
A_s(z) = A(z) + \frac23 \Phi(z) \,. 
\label{WarpFactorSF}
\eeqa
Consider a static string living in the spacetime (\ref{MetricSF}), with endpoints attached to the boundary and separated by a distance $L$ in one of the boundary directions. As shown in \cite{Kinar:1998vq}, inspired by \cite{Maldacena:1998im}, this problem maps to a rectangular Wilson loop
describing in the large $L$ limit the potential energy of a heavy quark-antiquark pair. Solving the Nambu-Goto equations one finds that in the large $L$ limit the energy of the static string takes the form 
\beqa
E(L) = \mu  f(z_*) L + \dots \, , \label{StringEnergy}
\eeqa
where $\mu$ is the fundamental string tension, $f(z) = \exp (2 A_s)$,   and $z_*$ is the point where $f(z)$ has a minimum and the dots represent subleading terms in the large $L$ limit. Following ref. \cite{Kinar:1998vq}, the energy 
(\ref{StringEnergy}) maps to the quark-antiquark potential and confinement is achieved for  $f(z_*)>0$.
In this case, the quantity $\mu f(z_*)$ is identified with the confining string tension $\sigma$. 

Thus we conclude that confining backgrounds are those where the function $f(z) = 
\exp (2 A_s)$ has a non-zero minimum. Since we always consider backgrounds that 
are asymptotically AdS we have that $f(z) \to \infty$ in the UV $(z\to 0)$. We 
are interested in backgrounds where $0 < z < \infty$ 
so we conclude that in the IR $f(z \to \infty) > 0 $ to guarantee confinement. 
Taking a power
ansatz  for the dilaton $\Phi (z) = z^{\alpha}$ we find from 
(\ref{IHQCDEqsv2})
that at large $z$ 
\beqa
A(z) = - \frac23 \Phi (z) + \frac12 \log | \Phi' (z) | + \dots \, , 
\label{IRWarpFactor}
\eeqa
so that 
\beqa
A_s(z) = \frac{\alpha-1}{2} \log z + \dots \, .
\eeqa
Then the confinement criterion becomes the condition $\alpha \ge 1$. From (\ref{RicciscalarIHQCD}) and (\ref{IRWarpFactor}) we find that the condition $\alpha \ge 1$ implies the existence of a curvature singularity at $z \to \infty$. Interestingly, a WKB analysis of the glueball spectrum \cite{Gursoy:2007er} leads to the stronger restriction $\alpha =2$ that 
corresponds to asymptotically linear Regge trajectories $m_n^2 \sim n$. In this work we will take a quadratic dilaton $\Phi(z) = z^2$ in the IR to guarantee confinement and an approximate linear glueball spectrum. The UV, on 
the other hand, will differ significantly from the proposal of 
\cite{Gursoy:2007cb,Gursoy:2007er} where instead of imposing asymptotic freedom we will consider a CFT deformation, inspired by the work of \cite{Gubser:2008ny,Gubser:2008yx}. But first we will finish this section by reviewing below how the Schr\"{o}dinger equations, that determine the glueball spectrum, arise from the linearised Dilaton-Gravity equations.

\subsection{Linearised Dilaton-Gravity equations}
\label{SubSec:LinDilGrav}

The linearised version of the Dilaton-Gravity equations are obtained by
expanding at first order both sides of (\ref{KGeq}) and (\ref{RicciFormEqs}),
with $\Phi \rightarrow \Phi + \chi$ and $g_{mn}\rightarrow 
e^{2A}(\eta_{mn}+h_{mn})$,
where $\chi$ and $h_{mn}$ are first-order perturbations in the dilaton and
the reference background metric, respectively. The resulting equations take 
the form 
\beq
R_{mn}^{(1)} =\frac83 \partial_{(m} \Phi \,\partial_{n)} \chi -
\frac13 e^{2A} \left[ V h_{mn} +  \left(\partial_{\Phi} V\right)
 \chi \, \eta_{mn}  \right],
\label{EinsteinLinear}
\eeq
\beq
\frac43 (\nabla^2 \Phi)^{(1)} = - \frac12 \left(\partial_{\Phi}^2 V\right)\chi 
\, ,
\label{KGLinear} 
\eeq
where $R_{mn}^{(1)}$ is given by (\ref{Ricci1storderv2}),
\beq
(\nabla^2 \Phi)^{(1)} = e^{-2A} \Big \{ \left [ \partial_z^2 +
3 A' \partial_z + \Box \right ] \chi   - \Big  [ \phi' + \partial_{\mu} 
\mathcal{A}^{\mu}
- \frac12 h' + 2 \phi ( \partial_z + 3 A') \Big  ] \Phi' \Big \} \, ,
\eeq
and recall that $h_{zz} = 2 \phi$, $h_{z \mu} = \mathcal{A}_{\mu}$ and
$h^{\mu}_{\, \, \mu} = h$. Taking the components $(zz , z \mu , \mu \nu)$ of
the linearised Einstein equations (\ref{EinsteinLinear}),
the system (\ref{EinsteinLinear})-(\ref{KGLinear}) becomes 
\begin{align}
\left [ \partial_z + A' \right ] ( \partial_{\mu} \mathcal{A}^{\mu} - \frac12 
h' ) +
4 A' \phi' - \Box \phi - \frac83 \Phi' \chi' &   \cr 
+ 2 \phi (A'' + 3 A'^{\, 2}) - \frac89 \chi \left [ \partial_z + 3 A' \right ]
\Phi' &= 0 \, , \label{LinEinstzz} \\
\frac12 \partial_{\nu} {h'}_{\mu}^{\, \, \nu} - \frac12  \partial_{\mu} h' +
\frac12 \partial_{\nu} \mathcal{F}_{\mu}^{\, \, \nu} + 3 A' \partial_{\mu} 
\phi -
\frac43 \Phi' \partial_{\mu} \chi &= 0 \,, \label{LinEinstzmu} \\
 \left [ \partial_z + 3 A' \right ] \partial_{(\mu} \mathcal{A}_{\nu)} - 
\frac12
\left [ \partial_z^2 + 3 A' \partial_z  + \Box \right ] h_{\mu \nu}  
+ \partial_\rho \partial_{( \mu } h_{\nu )}^{\, \, \rho} 
- \frac12 \partial_\mu \partial_\nu (2 \phi + h) & \cr 
+  A' \left [ \phi' + \partial_\rho \mathcal{A}^{\rho} - \frac12 h' \right ]
\eta_{\mu \nu} + 2 (A'' + 3 A'^{\, 2})  \phi \,  \eta_{\mu \nu} - \frac89 \chi
\left [ \partial_z + 3 A' \right ] \Phi' \eta_{\mu \nu} &= 0  ,
\label{LinEinstmunu} \\
\left [ \partial_z^2 + 3 A' \partial_z + \Box \right ] \chi   -
\Big  [ \phi' + \partial_{\mu} \mathcal{A}^{\mu} - \frac12 h' + 2 \phi
( \partial_z + 3 A') \Big  ] \Phi' + \frac38 e^{2A} (\partial_{\Phi}^2 V)
\chi &= 0  \, , 
\label{LinKG}
\end{align}
where we have used the result (\ref{Ricci1storderv2}) for the Ricci tensor and we have also used the following background relations 
\beqa
\frac13 e^{2A} V =  A'' + 3 A'^{\,2} \,, \quad\quad 
\frac13 e^{2A} \partial_{\Phi} V = - \frac89
\left [  \partial_z + 3 A' \right ] \Phi' \, . 
\label{background_rel2}
\eeqa
As explained in \cite{Kiritsis:2006ua}, the next step is to decompose the four-vector
$\mathcal{A}_{\mu}$ and the symmetric tensor $h_{\mu \nu}$ into
irreducible representations of the Lorentz group, i.e, 
\beqa
\mathcal{A}_{\mu} &=& \mathcal{A}_{\mu}^{\scriptscriptstyle{T}} + 
\partial_{\mu} \mathcal{W}  \, , \cr
h_{\mu \nu} &=& h_{\mu \nu}^{\scriptscriptstyle{TT}} + 2 \partial_{ ( \mu} 
\mathcal{V}_{\nu )}^{\scriptscriptstyle{T}}
+ 2\partial_{\mu} \partial_{\nu} \mathcal{E} + 2 \psi \eta_{\mu \nu} \,,
\label{LorentzDecomp} 
\eeqa
where $\mathcal{A}_{\mu}^{\scriptscriptstyle{T}}$ and 
$\mathcal{V}_{\mu}^{\scriptscriptstyle{T}}$ are
divergenceless vectors, $h_{\mu \nu}^{\scriptscriptstyle{TT}}$ is a traceless and divergenceless tensor and $\mathcal{W}$, $\mathcal{E}$, $\psi$ are Lorentz-scalars.
Applying the decomposition (\ref{LorentzDecomp}) into the Dilaton-Gravity equations (\ref{LinEinstzz})-(\ref{LinKG}),
we find one tensorial equation  
\beqa
\left [ \partial_z^2 + 3 A' \partial_z + \Box \right ] h_{\mu 
\nu}^{\scriptscriptstyle{TT}} = 0 \,,
\label{Spin2eq}
\eeqa
two vectorial equations 
\begin{align}
\left [\partial_z + 3 A'  \right ] \left ( 
\mathcal{A}_{\mu}^{\scriptscriptstyle{T}} - 
\mathcal{V}_{\mu}^{\scriptscriptstyle{T}\,'} \right )
&= 0 \, , \label{Spin1eq1} \\
\Box  \left ( \mathcal{A}_{\mu}^{\scriptscriptstyle{T}} - 
\mathcal{V}_{\mu}^{\scriptscriptstyle{T}\,'} \right ) &= 0 \, , 
\label{Spin1eq2}
\end{align}
and five scalar equations
\begin{align}
\left [ \partial_z + 3 A' \right ] \left (\mathcal{W} - \mathcal{E}' \right ) 
- 2 \psi - \phi &= 0
\, , \label{Spin0eq1}\\
- \left [ \partial_z^2 + 3 A' \partial_z + \Box \right ] \psi + A'
\left [ \phi' - 4 \psi' + \Box (\mathcal{W} - \mathcal{E}') \right ] & \cr
+ 2 \phi (A'' + 3 A'^{\, 2}) - \frac89 \chi \left [ \partial_z + 3 A' \right ]
\Phi' &= 0 \, , \label{Spin0eq2} \\
- 3 \psi' + 3 A' \phi - \frac43 \Phi' \chi &= 0 \, , \label{Spin0eq3}
\end{align}
\begin{align}
\left [ \partial_z + A' \right ] \left [ \Box (\mathcal{W} - \mathcal{E}') - 4 
\psi' \right ] +
4 A' \phi' - \Box \phi & \cr
- \frac83 \Phi' \chi' + 2 \phi (A'' + 3 A'^{\, 2}) - \frac89 \chi
\left [ \partial_z + 3 A' \right ] \Phi' &= 0 \, , \label{Spin0eq4} \\
\left [ \partial_z^2 + 3 A' \partial_z + \Box \right ] \chi 
- \Big  [ \phi' + \Box (\mathcal{W}- \mathcal{E}')- 4 \psi' + 2 \phi  \left ( 
\partial_z +
3 A'  \right ) \Big ] \Phi' & \cr 
+ \frac38 e^{2A} \left(\partial_{\Phi}^2 V\right)\chi &= 0 \,. 
\label{Spin0eq5}
\end{align}
The tensorial equation (\ref{Spin2eq}) leads to the spectrum of spin 2 glueballs. The vectorial equations (\ref{Spin1eq1})-(\ref{Spin1eq2}) do not lead to normalizable modes so we can set 
$\mathcal{A}_{\mu}^{\scriptscriptstyle{T}}
= \mathcal{V}_{\mu}^{\scriptscriptstyle{T}}=0$. 
From the five scalar equations (\ref{Spin0eq1})-(\ref{Spin0eq5}) only one combination decouples from the rest and describes the spectrum of spin 0 glueballs. Below we describe how this equation can be obtained. Subtracting (\ref{Spin0eq2}) from (\ref{Spin0eq4}) and using (\ref{Spin0eq1}),
we obtain the equation 
\beqa
\left [ \partial_z^2 - A' \partial_z - \Box \right ] \psi - A' \phi' +
A' \Box (\mathcal{W} -\mathcal{E}') + \frac89 \Phi' \chi' = 0 \,. 
\label{Spin0auxeq}
\eeqa
This equation can be combined with (\ref{Spin0eq5}) to get rid of the term $\Box (\mathcal{W} - \mathcal{E}')$ and to find 
\begin{align}
\left [ \partial_z^2 + 3 A' \partial_z - \Box \right ] \psi + \frac{1}{3 X}
\left [ \partial_z^2 + 3 A' \partial_z + \Box \right ] \chi & \cr
- 2 A' \phi' - 2 \frac{\phi}{X} \left [ \partial_z + 3A' \right ] (A'X) 
+ \frac83 A' X \chi' + \frac{1}{8X} e^{2A} \left(\partial_{\Phi}^2 
V\right)\chi &= 0 \,, 
\end{align}
where $X$ was defined in \eqref{xEq}. Using equation (\ref{Spin0eq3}) to replace 
$\phi$ in terms of $\psi$ and $\chi$ and the background relations 
(\ref{IHQCDEqsv2}) and (\ref{background_rel2}), we arrive at the decoupled 
equation 
\beqa
\xi'' + \left ( 3 A' + 2 \frac{X'}{X} \right ) \xi' + \Box \xi = 0 \, , 
\label{Spin0DecEq}
\eeqa
where the field $\xi$ is defined by 
\beqa
\xi = \psi - \frac{\chi}{3X} \,. 
\eeqa
The solutions of equation (\ref{Spin0DecEq}) lead to the spectrum of spin 0 
glueballs.

\subsubsection{Schr\"{o}dinger like equation - scalar sector}
\label{schroscaeq}
It is possible to rewrite equation  (\ref{Spin0DecEq}) in a
Schr\"odinger like form.  To do so, we define an auxiliary 
function $B_{s}$ as
\begin{equation}\label{Bsca}
B_{s}=\frac{3}{2}A+ \log X.
\end{equation}
Substituting $\xi=e^{-B_{s}}\psi_{s}$ and introducing the Fourier transform
($\Box\rightarrow m_{s}^2$) in equation  (\ref{Spin0DecEq}), we get
 \begin{equation}\label{schrodingerscaeq}
 -\psi_{s}'' +V_{s}\,\psi_{s}=m_{s}^{2}\psi_{s},
 \end{equation}
where the potential is defined as
 \begin{equation}\label{potentialsca}
 V_{s} =\left(B_{s}'\right)^{2}+B_{s}''.
 \end{equation}

\subsubsection{Schr\"{o}dinger like equation - tensor sector}
\label{schroteneq}
Following the same procedure as above in the tensor 
sector we use the auxiliary function 
 \begin{equation}\label{Bten}
 B_{t}=\frac{3}{2}A,
 \end{equation}
to rewrite (\ref{Spin2eq}) in the Schr\"{o}dinger like form
 \begin{equation}\label{schrodingerteneq}
 -\psi_{t}''+V_{t}\,\psi_{t}=m_{t}^{2}\psi_{t},
 \end{equation}
 where the potential is defined as
 \begin{equation}\label{potentialtens}
 V_{t}=\left(B_{t}'\right)^{2}+B_{t}''.
 \end{equation}
Interestingly, the difference between the spin $0$ and spin $2$ sectors lies in the term $B_{s}-B_{t}=\log(X)$. This is an important feature when calculating the glueball spectrum that explains the non-degeneracy of scalar and tensor glueballs.

 \section{Effective holographic QCD}
 \label{Sec:AnomConf}

As anticipated in the previous sections, our work will explore the idea of a massive term for the dilaton in the UV as the dual of a CFT deformation $\delta {\cal L} = \phi_{0}\, {\cal O}$. The coupling $\phi_0$ and the  operator ${\cal O}$ have conformal dimensions $\epsilon$ and $4-\epsilon$, respectively. The latter  will be related to the QCD operator ${\rm Tr} \, F^2$. The idea of a CFT relevant deformation was proposed in \cite{Gubser:2008ny,Gubser:2008yx} when constructing a holographic model for QCD at finite temperature. Since the CFT deformation takes place at a particular energy scale $E^{*}$, which becomes an upper energy bound for the 4-$d$ theory, we dub this approach effective holographic QCD. 

In this work we will show that this type of UV asymptotics, at zero temperature, is compatible with (explicit) conformal symmetry breaking, confinement and the glueball spectrum. To achieve this we also constrain the IR in the way proposed in \cite{Gursoy:2007cb,Gursoy:2007er}, namely by considering a quadratic dilaton. Below we describe in more detail the universal UV and IR asymptotic behaviour for the family of HQCD backgrounds considered in this work. We also describe specific models that interpolate smoothly between the UV and IR asymptotics. Then in the next section we will calculate the glueball spectrum and investigate how the model parameters evolve with the conformal dimension $\epsilon$.
 
 \subsection{CFT deformation in the UV}\label{secanoma}
In the pioneer work \cite{Csaki:2006ji}, Csaki and Reece identified a dynamical dilaton as the five-dimensional scalar field dual to the operator ${\rm Tr}\, F^2$. They 
initially considered a massless dilaton and obtained a nonzero gluon condensate. However, as found in \cite{Csaki:2006ji}, a massless dilaton leads to a Nambu-Goldstone boson in the spectrum of scalar glueballs, indicating that conformal symmetry was spontaneously broken. It is important to remark, however, that the background of \cite{Csaki:2006ji} requires additional boundary conditions at the singularity, which brings some ambiguities \cite{Gursoy:2007er}. As a mechanism for explicit conformal symmetry breaking, the authors of \cite{Csaki:2006ji} proposed the holographic implementation of asymptotic freedom in the UV. This was correctly implemented in the IHQCD background developed by Gursoy, Kiritsis and Nitti \cite{Gursoy:2007cb,Gursoy:2007er}. The background of \cite{Gursoy:2007cb,Gursoy:2007er} is also consistent with linear confinement in the IR and does not require additional boundary conditions at the singularity. 

As explained above, instead of implementing asymptotic freedom in the UV, we are interested in explicitly breaking conformal invariance through a UV mass term for the dilaton. According to the AdS/CFT dictionary, the mass $M$ of a 5-$d$ dilaton, responsible for deforming AdS space, is related to the conformal dimension $\Delta$ of a dual scalar operator ${\cal O}$, responsible for deforming the 4-$d$ CFT, by the equation $M^2 = \Delta (\Delta -4)$. Then by considering a relevant operator with dimension  $\Delta = 4 - \epsilon$, we end up with a nonzero mass for the dilaton. We will later relate this operator to the QCD operator ${\rm Tr} \, F^2$ and investigate the connection between the trace anomaly of deformed CFTs and the QCD trace anomaly. This in turn will shed some light on the role played by the conformal dimension $\epsilon$ in effective holographic QCD. 

Let us first consider an expansion of the dilaton potential around the UV minimum $\Phi=0$, which includes a constant term, associated with a negative cosmological contant,
and a nonzero mass term: 
\begin{equation}\label{asymptotics1}
V(\Phi)=12-\frac{4}{3}\,M^{2} \Phi^2 + \dots \,,
\end{equation}
where the ellipses denote higher powers of $\Phi$.
On basis of equations \eqref{IHQCDEqs}, one finds that the constant term in 
\eqref{asymptotics1} leads to the AdS asymptotics, with an AdS radius $\ell=1$, 
whereas the mass term implies the following near boundary behaviour for the 
dilaton:
\beqa\label{asympdila}
\Phi = \phi_{0}\, z^{\Delta_{-}}+ G \,z^{\Delta_{+}} + \dots\,,
\eeqa
where $\Delta_{+}= \Delta = 4 - \epsilon$ and $\Delta_{-}=4-\Delta_{+}= \epsilon$,
both related to the 5-$d$ mass by $M^2 = - \Delta_{+} \Delta_{-}$. Following the AdS/CFT dictionary, we will interpret the coefficient $\phi_0$ as the source for the dual operator ${\cal O}$. The coefficient $G$ will be related to the VEV  of ${\cal O}$. We also remind the reader that the CFT deformation $\delta {\cal L}= \phi_{0}\, {\cal O}$ takes place at some cut-off energy $E^{*}$. 

Considering the perturbative expansion for $\Phi$, given by (\ref{asympdila}), 
we find from equation (\ref{zetaEq}) the asymptotic expansions for $\zeta(z)$ 
and the warp factor
$A(z)=-\log \zeta(z)$: 
\beqa
\label{asymptotics2}
\zeta(z)&=&
{z}\left[1+
\frac{2 \Delta_{-}}{9(1+2\Delta_{-})}\phi_{0}^{2} z^{2\Delta_{-}}+
\frac{2 \Delta_{-}\,\Delta_{+}}{45} \phi_{0} G z^4 +
\frac{2\Delta_{+}}{9(1+2\Delta_{+})} G^{2} z^{2\Delta_{+}}
+\dots\right], \cr 
A(z)&=&
-\log{z}-
\frac{2\Delta_{-}}{9(1+2\Delta_{-})}\phi_{0}^{2} z^{2\Delta_{-}}-
\frac{2\Delta_{-}\Delta_{+}}{45}\phi_{0}G z^4-
\frac{2 \Delta_{+}}{9(1+2\Delta_{+})}G^{2} z^{2\Delta_{+}}
-\dots\,.
\eeqa
If instead of a mass term for the dilaton, we followed the prescription of \cite{Gursoy:2007cb,Gursoy:2007er} and imposed asymptotic freedom, we would obtain
a UV asymptotics involving logarithmic terms to be consistent with the logarithmic
dependence of the 't Hooft coupling with the energy.

Since we know the asymptotic behaviour of the warp factor
(\ref{asymptotics2}) we are also able to find, from the second equation in 
(\ref{superpotEq}), the small $z$ behaviour of the superpotential,  
\begin{equation}\label{eqsuperpot2}
W(z)=\frac{9}{4}
+\frac{\Delta_{-}}{2} \phi_{0}^{2} 
z^{2 \Delta_{-}}
+\frac{\Delta_{-}\Delta_{+}}{2} \phi_{0} G
z^{4}
+\frac{\Delta_{+}}{2} G^{2}
z^{2\Delta_{+}}+\dots \, . 
\end{equation}
Alternatively, we can solve the differential equation (\ref{superpotEq}) for the 
superpotential $W(\Phi)$ for a dilaton potential $V(\Phi)$ given by 
(\ref{asymptotics1}) and find \cite{Bourdier:2013axa,Kiritsis:2016kog} 
\begin{equation}\label{eqsuperpot2p5}
W_{\pm}(\Phi)=\frac{9}{4}+\frac{\Delta_{\pm}}{2}\, \Phi^{2}+\dots \,,
\end{equation}
which is in agreement with equation (\ref{eqsuperpot2}). From the above results 
it is easy to find the asymptotic expansion for the field $X(z)$, defined in 
(\ref{xEq}). The expansion takes the form 
\begin{equation}
3 X(z)= - \Delta_{-} \phi_{0} z^{\Delta_{-}}
-\Delta_{+} G z^{\Delta_{+}} + \dots.
\end{equation}
The results above also tell us how the metric behaves near the boundary,
\begin{equation}
ds^{2}=\frac{1}{z^{2}}\, \left(1-
\frac{4 \Delta_{-}}{9(1+2\Delta_{-})}\phi_{0}^{2}z^{2\Delta_{-}}-
\frac{4 \Delta_{-} \Delta_{+}}{45} \phi_{0}G z^{4}
-\dots\right)\left [ dz^2 + \eta_{\mu \nu} dx^{\mu}dx^{\nu}\right] .
\end{equation}
We finish this subsection writing the asymptotic 
expansion of the dilaton (\ref{asympdila}) in a form that will be useful when implementing the numerical procedure,
\begin{equation}\label{dilaUV}
 \Phi(z)= \hat \phi_0 \, (\Lambda z)^{\epsilon}+(\Lambda z)^{4 - \epsilon}+\cdots\,,
 \end{equation}
where the parameters $\hat \phi_0$ and $\Lambda$ are related to $\phi_0$ and $G$ by 
\beq
\phi_0 = \hat \phi_{0}\, \Lambda^{\epsilon}, \quad \quad 
G = \Lambda^{4-\epsilon} \,. \label{DefParameters}
\eeq
The parameter $\Lambda$ has conformal dimension $1$ and plays a role similar to $\Lambda_{QCD}$ whereas the parameter $\hat \phi_0$ is the massless version of the coupling $\phi_0$. 
For all practical purposes $\hat \phi_0$, $\Lambda$ and $\epsilon$ will be the relevant parameters of the model. We will see later that, for fixed $\epsilon$, the parameters $\hat \phi_0$ and $\Lambda$
can be fit in order to reproduce the glueball spectrum.  

\subsection{Confinement in the IR} \label{confinement}
In ref. \cite{Gursoy:2007er}, Gursoy, Kiritsis and Nitti did a careful analysis that is
universal for holographic QCD backgrounds arising from a Dilaton-Gravity theory.
Specifically, considering the general confinement criterion of \cite{Kinar:1998vq} and also a WKB analysis
for the glueball spectrum, they found that a quadratic dilaton in the IR  guarantees confinement and an
approximate linear spectrum for glueballs. Remarkably, this quadratic dilaton had been already proposed
in the phenomenological soft-wall model \cite{Karch:2006pv}, in order to arrive at a linear spectrum for mesons. Interestingly, a quadratic dilaton in the IR also provides the $T^2$ correction to the stress tensor trace anomaly of a deconfined plasma \cite{Caselle:2011mn}. 
 
Motivated by the results of \cite{Gursoy:2007er}, we consider in this work the following dilaton asymptotic behaviour at large $z$:
\begin{equation}\label{dilaIR}
\Phi(z)=C\,z^{2} + \dots\,, 
\end{equation}
where the dots indicate terms that depend on negative powers of $z$. Using the IR asymptotic relation \eqref{IRWarpFactor} between $A(z)$ and $\Phi(z)$,
we construct the  asymptotic forms for the warp factor $A(z)$ and the function $\zeta(z)$:
\beqa\label{eqsuperpot3}
A(z)=-\frac{2}{3}C\,z^{2}+\frac{1}{2}\log{z}+\dots \, , \quad \quad
\zeta(z)=\frac{1}{\sqrt{z}} \exp  \left (\frac23 C z^{2} \right ) +\dots\,.
\eeqa
With this information at hand, we use the second equation in \eqref{superpotEq} to write down the superpotential,
\begin{equation}\label{eqsuperpot4}
W(z)=3\,C\,\sqrt{z} \exp  \left ( \frac23 C z^{2} \right ) +\dots\, .
\end{equation}
We get the asymptotic expansion
of $X$ using equations \eqref{dilaIR} and \eqref{eqsuperpot3}
and the definition \eqref{xEq}: 
\begin{equation}\label{xEq_new}
X=-\frac{1}{2}\left[1+\frac{3}{8\,Cz^2}+\dots\right]=
-\frac{1}{2}\left[1+\frac{3}{8}\frac{1}{\Phi}+\dots\right].
\end{equation}
Substituting this expression for $X(\Phi)$ in \eqref{xEq} and integrating the resulting equation
in $\Phi$, we obtain the asymptotic behaviour for the superpotential in terms of $\Phi$,
\begin{equation}\label{eqsuperpot5}
W(\Phi) \propto  
\Phi^{1/4}\, \exp \left ({\frac23 \Phi} \right ) +\dots\,, 
\end{equation} 
which is consistent with \eqref{eqsuperpot4}. Similarly, the asymptotic expression for the dilaton potential in the radial coordinate $z$
can be found from equations (\ref{IHQCDEqsv2}) and (\ref{eqsuperpot3}),
\beqa\label{eqdilapotIR}
V(z)=  8\,C z\, \exp \left (\frac43C z^2 \right ) +\dots\,, \quad , \quad
\frac{V'(z)}{V(z)} = \frac{8}{3}\,C z+\frac{1}{z}+\dots\,.
\eeqa
For completeness, we write down the dilaton potential as a function of $\Phi$. Substituting 
equation (\ref{eqsuperpot5}) into the differential equation (\ref{superpotEq}) and  taking
the leading term, we get
\begin{equation}\label{eqdilapotIR2}
V(\Phi) \propto
\Phi^{1/2}\, \exp \left (\frac43 \Phi \right ) +\dots\,, \qquad\quad 
\frac{\partial_{\Phi}V(\Phi)}{V(\Phi)}=\frac{4}{3}+\frac{1}{2 \Phi}+\cdots\,.
\end{equation} 
The results (\ref{eqsuperpot4}) and (\ref{eqdilapotIR}) satisfy the general criteria presented  in \cite{Gursoy:2007cb,Gursoy:2007er} to 
guarantee linear confinement in the IR. It is interesting to 
point out the difference between this IR asymptotics and 
the one considered in
\cite{Gubser:2008ny,Gubser:2008yx}. In 
that case the ratio $\partial_{\Phi}V(\Phi)/V(\Phi)$ is a constant
because the potential goes like $\exp (\gamma \Phi)$, 
where $\gamma$ is a  constant. Here we have
considered the IR asymptotics of \cite{Gursoy:2007cb,Gursoy:2007er},
where the ratio $\partial_{\Phi}V(\Phi)/V(\Phi)$ has a 
subleading term that decreases as $1/z^{2}$
because the dilaton is quadratic in the IR.

From the result (\ref{eqsuperpot3}), we see that the metric shrinks to zero at large $z$ as
\begin{equation}
ds^{2}=z \,\exp \left (- \frac43 C z^2 + \dots \right )\left[dz^{2}
+\eta_{\mu\nu}dx^{\mu}dx^{\nu}\right].
\end{equation} 
As explained above, this leads to a curvature singularity at $z \to \infty$, i.e. a divergent Ricci scalar.  

The above results were obtained in the Einstein frame, where the warp factor decreases monotonically. If we calculated the warp factor in the string frame, $A_s(z)$, we would see that it has a minimum, associated with the fundamental string tension \cite{Gursoy:2007er}.

Again, we write the IR dilaton asymptotics (\ref{dilaIR})  in a convenient form 
\begin{equation}\label{dilaIR2}
\Phi(z)=(\Lambda z)^{2}+\dots\,.
\end{equation}
Note that we are using the same coefficient $\Lambda$ that appeared already in the UV expansion.
This implies a relation between the IR coefficient $C$ at large $z$ and the UV coefficient $G$ at small $z$. In the next section, we will show that the parameter $\Lambda$ will be responsible for fixing the scale of the glueball masses when comparing the numerical results against lattice data.
 
\subsection{UV/IR interpolation}
\label{intersection}
In the pioneer soft-wall model \cite{Karch:2006pv}, a quadratic dilaton was introduced by hand to get the desired behaviour in the dual QCD-like theory, namely, the Regge-like
behaviour $m^{2}\propto n $. As explained before, starting from \cite{Csaki:2006ji} there have been interesting proposals for holographic QCD considering a dilaton field dynamically coupled  to the metric. This in turn leads to a nonzero gluon condensate and confinement. A particularly interesting proposal was considered recently in \cite{Li:2013oda,Chen:2015zhh} where an analytic function was used to interpolate the dilaton between a quartic form in the UV and a quadratic form in the IR. However, as explained before, a quartic dilaton in the UV necessary leads to an unacceptable massless mode in the scalar sector of glueballs \cite{Csaki:2006ji}.\footnote{This result is missing in Ref. \cite{Li:2013oda,Chen:2015zhh} because the authors did not describe scalar glueballs in terms of scalar perturbations in Dilaton-Gravity.}

In section \ref{SubSec:LinDilGrav} we described the process of linearising the Dilaton-Gravity equations to arrive at the  equations governing the dynamics of the scalar and tensor glueballs. A careful analysis of these equations suggest two possible solutions for the massless mode problem. The first solution, originally proposed in \cite{Csaki:2006ji} and beautifully realised in \cite{Gursoy:2007cb,Gursoy:2007er}, consists of introducing asymptotic freedom in the UV. However, the price to be paid when introducing asymptotic freedom is the presence of logarithmic corrections in the warp factor and the dilaton which make the AdS/CFT dictionary more involved\cite{Papadimitriou:2011qb,Kiritsis:2014kua}. The second solution, considered in this work, is implementing a CFT deformation of the form $\delta {\cal L}= \phi_{0}\, {\cal O}$, being ${\cal O}$ a relevant operator dual to the dilaton. The conformal dimension of ${\cal O}$ was set to $\Delta=4-\epsilon$ in equation (\ref{asympdila}), with $\epsilon$ small. Mapping this operator to the QCD operator ${\rm Tr} \, F^2$  signalizes the presence of a UV cut-off $E^{*}$ in the QCD-like theory, as advocated in \cite{Gubser:2008ny,Gubser:2008yx}. 

The UV asymptotics corresponding to the CFT deformation was described in subsection \ref{secanoma} whereas the IR asymptotics leading to linear confinement was described in subsection \ref{confinement}. In particular, the UV and IR asymptotics for the dilaton were given in (\ref{dilaUV}) and (\ref{dilaIR2}) respectively. There are two immediate options for interpolating between the UV and IR asymptotics. We can define an analytic function for the dilaton that interpolates from the UV to the IR and solve 
numerically for the warp factor and the potential $V(z)$. The models arising in this approach will be called models A. The second possibility is to interpolate the warp factor between the two regimes, so that we get numerically the dilaton and the potential $V(z)$. The models in this second approach will be called models B. In the following we present and analyse the first case (models A) and leave the analysis of models B for the appendix \ref{App:ModelB}.

 In section \ref{Sec:Spectrum} we will show how these effective holographic QCD models (models A and B) lead to a realistic spectrum for scalar and tensor glueballs. In section \ref{Sec:TraceAnomaly} we will take advantage of the fact that the effective holographic QCD approach allows the use of the standard AdS/CFT dictionary and calculate the VEV of the operator ${\cal O}$. We will relate this VEV to the gluon condensate $\langle {\rm Tr} \, F^2 \rangle$ and discuss the trace anomaly of deformed CFTs in connection with the QCD scale anomaly. A general discussion of the massless mode and its resolution is done in appendix \ref{App:massless}.


\subsubsection{Models A: analytic form for the dilaton field }
As explained above, in the models A we interpolate the dilaton field, from the UV asymptotics \eqref{dilaUV} to the IR asymptotics \eqref{dilaIR2}, in order to describe a CFT deformation in the UV and confinement in the IR. Among the many possibilities for interpolating the dilaton between the UV and IR asymptotics,  we choose two of them. 

The first interpolation (model A1) is constructed in terms of powers of the holographic coordinate: 
\begin{equation}\label{interp1}
\Phi (z)= \hat \phi_0 (\Lambda z)^{\epsilon} 
+\frac{(\Lambda z)^{4-\epsilon}}{1+(\Lambda z)^{2-\epsilon}}\,,
\end{equation}
where the parameters $\hat \phi_0$, $\Lambda $, and $ \epsilon$ were already 
defined below equation \eqref{dilaUV}. This is the simplest way of interpolating 
from the UV to the IR. 

The second interpolation (model A2) introduces a hyperbolic tangent function,
\begin{equation}\label{interp2}
\Phi (z)= \hat \phi_0 (\Lambda z)^{\epsilon} 
+(\Lambda z)^{2}\tanh{\left[(\Lambda z)^{2-\epsilon}\right]}.
\end{equation}
It is easy to see that both equations (\ref{interp1}) and (\ref{interp2}) 
recover the previous asymptotic expansions in the UV and IR, equations 
(\ref{dilaUV}) 
and (\ref{dilaIR2}), respectively. The warp factor, on the other hand, is 
obtained by solving the first differential equation in \eqref{IHQCDEqsv2}. 

As explained previously, the dilaton field $\Phi(z)$ and warp factor $A(z)$ 
will depend only on the parameters $\hat \phi_0$, $\Lambda$ and $\epsilon$.  
The parameter $\Lambda$ is used to fix the energy scale while the value of 
$\hat \phi_0$ plays the role of a dimensionless coupling. Our numerical 
strategy will be to fix the conformal dimension $\epsilon$ and fit the 
parameters $\hat \phi_0$ and $\Lambda$ using the masses of the first two scalar 
glueballs (taken from lattice QCD). This analysis will be developed in section 
\ref{Sec:Spectrum}.  

\subsubsection{Numerical analysis of the background}
\label{numebackground}

Having specified the models  A, where the dilaton is an analytic function, we 
can solve numerically the Dilaton-Gravity equations (\ref{IHQCDEqsv2}), with the 
appropriate boundary conditions, and explore the evolution of the geometric 
quantities such as the warp factor $A(z)$, the field $X(z)$ and the 
superpotential $W(z)$. The parameters that we use to get the results and plot 
the figures in this  section are presented in Table \ref{taba-1} and will be 
justified in  section \ref{Sec:RunningParameters}. Our goal here is to show the 
nonsingular behaviour of these quantities. The dilaton field $\Phi(z)$, dual to 
the relevant operator ${\cal O}$, is shown  in Fig.  \ref{DilatonAB}. The effect 
of the conformal dimension $\epsilon$ is evident near the boundary where the 
dilaton field goes as $\sim \phi_0 z^{\epsilon}$ (see the box in the figure), 
with $\phi_0 = \hat \phi_{0}\, \Lambda^{\epsilon}$.  This is the dominant term 
responsible for the explicit breaking of conformal invariance. Figure
\ref{DilatonAB} also shows the  asymptotic behaviour of the dilaton $\sim 
\Lambda^2 z^2$ in the IR, responsible for confinement and a linear spectrum. The 
difference between the models A1 and A2 lies on the values of the parameters 
$\hat \phi_0$ and $\Lambda$, in addition to the interpolation form.
\TABLE{
\begin{tabular}{c|c|c|c|c|c}
\hline 
\hline
 \text{Models} & $\hat \phi_0\, $&
$\Lambda\, (\text{MeV})$&$\phi_{0}\, (\text{MeV}^{\epsilon})$
 &$G\, (10^{11}\text{MeV}^{4-\epsilon})$ & 
 $C\, (10^{5}\text{MeV}^{2})$   \\
\hline 
 Model A1 & $5.59$ & $742.75$& $10.83$& $1.57$  & $5.52$   \\
 Model A2 & $5.33$ & $677.98$& $10.23$& $1.10$  & $4.60$ \\
 \hline\hline
\end{tabular}
\caption{The values of the parameters for models 
A1 and A2 used to get the results presented in subsection \ref{numebackground} 
for $\epsilon=0.1$.}
\label{taba-1}
}

%
\FIGURE{
\centering
\includegraphics[width=10cm,angle=0]{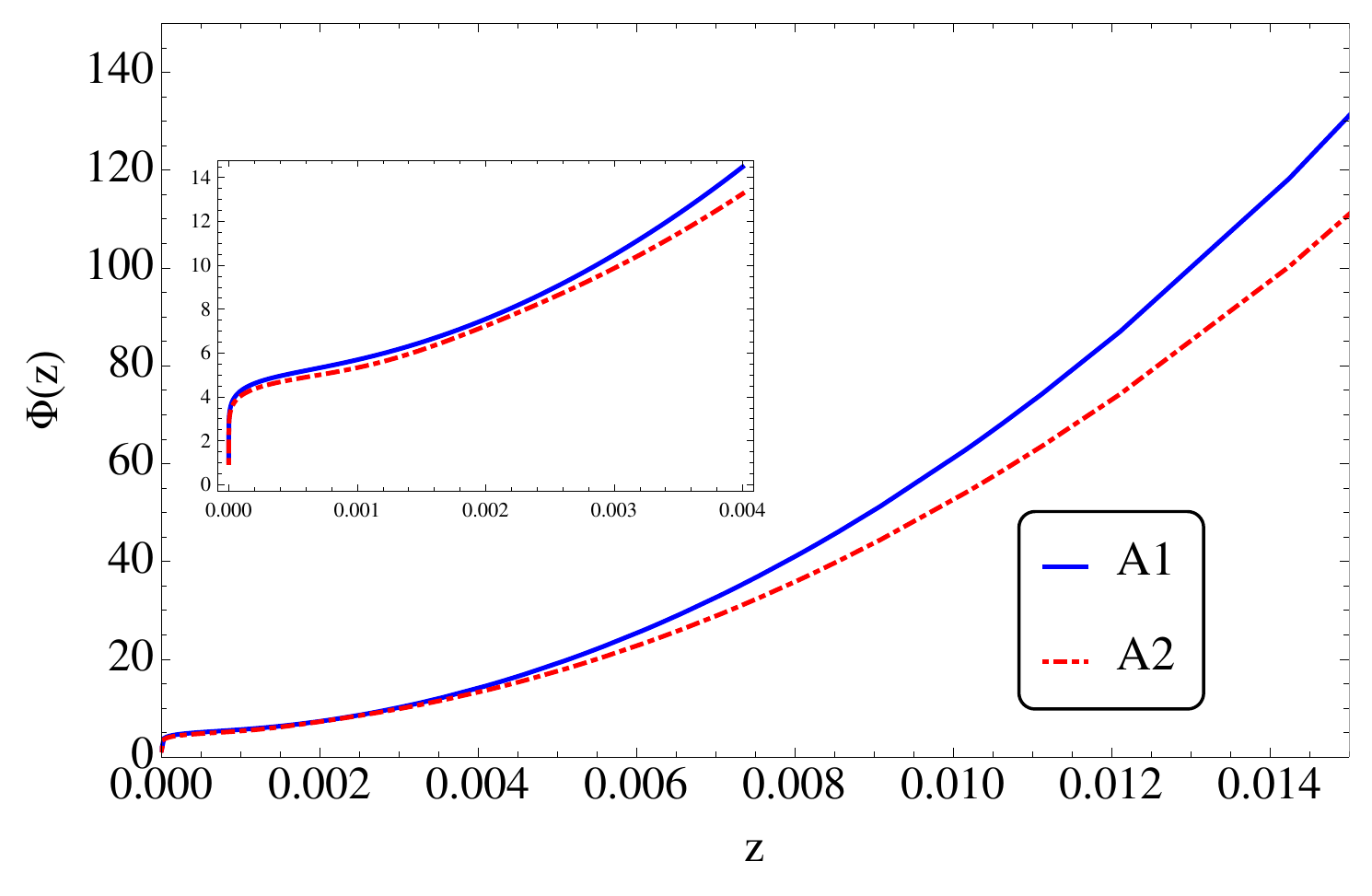} 
\caption{Dilaton profiles for models A1 and A2, defined in \eqref{interp1} and \eqref{interp2} respectively, for $\epsilon=0.1$ and the parameters given in Table \ref{taba-1}. 
}
      \label{DilatonAB}}

On the left panel of Fig. \ref{AAsAB}, we plot the numerical solution for the 
Einstein-frame warp factor $A(z)$ in models A1 and A2. These results are 
consistent with the UV and IR asymptotics, given by equations 
(\ref{asymptotics2}) and (\ref{eqsuperpot3}), respectively. The difference 
between models A1 and A2 lies in the region for  large z. As we shall see in 
section \ref{Sec:Spectrum}, the effect of this difference is realized  in the 
glueball spectrum. The right panel in Fig. \ref{AAsAB} shows the string-frame 
warp factor $A_s(z)$, obtained from equation \eqref{WarpFactorSF}. This function 
has a minimum at some $z=z^{*}$, which is consistent with the confinement 
criterion described in subsection \ref{Sec:Dilaton-Gravity}.

%
\FIGURE{
\begin{tabular}{*{2}{>{\centering\arraybackslash}p{.5\textwidth}}}
        \includegraphics[width=7.2cm,angle=0]{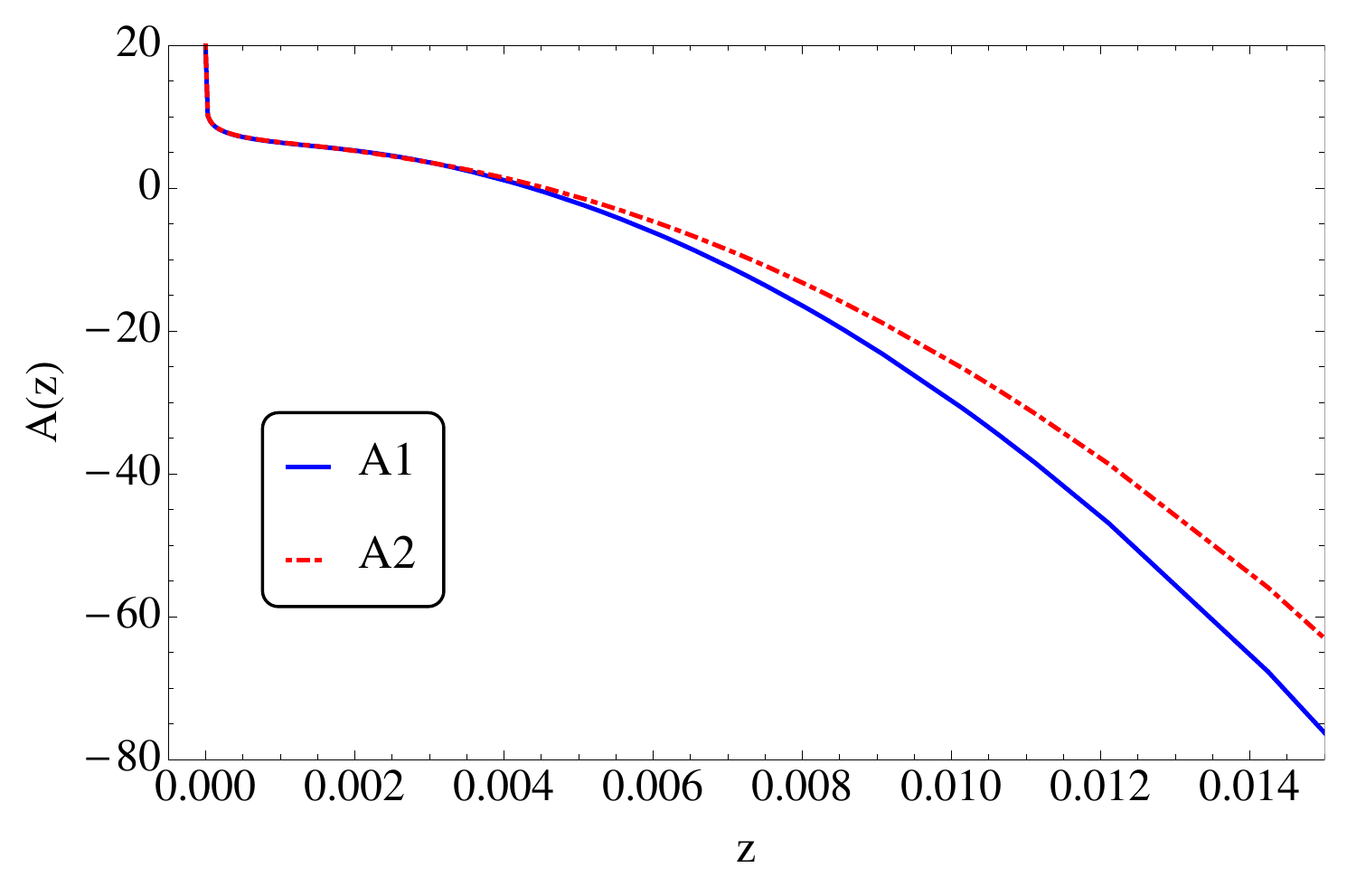} 
        & 
 \hspace{-1cm} \includegraphics[width=7.2cm,angle=0]{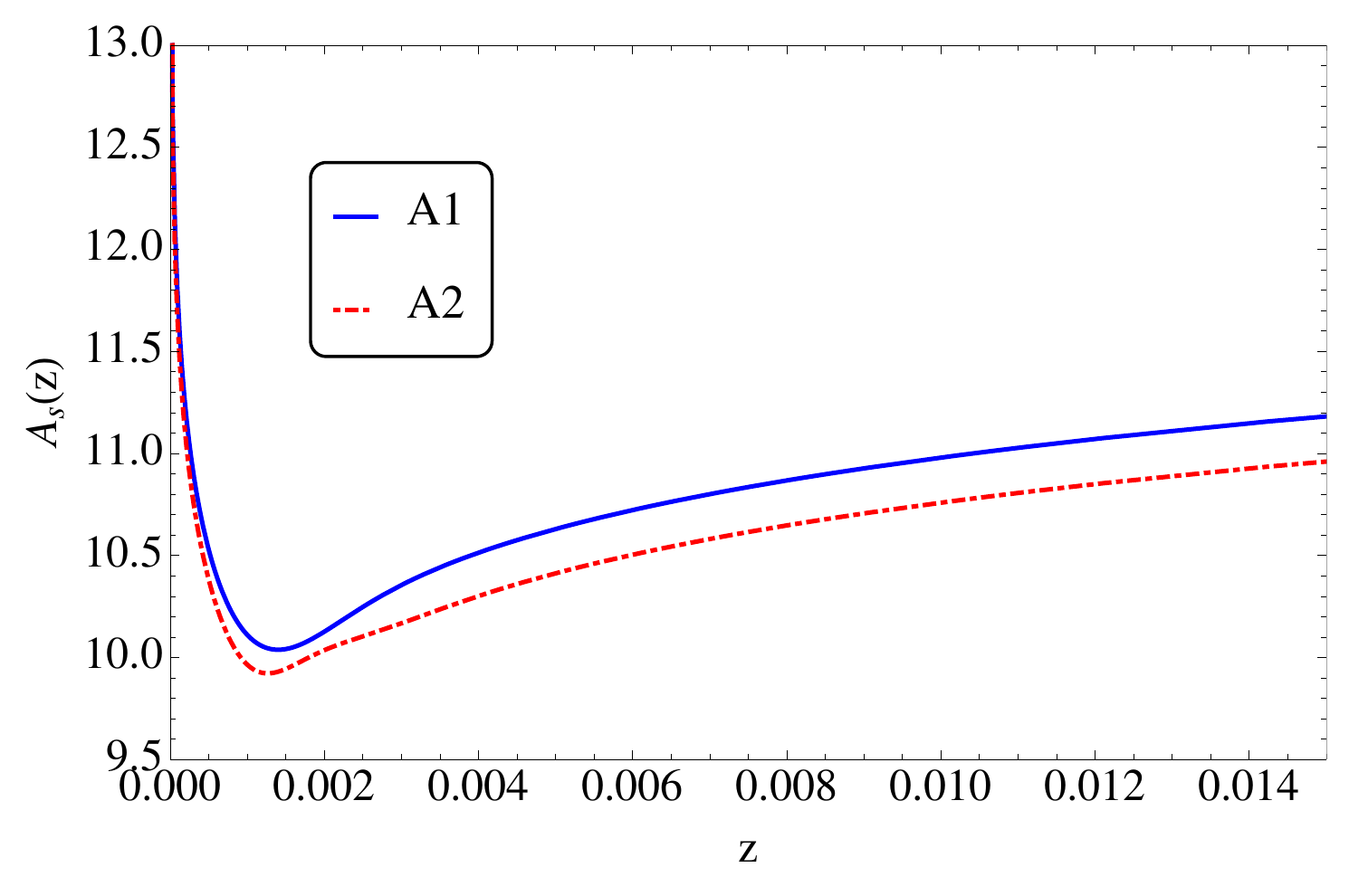} 
\\
\end{tabular}
    \caption{{\bf Left Panel:} The Einstein-frame warp factor  $A(z)$ for models A1 and A2, obtained by solving numerically the Dilaton-Gravity equations (\ref{IHQCDEqsv2}). {\bf Right Panel:}
The string-frame warp factor  $A_s(z)$ for models A1 and A2, obtained from 
equation \eqref{WarpFactorSF}. The results shown in this figure correspond to 
$\epsilon=0.1$ and the parameters given in Table \ref{taba-1}.}
     \label{AAsAB}}
%

Another important geometric quantity is the field  $X(z)$, defined in equation (\ref{xEq}). As shown on the left panel  in Fig. \ref{XWAB}, this quantity
has the same asymptotic behaviour for both models A1 and A2,  and the relevant  difference lies in the intermediate region. The presence of confinement in the IR is consistent with $X(z)$ approaching a constant value for large $z$. As discussed in \cite{Bourdier:2013axa, Megias:2014iwa}, the field $X$ can be interpreted as a bulk effective  $\beta$-function associated with the 4-$d$ RG-flow driven by the operator ${\cal O}$. 

%
\FIGURE{
\begin{tabular}{*{2}{>{\centering\arraybackslash}p{.5\textwidth}}}
        \includegraphics[width=7.2cm,angle=0]{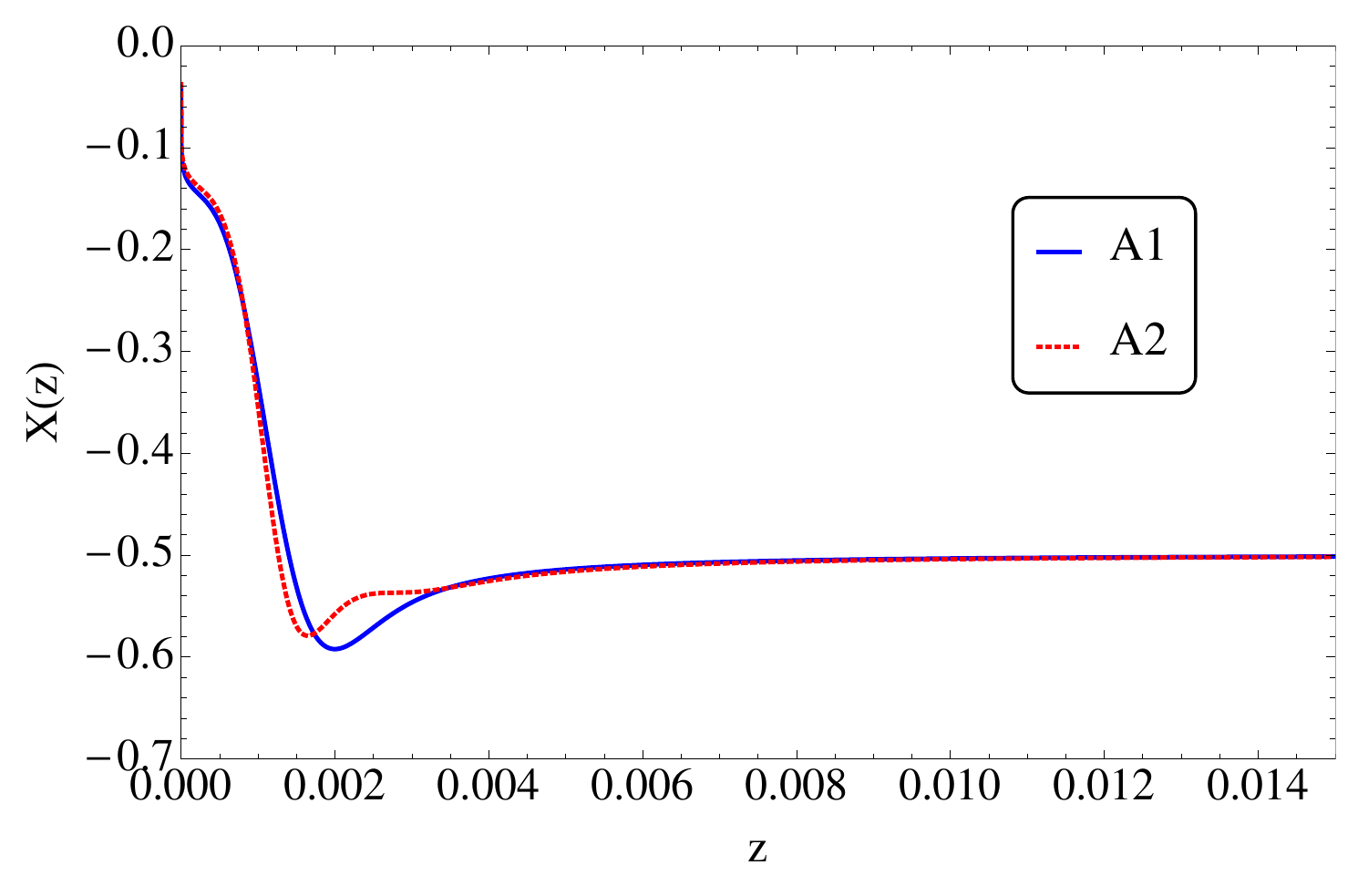} 
        & 
 \hspace{-1cm} \includegraphics[width=7.2cm,angle=0]{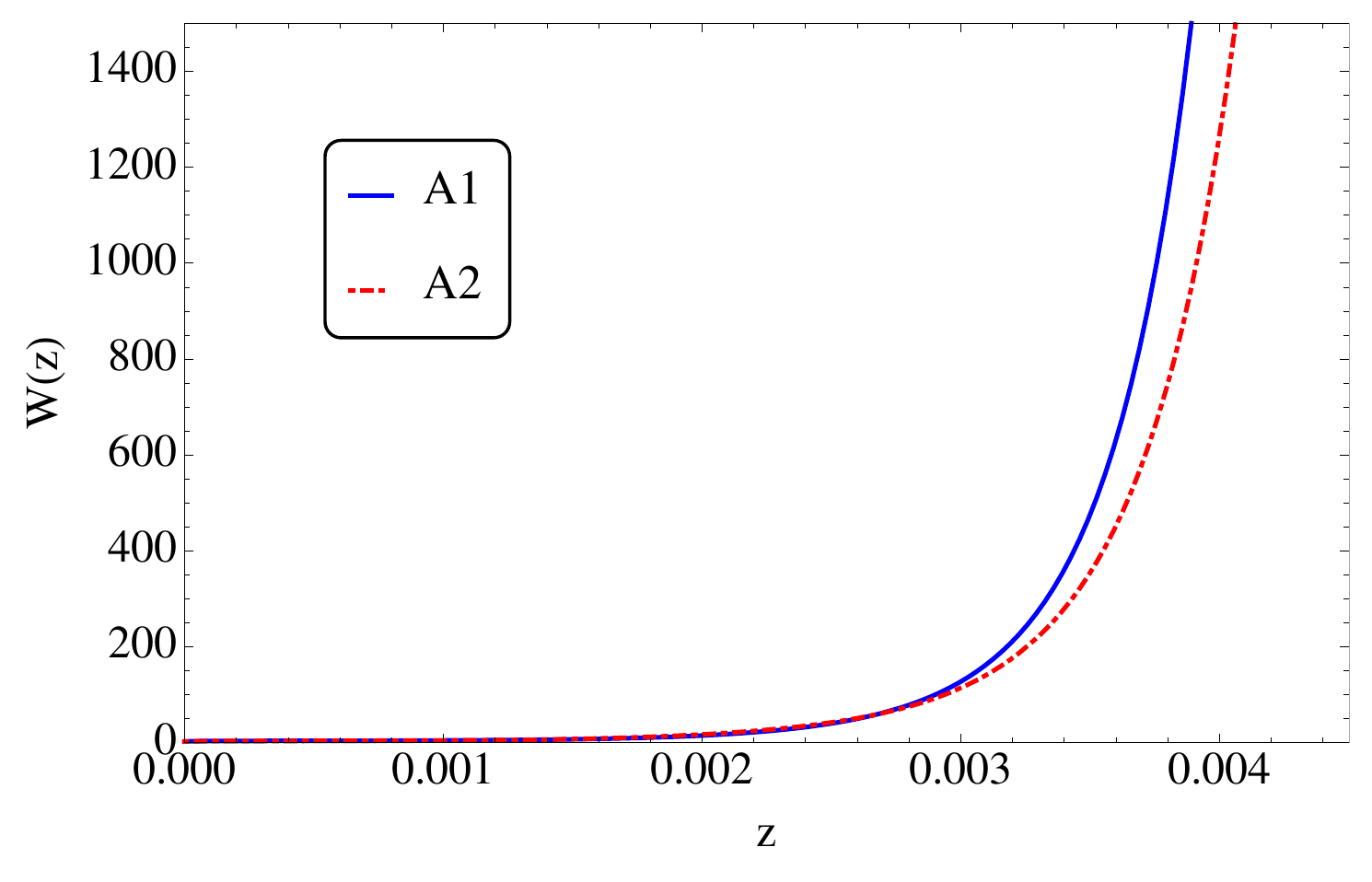} 
\\
\end{tabular}
    \caption{{\bf Left Panel:} The field $X(z)$ for models A1 and A2, obtained 
from equation (\ref{xEq}). {\bf Right Panel:} The superpotential $W(z)$ for 
models A1 and A2, obtained from the second equation in (\ref{superpotEq}). The 
results shown in this figure correspond to $\epsilon=0.1$ and the parameters 
given in Table~\ref{taba-1}.}
     \label{XWAB}}
%

The breaking of conformal invariance can be appreciated by looking at the 
superpotential $W(z)$, obtained from the second equation in (\ref{superpotEq}). 
As a consequence of the CFT deformation in the UV, the superpotential raises 
rapidly with the coordinate $z$, as shown on the right panel in Fig. \ref{XWAB}. 
This behaviour was 
previously noticed for the case of nearly marginal operators \cite{Megias:2014iwa}. In contrast with $X(z)$, the difference in $W(z)$ between the models A1 and A2 lies in the IR. Those differences play a role in the spectrum of glueballs, that will be calculated in section \ref{Sec:Spectrum}. 


\subsection{The vacuum energy density}
\label{subsec:vacuumenergy}

A very important quantity in holographic QCD is the vacuum energy density $\langle T^{00} \rangle$. In the absence of a CFT deformation, we expect $\langle T^{00} \rangle$ to vanish since the 4-$d$ theory lives in Minkowski spacetime.\footnote{When the 4-$d$ CFT lives in a curved background one usually gets a nonvanishing vacuum energy density.} In our case, the CFT deformation $\delta {\cal L} = \phi_{0}\, {\cal O}$ will generate a nontrivial negative $\langle T^{00} \rangle$ that will be interpreted as the QCD vacuum energy density. 

In holography the vacuum energy of a 4-$d$ theory corresponds to  the on-shell Euclidean action associated with the 5-$d$ gravitational background. In Dilaton-Gravity the action is given by
\begin{equation}\label{QCDEnergy}
S=S_E+S_{GH},
\end{equation}
where 
\begin{equation}
S_E=
-M^3N_c^2\int d^4x\int_{z_0}^{\infty}dz
\sqrt{g}\left(R-\frac{4}{3} g^{mn} \partial_m\Phi \partial_n \Phi+V(\Phi)\right) \, , 
\end{equation}
and 
\begin{equation}
S_{GH}=
M^3N_c^2\int d^4x\sqrt{\gamma}\,2K\bigg{|}_{z= z_0}
\label{GibbonsHawking}
\end{equation}
is the Gibbons-Hawking boundary term. In holographic renormalization one first defines the boundary   
at $z=z_0$ and takes the limit $z_0 \to 0$ only at the very end of the renormalization process.
We have introduced in \eqref{GibbonsHawking} the induced metric $\gamma_{\mu\nu}= \exp (2 A) \eta_{\mu \nu}$ and the trace of the extrinsic curvature $K$, given by
\beq
K = \nabla_m \eta^m  \, , \qquad \eta^m = \zeta(z) \delta^m_z \, , 
\eeq
where $\zeta(z)$ is defined in \eqref{zetadef}.
%
%
As shown in \cite{Gursoy:2007er}, using the equations of motion 
(\ref{IHQCDEqs}) the on-shell (o-s) action densities take the form
\begin{equation}\label{OnShellAct}
\begin{split}
\mathcal{S}_E^{{\rm o-s}}&=\frac{S_E^{{\rm o-s}}}{V_4}=
-2M^3N_c^2e^{3A(z_0)}A'(z_0)\,,\\
\mathcal{S}_{GH}^{{\rm o-s}}&=\frac{S_{GH}^{{\rm o-s}}}{V_4}=
8M^3N_{c}^{2}e^{3A(z_0)}A'(z_0)\,,
\end{split}
\end{equation}
where $V_4 $ is the four dimensional volume. 
Thus we find the bare energy density
\begin{equation}
\langle T^{00} \rangle = \mathcal{S}_E^{{\rm o-s}} + \mathcal{S}_{GH}^{{\rm o-s}} = 6M^3N_c^2e^{3A(z_0)} A'(z_0). \label{BareEnergy}
\end{equation}
The near boundary asymptotic behaviour of the warp factor $A(z)$ was obtained in \eqref{asymptotics2}. Using that result  in \eqref{BareEnergy} one finds 
 terms that diverge when $z_0$ goes to zero. We consider a minimal subtraction (MS) scheme where the divergent terms are eliminated by adding the appropriate counterterms.  After this renormalization process one finally takes the limit $z_0 \rightarrow 0$.  Most of the non-divergent terms vanish in this limit and the surviving finite piece becomes  the renormalized vacuum energy density:
\begin{equation}\label{RNEnergyEq}
\mathcal{E}^{ren}_{\text{QCD}}=\langle T^{00} \rangle^{ren}=
-\frac{4}{15} M^3N_c^2\epsilon (4 - \epsilon) \phi_0 G  \, ,
\end{equation}
where the superscript ``$ren$'' means renormalized.  The numerical results for this renormalized vacuum energy density will be shown in the next section. Here we just notice that the vacuum energy density is negative and therefore lower than the (zero) energy of the CFT vacuum. We remark, however, that the result \eqref{RNEnergyEq} may change if we use a different renormalization scheme. Note also from \eqref{DefParameters} that $\phi_{0}\,G=\hat{\phi}_{0}\,\Lambda^4$ has conformal dimension $4$, which is indeed the protected conformal dimension of $T^{\mu \nu}$.


\section{Glueball spectrum}
\label{Sec:Spectrum}
Glueballs are bound states of gluons predicted by QCD. So far, they have not been detected although
there is a recent claim that the $f_0(1710)$ scalar particle may actually be the scalar glueball state $0^{++}$ \cite{Janowski:2014ppa,Brunner:2015yha}. Furthermore, it was also recently proposed that the odd glueball (oddball) $0^{--}$ could be detected soon by the experiments BESIII, BELLEII, Super-B,
PANDA, and LHCb \cite{Qiao:2014vva}, although there is some controversy on this prediction \cite{Pimikov:2017xap}. Other interesting glueball states, as for example, $2^{++}$, $0^{-+}$
and $1^{--}$  are under investigation and have candidates in the particle spectrum \cite{Mathieu:2008me}. 

In this work, we are particularly interested in the scalar $0^{++}$ and tensor $2^{++}$ glueball states, as well as their radial excitations. The investigation of those glueball states have been made in lattice QCD and other non-perturbative approaches. For a review, see for instance \cite{Mathieu:2008me}. Previous holographic approaches to the glueball spectrum include the Witten's model \cite{Witten:1998zw, Csaki:1998qr, Brower:2000rp}, the Klebanov-Strassler model \cite{Caceres:2000qe,Dymarsky:2007zs,Elander:2017hyr}, the Maldacena-Nunez model \cite{Berg:2005pd,Berg:2006xy}, the hardwall model \cite{BoschiFilho:2002vd, BoschiFilho:2005yh, Capossoli:2013kb}, the soft-wall model \cite{Colangelo:2007pt} and extensions \cite{BoschiFilho:2012xr, Capossoli:2016ydo}, dynamical soft-wall models \cite{Li:2013oda,Capossoli:2016kcr,Capossoli:2015ywa, Chen:2015zhh} and improved holographic QCD models \cite{Gursoy:2007er}. 

We will find the spectrum of scalar $0^{++}$ and tensor $2^{++}$ glueball states from solving the differential equations  (\ref{schrodingerscaeq}) and (\ref{schrodingerteneq}), respectively. In those Schr\"odinger like equations the glueball states are represented by wave functions $\psi_s$ and $\psi_t$. Under suitable boundary (asymptotic) conditions on these wave functions, the mass spectrum of the respective sector is found.
As explained in the previous section, the parameters in our models are $\hat \phi_0$, $\Lambda$ and $\epsilon$ and the strategy is the following: for each value of $\epsilon$ the parameters $\hat \phi_0$ and $\Lambda$ will be fixed using as input the lattice QCD results for the first two scalar glueballs \cite{Meyer:2004gx}. At the end of the numerical calculation, we compare our results for all the other glueball states against lattice QCD data \cite{Meyer:2004gx}, and the results obtained in the improved holographic QCD model \cite{Gursoy:2007er}. 

In the numerical calculation it is convenient to rewrite the interpolations (\ref{interp1}) and  (\ref{interp2}) in terms of a dimensionless coordinate $u=\Lambda z$:
\begin{equation}\label{interp1Coordu}
\Phi (z)= \hat \phi_0 u^{\epsilon} 
+\frac{u^{4-\epsilon}}{1+u^{2-\epsilon}}\,,
\end{equation}
\begin{equation}\label{interp2p5Coordu}
\Phi (u)=\hat \phi_0\, u^{\epsilon}+u^{2}\tanh{\left(u^{2-\epsilon}\right)}.
\end{equation}
Notice that the parameter $\Lambda$ has disappeared in \eqref{interp1Coordu} and \eqref{interp2p5Coordu}. This is because the $u$ coordinate is dimensionless and the Schr\"odinger like equation in this coordinate leads to a spectrum where the masses are given in units of $\Lambda$. We remind the reader that fixing $\Lambda$ will fix also the VEV coefficient 
$G=\Lambda^{4-\epsilon}$ in the UV as well as the IR coefficient $C=\Lambda^{2}$ characterizing confinement. 

 \subsection{Analysis of the effective potentials}
 \label{effectivepot}

The spectrum of glueballs will depend on the form of the effective potentials $V_s$ and $V_t$, which appear in the Schr\"odinger like equations (\ref{schrodingerscaeq}) and (\ref{schrodingerteneq}). Here we present an analysis of those potentials. 
 
Let us start with the effective potential $V_s$ for the scalar sector, defined in equation \eqref{potentialsca}. In
terms of the dimensionless variable $u$, this potential takes the form
%
%
%
\beqa \label{scalarpot}
\frac{V_{s}(u)}{\Lambda^{2}}=  \left [ \partial_u B_s(u) \right ]^2 + \partial_u^2 B_s(u)  \, , 
\eeqa
with
\beqa
B_s(u) = \frac32 A(u) + \log [ X(u)]. 
\eeqa
From equations (\ref{asymptotics2}) and (\ref{dilaUV}),
we know how $A(u)$ and $\Phi(u)$ behave in the UV.
Using those results and equation \eqref{xEq}, we get the 
UV asymptotic behaviour of $X(u)$ and $B_{s}(u)$:
\begin{equation}
\begin{split}
3 X(u)&=- \epsilon \hat \phi_0 \,  u^{\epsilon}-(4-\epsilon) u^{4-\epsilon}+\cdots,\\
B_{s}(u)&= \left(-\frac{3}{2}+\epsilon\right)\log u +\cdots.
\end{split}
\end{equation}
Our hypothesis is that the conformal dimension $\epsilon$ is small. Then the leading term of the scalar potential (\ref{scalarpot}) takes the form 

\begin{equation}\label{uvscalarpot}
 \frac{V_{s}(u)}{\Lambda^{2}}
 =\left(\frac{15}{4}+M^{2}\right)\frac{1}{u^{2}}\,,
\end{equation}
where we have introduced the dilaton mass term $M^{2}=\epsilon(\epsilon-4)$. This term is responsible in the UV for the explicit break of conformal symmetry.
 
Now we turn attention to the effective potential $V_t$ for the tensor sector, defined in equation \eqref{potentialtens}. In terms of the $u$ coordinate it takes the form 
 \begin{equation}\label{tensorpot}
 \frac{V_{t}(u)}{\Lambda^{2}}=  \left [ \partial_u B_t(u) \right ]^2 + \partial_u^2 B_t(u)  \, , 
 \end{equation}
with 
\beqa
B_t(u) = \frac32 A(u) \,. 
\eeqa
The UV asymptotic behaviour of this potential is obtained from the asymptotic 
behaviour of $A(u)$, given by equation (\ref{asymptotics2}). The result is 
simply
 \begin{equation}\label{uvtensorpot}
 \frac{V_{t}(u)}{\Lambda^{2}}=\frac{15}{4u^{2}}\,.
 \end{equation}
Notice that the conformal dimension $\epsilon$ does 
not affect the UV asymptotic behaviour of the tensor potential. 

In the IR regime, at large $u$, the asymptotic 
behaviour for the warp factor and dilaton are given 
by equations (\ref{eqsuperpot3}) and (\ref{dilaIR2})
respectively. Then, $X(u)$ and $B_{s,t}(u)$ have the asymptotic form
 \begin{equation}
X(u)=-\frac{1}{2}+\cdots,\qquad
B_{s,t}(u)=-u^{2}+\cdots\,.
\label{irscalarpot}
 \end{equation}
Therefore, the IR asymptotic behaviour of the effective potentials (for both sectors) take the form 
 \begin{equation}\label{irpots}
 \frac{V_{s,t}(u)}{\Lambda^{2}}= 4u^{2}\,.
 \end{equation}

 In figure \ref{potentials} we show the effective potentials, obtained numerically, for models A1 and A2 at $\epsilon=0.01$ and  $\hat \phi_0=50$. The plots are consistent with the  asymptotic  results (\ref{uvscalarpot}), (\ref{uvtensorpot}) and (\ref{irpots}). As expected, the difference between models A1 and A2 lies in the intermediate region.

%
\FIGURE{
\begin{tabular}{*{2}{>{\centering\arraybackslash}p{.5\textwidth}}}
\includegraphics[width=7.2cm,angle=0]{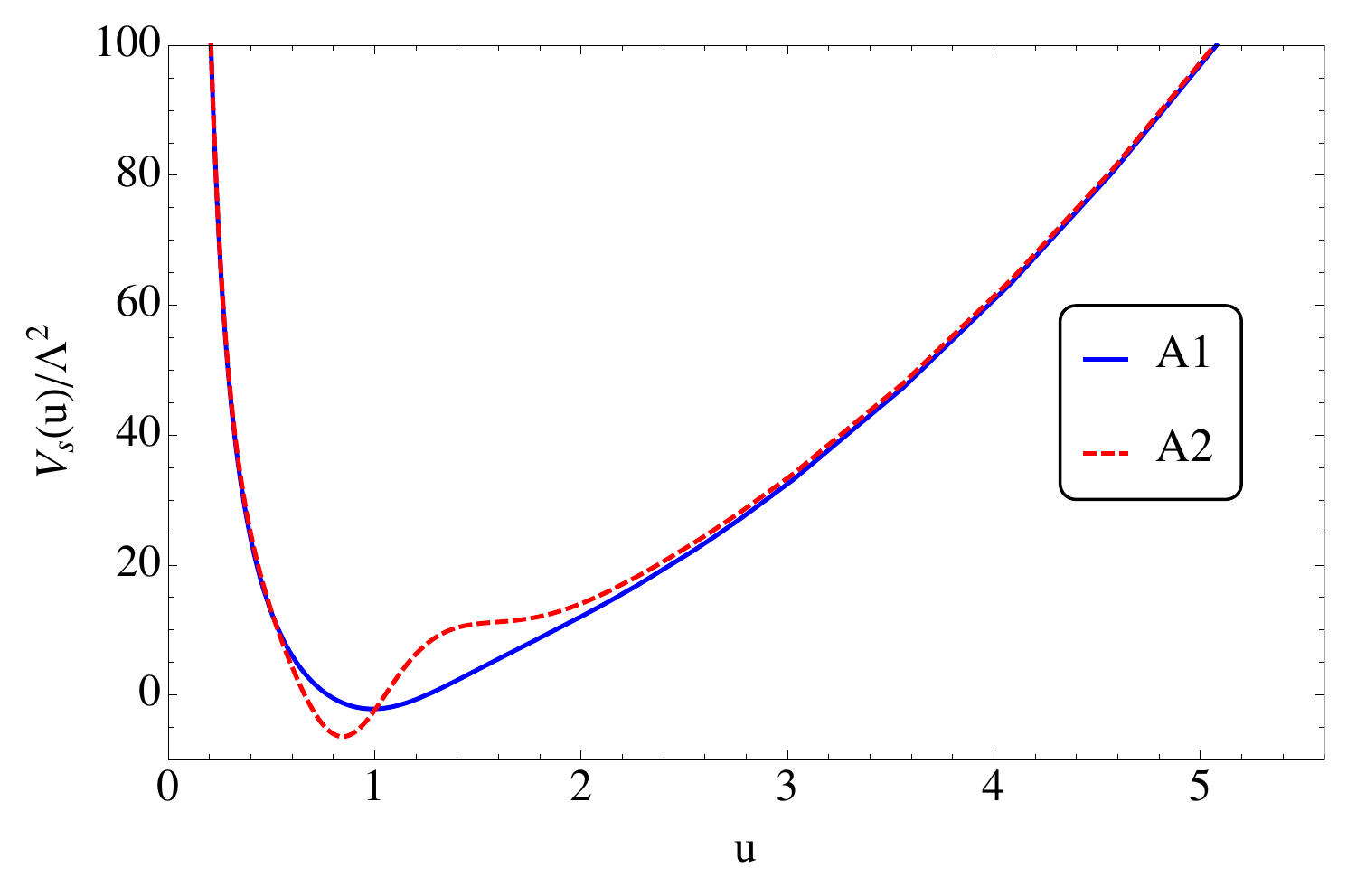} 
& 
\hspace{-1cm}  
        \includegraphics[width=7.2cm,angle=0]{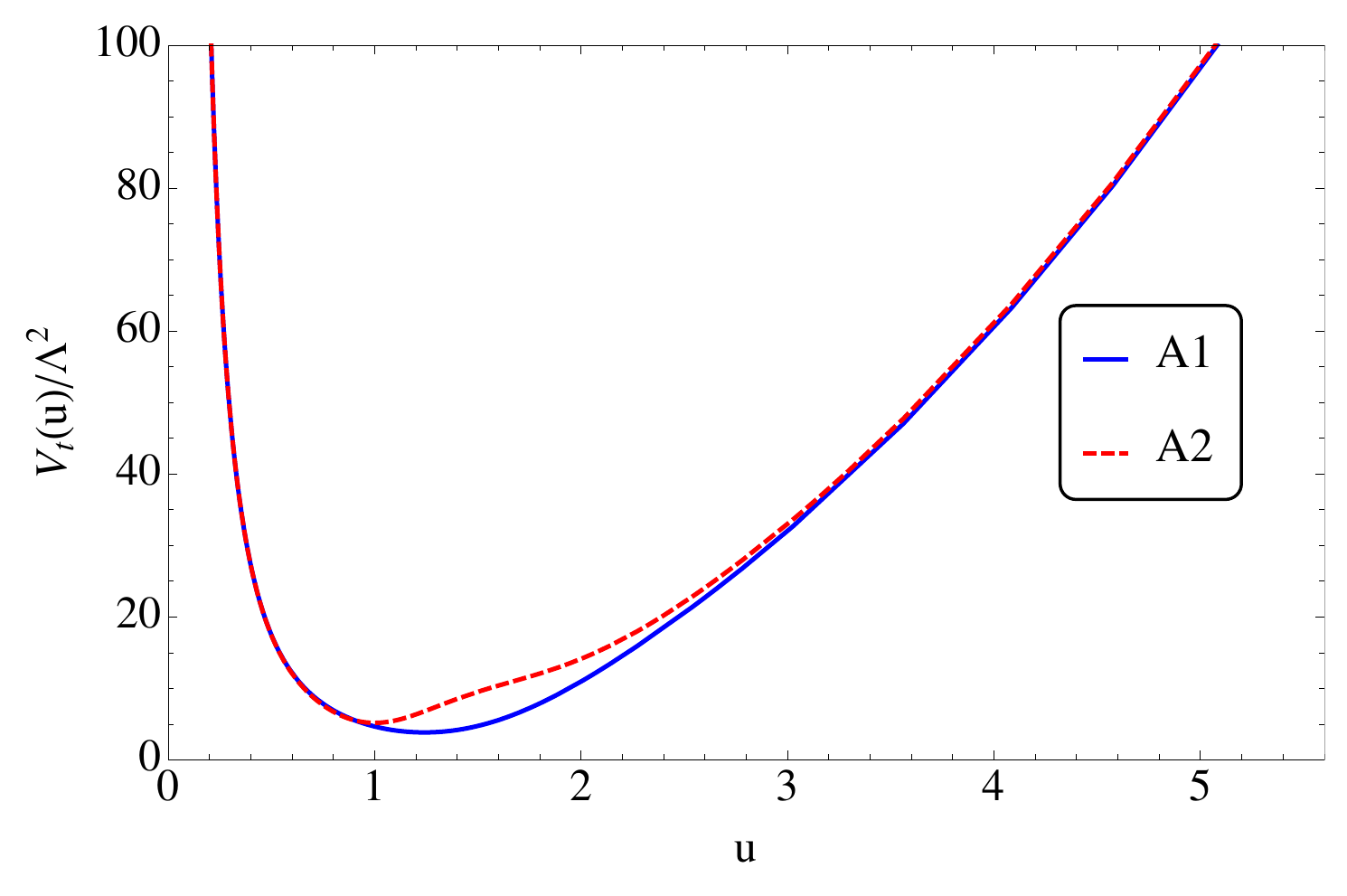} 
\end{tabular}
    \caption{The scalar ({\bf left panel}) and tensor ({\bf right panel}) 
        effective potentials for models A1 and A2  at $\epsilon=0.01$,  $\hat \phi_0=50$. An analogous figure for models B is shown in Appendix \ref{App:ModelB}.}
     \label{potentials}}

\subsection{UV and IR asymptotic solutions for the wave functions}
 
 In this section we will find the UV and IR asymptotic solutions for the  wave functions $\psi_s$ and $\psi_t$ in the Schr\"odinger like equations (\ref{schrodingerscaeq}) and   (\ref{schrodingerteneq}). 
 
 First we look for the UV asymptotic solution for the wave function $\psi_s$ of the scalar sector. Substituting the UV asymptotic form (\ref{uvscalarpot}) of the potential $V_{s}$ in the differential equation (\ref{schrodingerscaeq}), we find
\begin{equation}\label{UVDiffEqSca}
-\psi_{s}''(u)+\left(\frac{3}{2}-\epsilon\right)\left(\frac{5}{2}-
\epsilon\right)\frac{1}{u^{2}}\psi_{s}(u)=\hat{m}_{s}^{2}\psi_{s}(u),
\end{equation}
where $\hat{m}_{s}=m_{s}/\Lambda$ parametrizes the 4-$d$ masses of scalar glueballs in units of $\Lambda$. Near the boundary the wave 
function $\psi_{s}$ behaves as a power law,
$\psi_{s}(u)=u^{\alpha_{1}}$. Substituting this ansatz into the differential equation (\ref{UVDiffEqSca}), and solving the resulting indicial equation, we find two solutions for $\alpha_1$, namely,  $\alpha_1^{-}=-\epsilon+5/2$ and 
$\alpha_1^{+}=\epsilon-3/2$, which means that the asymptotic
general solution is of the form
\begin{equation}\label{asympsolescauv}
\psi_{s}(u)=c_{1}u\,^{-\epsilon+\frac{5}{2}}+c_{2}u\,^{\epsilon-\frac{3}{2}}. 
\end{equation}
The coefficient $c_{2}$ is set to zero because we are looking for a nor\-ma\-li\-za\-ble solution. 

We perform a similar analysis to obtain the UV asymptotic behaviour of the wave function $\psi_t$ of the tensor sector. Using the UV asymptotic form of the potential (\ref{uvtensorpot}), the differential  equation (\ref{schrodingerteneq}) becomes
\begin{equation}
-\, \psi_{t}''(u)+\frac{15}{4u^{2}}\psi_{t}(u)=\hat{m}_t^{2}\, \psi_{t}(u)\,,
\end{equation}
where $\hat{m}_t=m_t/\Lambda$ parametrizes the masses of tensor glueballs in units of $\Lambda$. Again, we select the normalizable solution 
\begin{equation}\label{asympsoltenuv}
\psi_{t}(u)=c_{3}u\,^{5/2}.
\end{equation}

Finally we look for the IR asymptotic behaviour of $\psi_s$ and $\psi_t$. As 
shown in equation \eqref{irpots}, 
the effective potentials of the scalar and tensor sectors have the same asymptotic behaviour in the
IR. Then both Schr\"odinger like equations assume the form
\begin{equation}
-\, \psi_{s,t}''(u)+4u^{2}\psi_{s,t}(u)=\hat{m}_{s,t}^{2}\, \psi_{s,t}(u).
\end{equation} 
Solving this equation, we find the IR asymptotic solutions that converge at infinity can be written as
\begin{equation}\label{asympsolir}
\psi_{s,t}(u)=c_4\,u^{(\hat{m}_{s,t}^{2}-2)/4}\,e^{-u^2}.
\end{equation}

 \subsection{Glueball spectrum at fixed $\epsilon$}
 \label{Sub:Spectrum}
 The task now is to solve the eigenvalue problem for the differential equations 
(\ref{schrodingerscaeq}) and (\ref{schrodingerteneq}). We solve this problem 
numerically using a shooting method, which was implemented
in a Mathematica code. 
 
In order to find a unique solution to a second order differential equation, we 
need two boundary conditions.   
There are two typical ways of doing this. We can use the asymptotic UV solutions 
(\ref{asympsolescauv}) and (\ref{asympsoltenuv}) and its derivatives, 
respectively for the scalar and tensor sectors, as boundary conditions at some 
$u=u_{min}$, with $u_{min}$ very small. Then, we integrate numerically from 
small $u$ to large $u$ and require that the wave function at large $u$ should 
behave as in \eqref{asympsolir}. Using those conditions and fixing the 
parameters $\epsilon$ and $\hat \phi_0$, we get a discrete spectrum (in units of 
$\Lambda$). Alternatively, we may take the asymptotic IR solutions 
(\ref{asympsolir}) and its derivatives as initial conditions at some 
$u=u_{max}$, with $u_{max}$ very large, and integrate numerically from large $u$ 
to small $u$ requiring the numerical solutions for $\psi_s$ and $\psi_t$ at 
small $u$ to behave as \eqref{asympsolescauv} and \eqref{asympsoltenuv}, 
respectively. 

The numerical results presented in this section were obtained using the first 
procedure described above.  Here we present our results for the models A 
(introduced in the previous section) as well as for the models B1 and B2, 
introduced in appendix \ref{App:ModelB}).

In this subsection we present the results for the glueball spectrum at a fixed 
value of the conformal dimension, namely $\epsilon =0.01$. At fixed $\epsilon$ 
the parameter $\hat \phi_0$ can be fixed by using as input the ratio between 
the first two scalar glueballs  
 \begin{equation} \label{ratio}
 R_{00}=\frac{m_{0^{++*}}}{m_{0^{++}}} = \frac{\hat m_{s,1}}{\hat m_{s,0}}\,, 
 \end{equation}
where $\hat m_{s,0}$ and $\hat m_{s,1}$ represent the first two scalar masses 
and $m_{0^{++}}$ and $m_{0^{++*}}$ are extracted from lattice QCD data 
\cite{Meyer:2004gx}. Once $\hat \phi_0$ is determined from the ratio $R_{00}$, 
we also fix the parameter $\Lambda$ by comparing the first scalar mass $m_{s,0} 
= \Lambda\, \hat m_{s,0}$ with the first glueball state $m_{0^{++}}$, extracted 
from lattice QCD data \cite{Meyer:2004gx}. 
Below we describe the results for the glueball spectrum obtained for each one 
of models A and B.

\subsubsection{Models A}

Implementing the procedure described above for the model A1, where the dilaton 
is given by equation (\ref{interp1}), we find for $\epsilon=0.01$ that  $\hat 
\phi_0=53.62$ and  $\Lambda=737 \, {\rm MeV}$. Any other parameter is defined in 
terms of $\epsilon$, $\hat \phi_0$ and $\Lambda$ and most of them are shown in 
Table \ref{taba0}.   
 
The results for the spectrum of scalar and tensor glueballs in model A1 are 
shown in the second column of Table \ref{taba01}. These results are in good 
agreement with the lattice QCD calculations \cite{Meyer:2004gx}, and 
also with the IHQCD model \cite{Gursoy:2007er}. The largest difference between 
our results and lattice QCD data is about $4.2\%$ in the case of $m_{0^{++**}}$. 
We remind the reader that the first two masses in Table \ref{taba01}, 
$m_{0^{++}}$ and $m_{0^{++*}}$, were used to fix $\hat \phi_0$ and $\Lambda$.  
Therefore, the predictions of the present models are the ones displayed from the 
third state ($0^{++**}$) and below in that table. 

\TABLE{
\begin{tabular}{c |c|c|c|c|c}
\hline 
\hline
 Model &  $\hat \phi_0$ &$\Lambda\left(\text{MeV}\right)$ & $\phi_0\left(\text{MeV}^\epsilon\right)$
  & $G\left(\text{MeV}^{4-\epsilon}\right)$ & $C\left(\text{MeV}^{2}\right)$  \\
\hline 
 A1 & 53.79 & 736 & 57.46 & $2.75\times 10^{11}$ & $5.42\times 10^{5}$   \\
 A2 & 49.41 & 682 & 52.75 & $2.03\times 10^{11}$ & $4.65\times 10^{5}$    \\
 B1 & 48.40 & 668 & 51.65 & $1.86\times 10^{11}$ & $4.46\times 10^{5}$   \\
 B2 & 46.57 & 709 & 49.73 & $2.36\times 10^{11}$ & $5.02\times 10^{5}$  \\
 \hline\hline
\end{tabular}
\caption{The values of the 
parameters we use to get the spectrum for the glueballs with
$\epsilon=0.01$.}
\label{taba0}}

The numerical results for the glueball spectrum are well fitted by linear 
trajectories. For the scalar sector we find the linear trajectory
\begin{equation}\label{fitscainter1}
m_{s,n}^{2}=\Lambda^2\,(8.65\, n+4.85), \quad n=0,1,2,...,
\end{equation}
while for the tensor sector we obtain
\begin{equation}\label{fitteninter1}
m_{t,n}^{\,2}=\Lambda^2\,(8.13\, n+7.92), \quad n=0,1,2,...
\end{equation}
The largest difference between the masses obtained with these fits and the 
lattice QCD results occurs for the state $0^{++}$ and is of about $10\%$.  

The same procedure was done for the model A2, where the dilaton is given by 
equation  (\ref{interp2}), and found for $\epsilon = 0.01$ the values $\hat 
\phi_0=49.41$ and $\Lambda=682 \, {\rm MeV}$. The other parameters are  
displayed in table \ref{taba0}.  

The mass spectrum obtained in model A2 is shown in the third column of table 
\ref{taba01}. We find again a good agreement between our results and those of 
the lattice QCD \cite{Meyer:2004gx} and the IHQCD model \cite{Gursoy:2007er}.  
In comparison to the lattice QCD masses, the largest difference is of about 
$2.4\%$ in the case of $m_{0^{++***}}$. 

The spectrum of the model A2 is also well approximated by linear fits. The 
linear trajectories in this case are 
\begin{equation}\label{fitscainter2}
m_{s,n}^{2}=\Lambda^2\,(8.62\, n+6.37), \quad n=0,1,2,...
\end{equation}
for the scalar sector and 
\begin{equation}\label{fitteninter2}
m_{t,n}^{\,2}=\Lambda^2\,(7.89\, n+10.16), \quad n=0,1,2,...
\end{equation}
for the tensor sector. When compared to the
the lattice QCD results, the maximum error obtained
for these fits is of about $16.8\%$, and it occurs for
the state  $0^{++}$.

\TABLE{
\begin{tabular}{l |c|c|c|c|c|l}
\hline 
\hline
 $n$&A1&
A2&B1 &B2 &
 IHQCD \cite{Gursoy:2007er}&
Lattice \cite{Meyer:2004gx} \\
\hline 
 $0^{++}$ & 1475 & 1475 & 1475 & 1475 & 1475   & 1475(30)(65)  \\
 $0^{++*}$ & 2755 & 2755 & 2755 & 2755 & 2753  & 2755(70)(120)  \\
 $0^{++**}$ & 3507 & 3376 & 3361 & 3449 & 3561  & 3370(100)(150) \\
 $0^{++***}$ & 4106 & 3891 & 3861 & 4019 & 4253  & 3990(210)(180)  \\
 $0^{++****}$ & 4621 & 4349 & 4313 & 4514 & 4860  &   \\
 $0^{++*****}$ & 5079 & 4762 & 4721 & 4956 & 5416   &  \\
 $2^{++}$ & 2075 & 2180 & 2182 & 2130  & 2055  & 2150(30)(100) \\
 $2^{++*}$ & 2945 & 2899 & 2887 & 2943 & 2991 & 2880(100)(130)  \\
 $2^{++**}$ & 3619 & 3468 & 3444 & 3568 & 3739 & \\
 $2^{++***}$ & 4185 & 3962 & 3928 & 4102  & 4396 &  \\
 $2^{++****}$ & 4680 & 4404 & 4365 & 4576  & 5530 &  \\
 $2^{++*****}$ & 5127 & 4807 & 4763 & 5006  &  &  \\
 \hline\hline
\end{tabular}
\caption{The glueball masses (in MeV) obtained in our model, compared against the results of 
IHQCD \cite{Gursoy:2007er} and Lattice QCD \cite{Meyer:2004gx}. The first two values of 
masses for $0^{++}$ and $0^{++*}$ are used as input data in our procedure.
The results here were obtained with $\epsilon=0.01$.}
\label{taba01}}

\subsubsection{Models B}

The two models we named B1 and B2 are described in appendix \ref{App:ModelB} 
and correspond to the case where the warp factor $A(z)$ has an analytic form, 
while the dilaton is solved numerically from the first equation in 
\eqref{IHQCDEqsv2}.  

The parameters obtained for the models B1 and B2 at $\epsilon=0.01$ are shown in 
the third and fourth row of Table \ref{taba0}. The glueball spectra obtained for 
the models B1 and B2 are shown in the fourth and fifth column of Table 
\ref{taba01}. The numerical results for models B1 and B2 are also in good 
agreement with the Lattice QCD and IHQCD model. The approximate linear 
trajectories for model B1 are 
\begin{equation}\label{fitB1}
m^{2}_{s,n}=\Lambda^2\,(8.80\,n+6.74),\qquad 
m^{2}_{t,n} =\Lambda^2\,(8.04\,n+10.62), \qquad n=0,1,2,\cdots\,,
\end{equation}
while for model B2 we obtain 
\begin{equation}\label{fitB2}
m^{2}_{s,n}=\Lambda^{2}\,(8.80\,n+5.47),\qquad 
m^{2}_{t,n}=\Lambda^2\,(8.17\,n+9.03), \qquad n=0,1,2,\cdots\,.
\end{equation}

In figure \ref{fitinter2} we plot the results for the spectra of scalar and 
tensor glueballs obtained in the 4 models considered in this work (A1, A2, B1 
and B2) and, for comparison, we include the data of lattice QCD 
\cite{Meyer:2004gx}. The results for models A1 and B1 are shown in figure 
\ref{scalartensorall}. The plots show clearly the pattern $m_{s,n} < m_{t,n}$, 
also observed in the IHQCD model \cite{Gursoy:2007er}. Notice that the 
difference between the scalar and tensor glueball masses decreases as $n$ 
increases. This indicates a degeneracy of the scalar and tensor glueballs at 
very large $n$. 
\FIGURE{
\begin{tabular}{*{2}{>{\centering\arraybackslash}p{.5\textwidth}}}
        \includegraphics[width=7.2cm,angle=0]{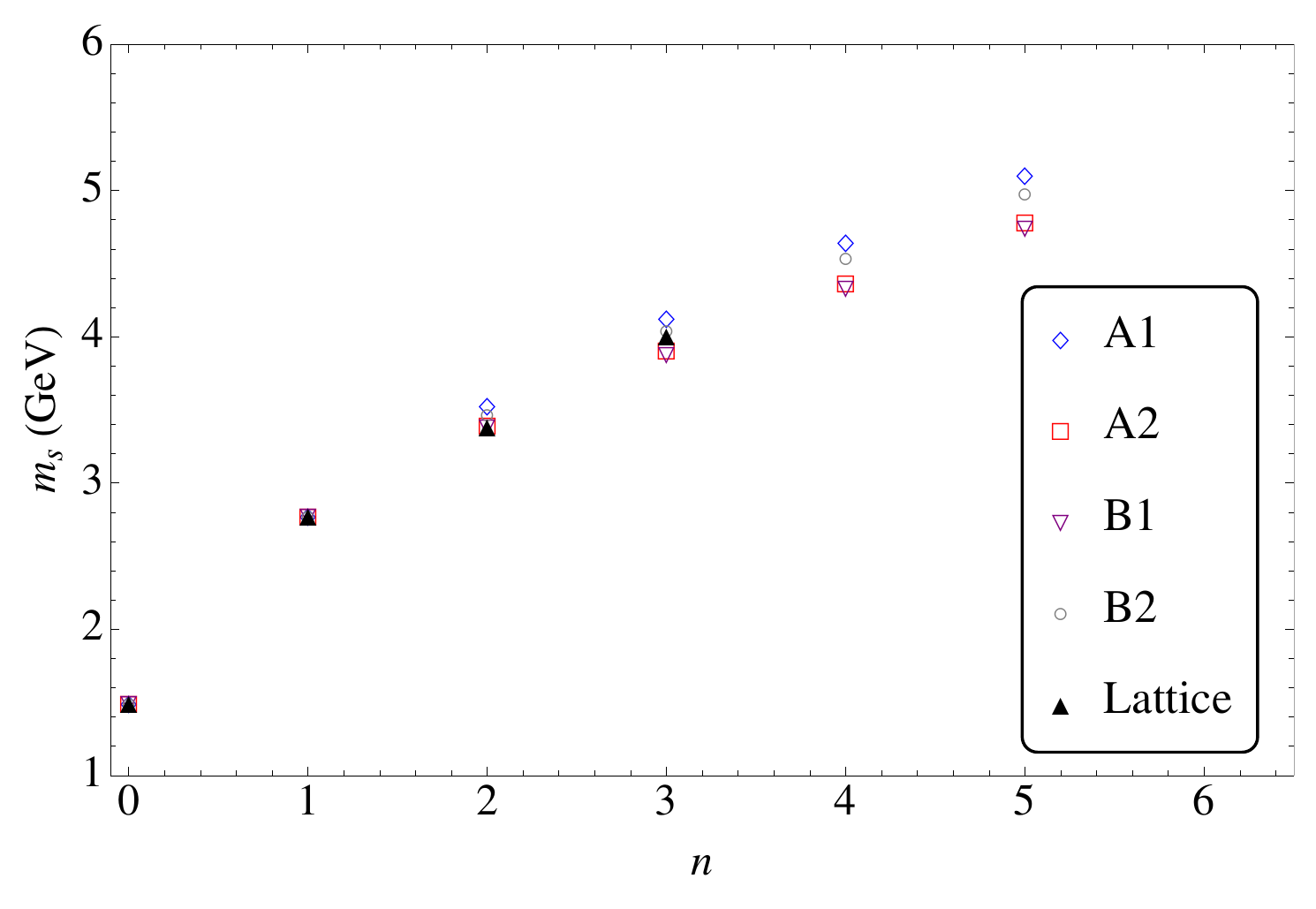} 
& 
\hspace{-1cm} \includegraphics[width=7.2cm,angle=0]{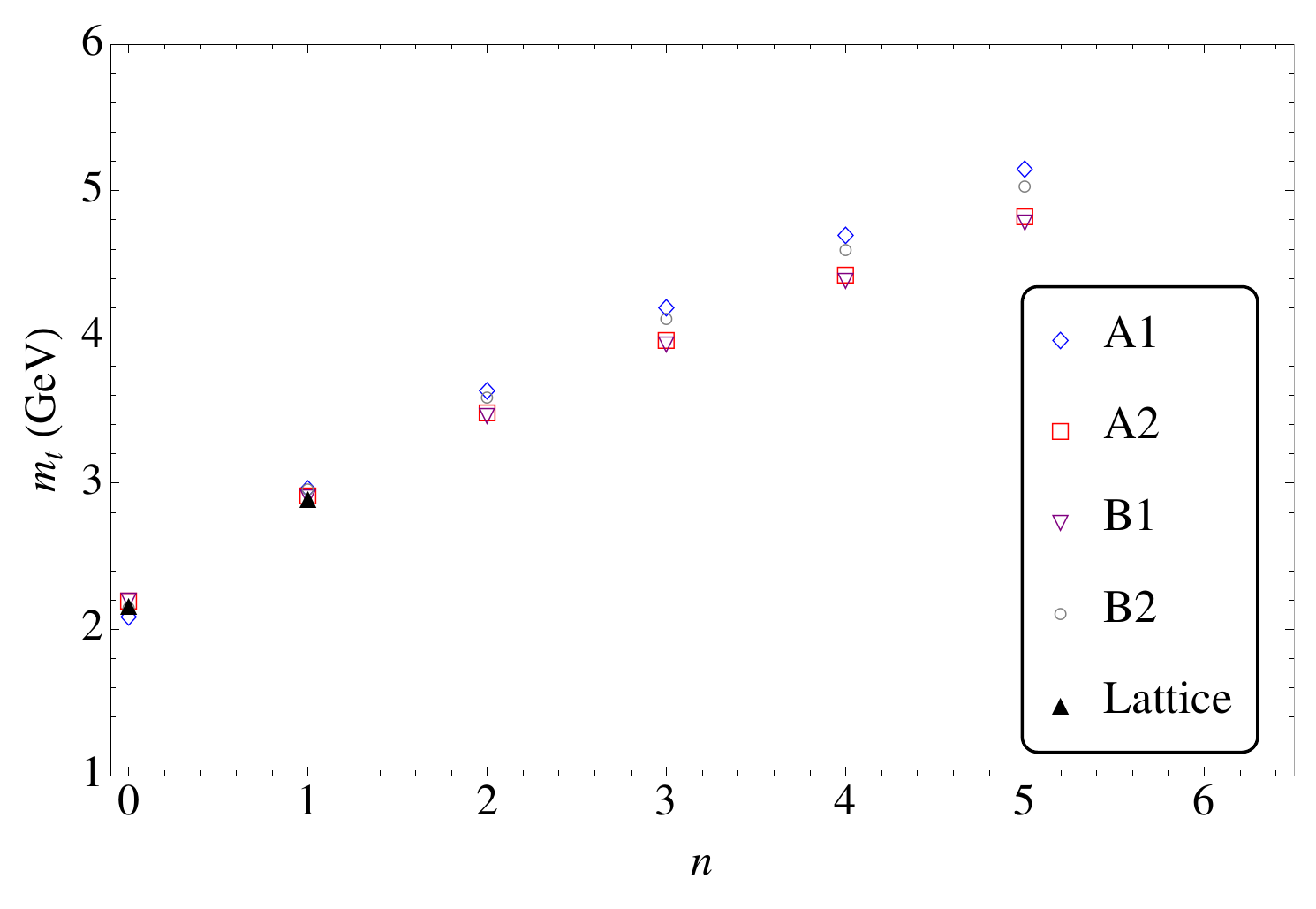} 
\end{tabular}
    \caption{The glueball spectrum for the scalar ({\bf left panel}) and tensor ({\bf right 
panel}) sectors obtained in models A1, A2, B1, and B2 at $\epsilon=0.01$, 
compared against lattice QCD data \cite{Meyer:2004gx}.}
     \label{fitinter2}}
\FIGURE{
\centering
\includegraphics[width=9.2cm,angle=0]{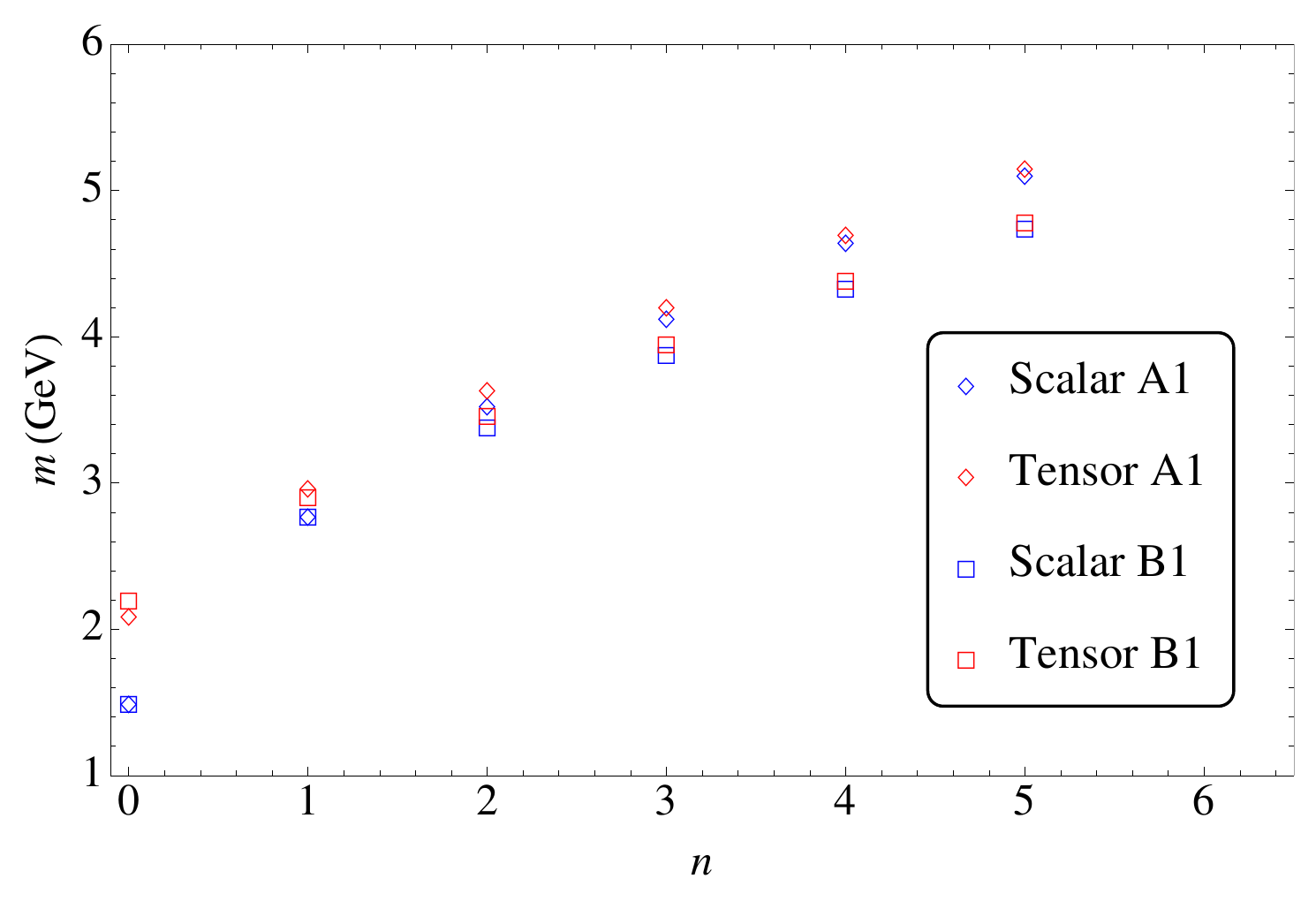} 
\caption{The scalar and tensor glueball spectrum for models A1 and B1 at $\epsilon=0.01$.}
      \label{scalartensorall}}

\subsection{Running parameters}
\label{Sec:RunningParameters}

So far in this section, all the calculations were done for a specific value of 
the conformal dimension, $\epsilon=0.01$. But what happens when this parameter 
varies? Here we find the evolution of the parameters $\hat \phi_0$ and $\Lambda$ 
with the conformal dimension $\epsilon$ for the models A1 and A2. As explained 
previously in this section, for any given $\epsilon$ we use the masses of the 
first two scalar glueballs, extracted from lattice QCD, as an input for fixing 
$\hat \phi_0$ and $\Lambda$.
 
The results for $\hat \phi_0$ are displayed on the left panel of figure 
\ref{lambdaRunning}. A  numerical fit shows  that $\hat \phi_0$ diverges as
$1/\epsilon$ as the parameter $\epsilon$ goes to zero. The evolution of 
$\Lambda$ with $\epsilon$ is shown on the right panel of figure 
\ref{lambdaRunning}. The evolution is very slow suggesting that $\Lambda$ should 
be approximated by a constant. Linear fits for these results give 
$\Lambda=735.18+75.48\, \epsilon$ for the model A1 and $\Lambda=682.27-43.22\, 
\epsilon$
 for the model A2, both in MeV units.  
 
Additionally, we find the evolution of the parameters $\phi_0$ and $G$, related 
to $\hat \phi_0$ and $\Lambda$ by equation \eqref{DefParameters}. The results 
for $\phi_0$ are 
 shown on the left panel of figure \ref{phiRunning}. A fit for the model A1 gives  
 $\phi_0=4.47+0.53/\epsilon$, while for the model A2 one finds
 $\phi_0=4.42+0.49/\epsilon$ (both in 
 MeV$^{\epsilon}$ units). The evolution of the parameter $G$ is displayed
 on the right panel of figure \ref{phiRunning}. A numerical fit of the 
data corresponding to such figures, shows that, when $\epsilon$ goes to zero, 
$G$ reaches a finite value: $2.92\times 10^{11}\text{MeV}^{4}$ for the model A1 
and $2.16\times 10^{11}\text{MeV}^{4}$ for the model A2.

\FIGURE{
\begin{tabular}{*{2}{>{\centering\arraybackslash}p{.5\textwidth}}}
\hspace{-1cm}   \includegraphics[width=7.5cm,angle=0]{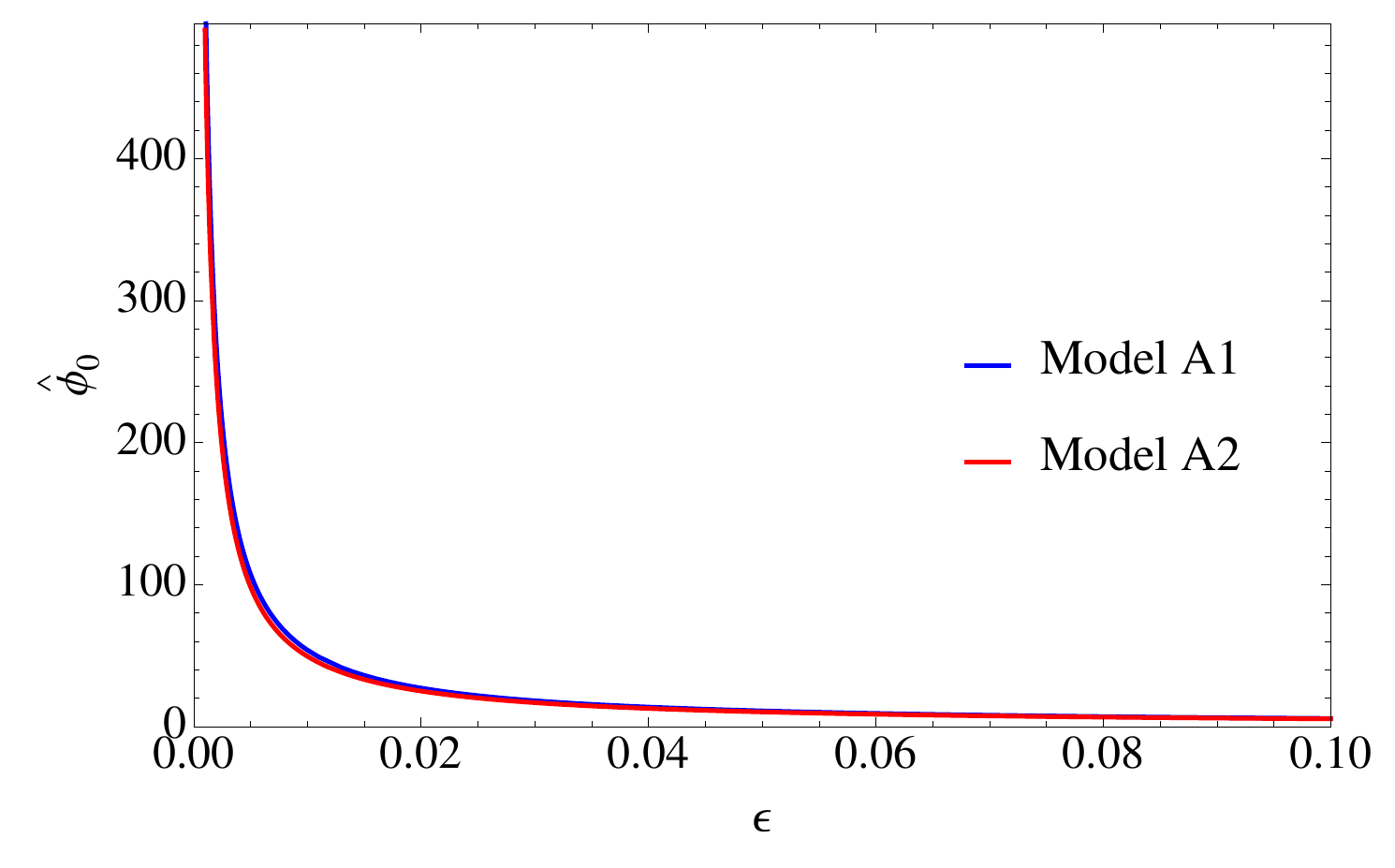} 
& 
\hspace{-1cm}  \includegraphics[width=7.5cm,angle=0]{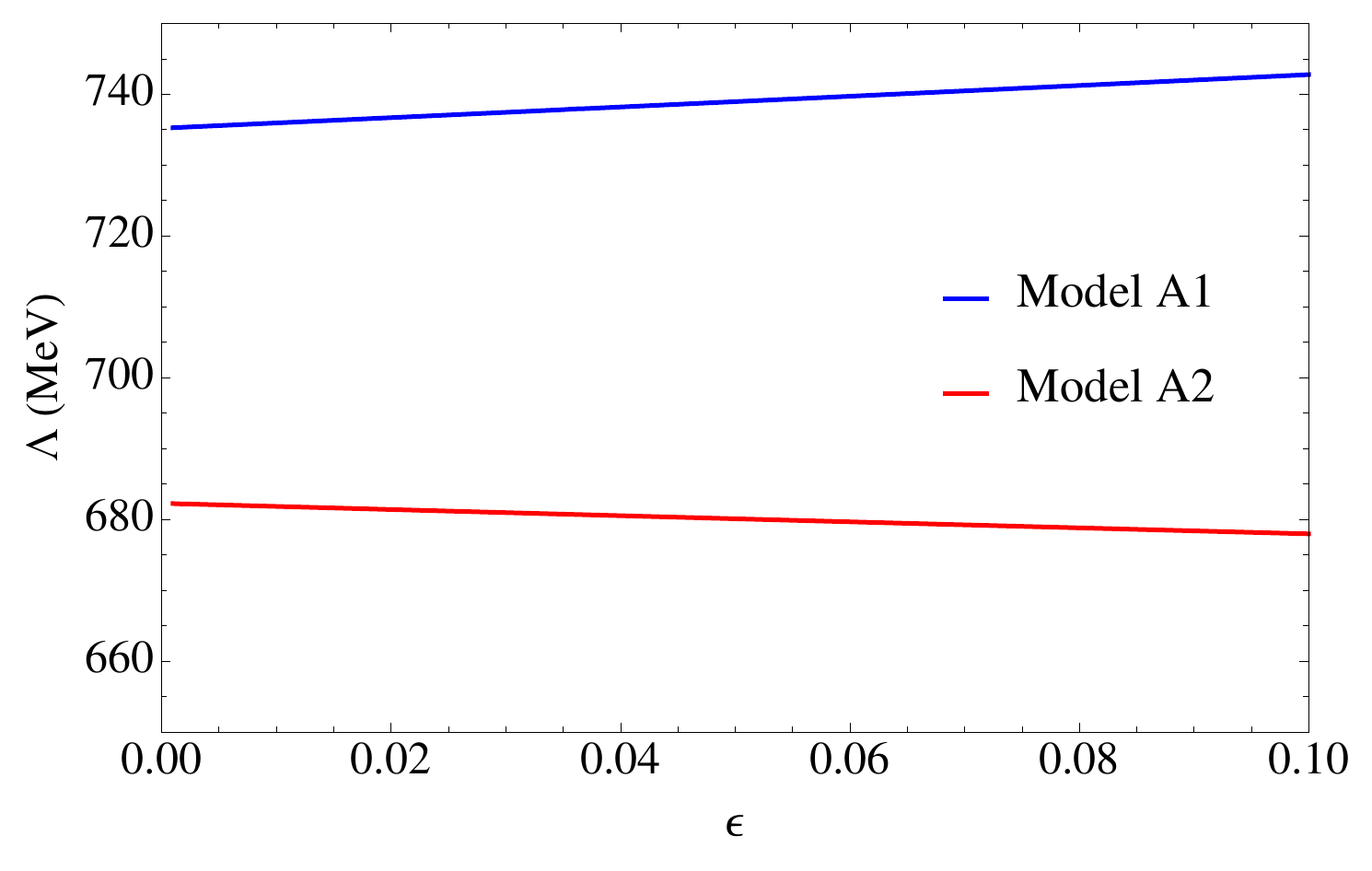}
\\
\end{tabular}
    \caption{Evolution of the parameters $\hat \phi_0$ ({\bf left panel})
    and $\Lambda$ ({\bf right panel}) with the conformal
    dimension $\epsilon$ for models A1 and A2.}
     \label{lambdaRunning}}
%
\FIGURE{
\begin{tabular}{*{2}{>{\centering\arraybackslash}p{.5\textwidth}}}
\hspace{-1cm}        \includegraphics[width=7.5cm,angle=0]{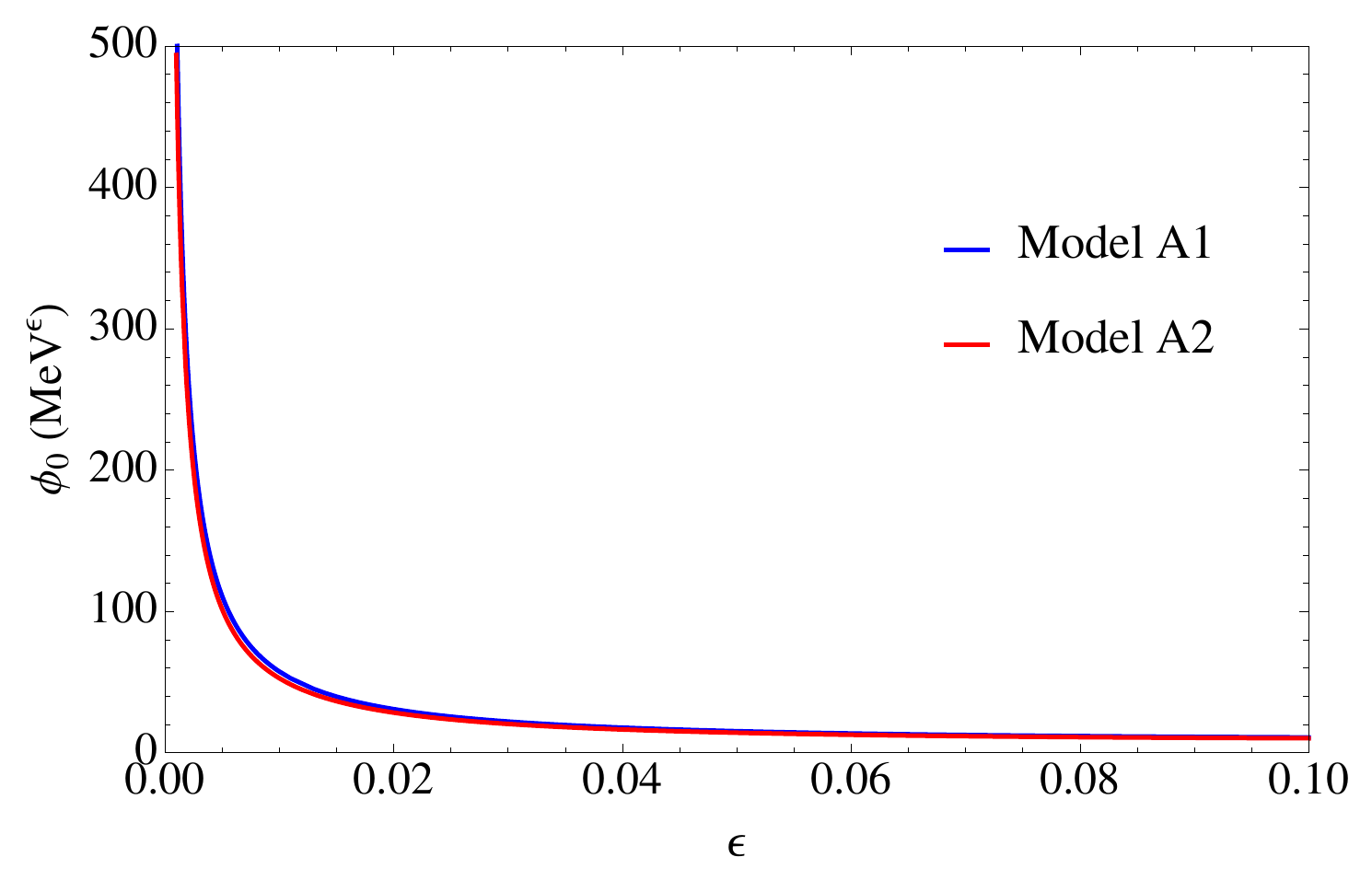}
& 
\hspace{-1cm}  \includegraphics[width=7.5cm,angle=0]{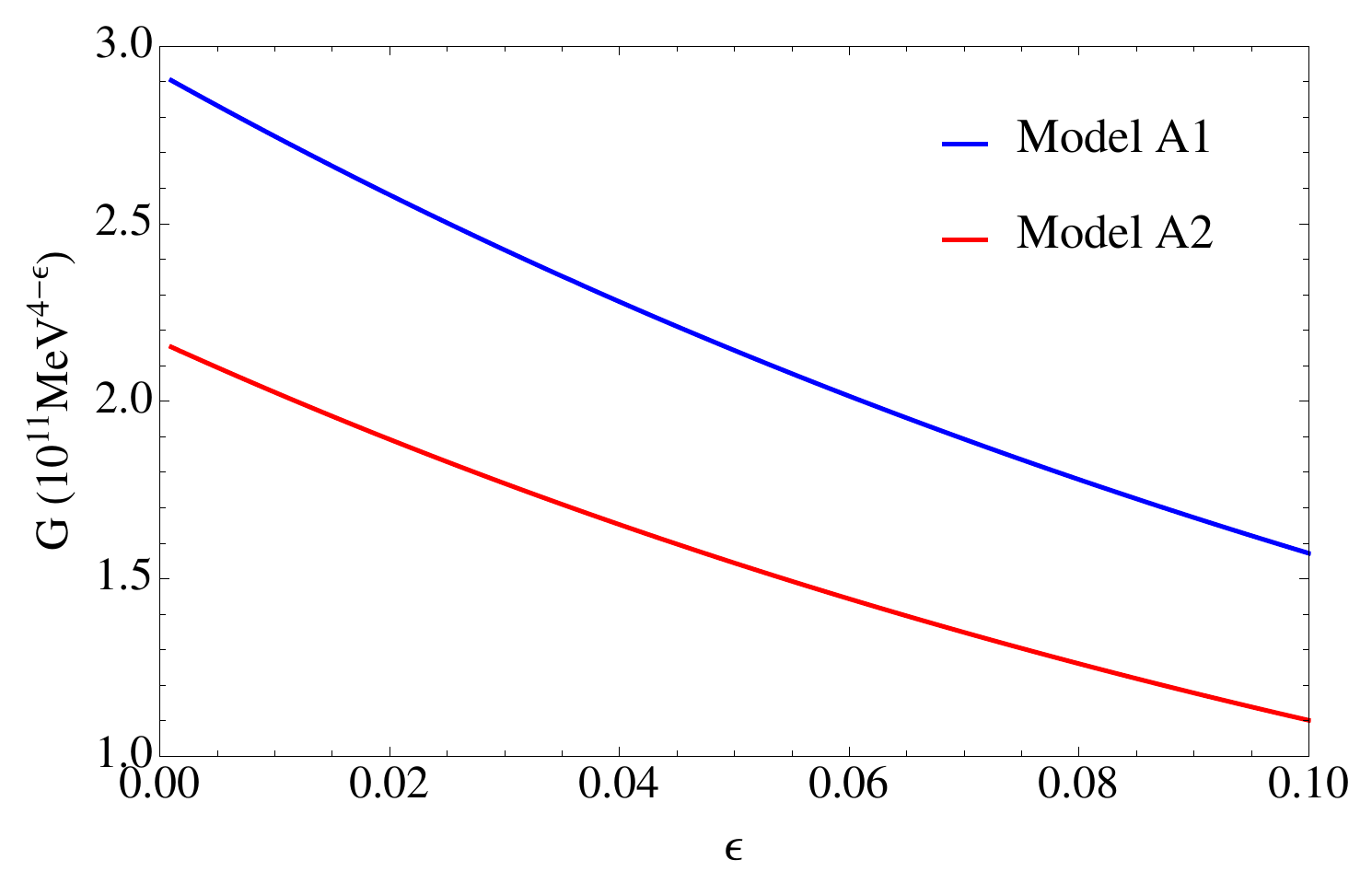} 
\\
\end{tabular}
    \caption{Evolution of the parameters $\phi_0$ ({\bf left panel})
    and $G$ ({\bf right panel}) with the conformal
    dimension $\epsilon$ for models A1 and A2.}
     \label{phiRunning}}

We have also obtained numerically the vacuum energy density, given by 
\eqref{RNEnergyEq}. The results are shown in figure \ref{EQCDRunning}. One 
clearly sees that the vacuum energy density converges to a finite value when 
$\epsilon$ goes to zero. This can be understood as follows. We showed previously 
that, when $\epsilon$ goes to zero, the source $\phi_0$ goes as 
$c_1+c_2/\epsilon$.  Therefore, the renormalized vacuum energy density 
($\mathcal{E}_{QCD}^{ren}\propto \epsilon\, \phi_0$) goes as $c_2+c_1\, 
\epsilon$. The numerical results shown in figure~\ref{EQCDRunning} indicate that 
the vacuum energy density evolves very slowly with the conformal dimension 
$\epsilon$, in both models A1 and A2, and can be approximated by a constant. 
Setting $M^3 N_c^2$ to unity, our predictions for $\mathcal{E}_{QCD}^{ren}$, in 
the limit $\epsilon \to 0$, are $-0.17  \, \text{GeV}^4$ in the model A1, and 
$-0.11 \, \text{GeV}^4$ in the model A2.  For comparison, an analysis 
made in \cite{Fraga:2012ev}, by considering the large $N_c$ limit, led to 
$\mathcal{E}_{QCD}=-c_{0}^{4}\,N_{c}^{2}\,\sigma_0^{2}\,$, where $\sigma_0$ is 
the QCD string tension and $c_0$ is a constant of order one.  Considering a 
phenomenological value for the string tension, e.g. $\sigma_0 =  \left ( 0.44 \, 
{\rm GeV} \right )^2$ \cite{Lucini:2013qja}, and taking $N_c=3$ and $c_0=1$ one 
gets $\mathcal{E}_{QCD} \approx - 0.34 \, \text{GeV}^4$. 
\FIGURE{
\centering
\includegraphics[width=10cm,angle=0]{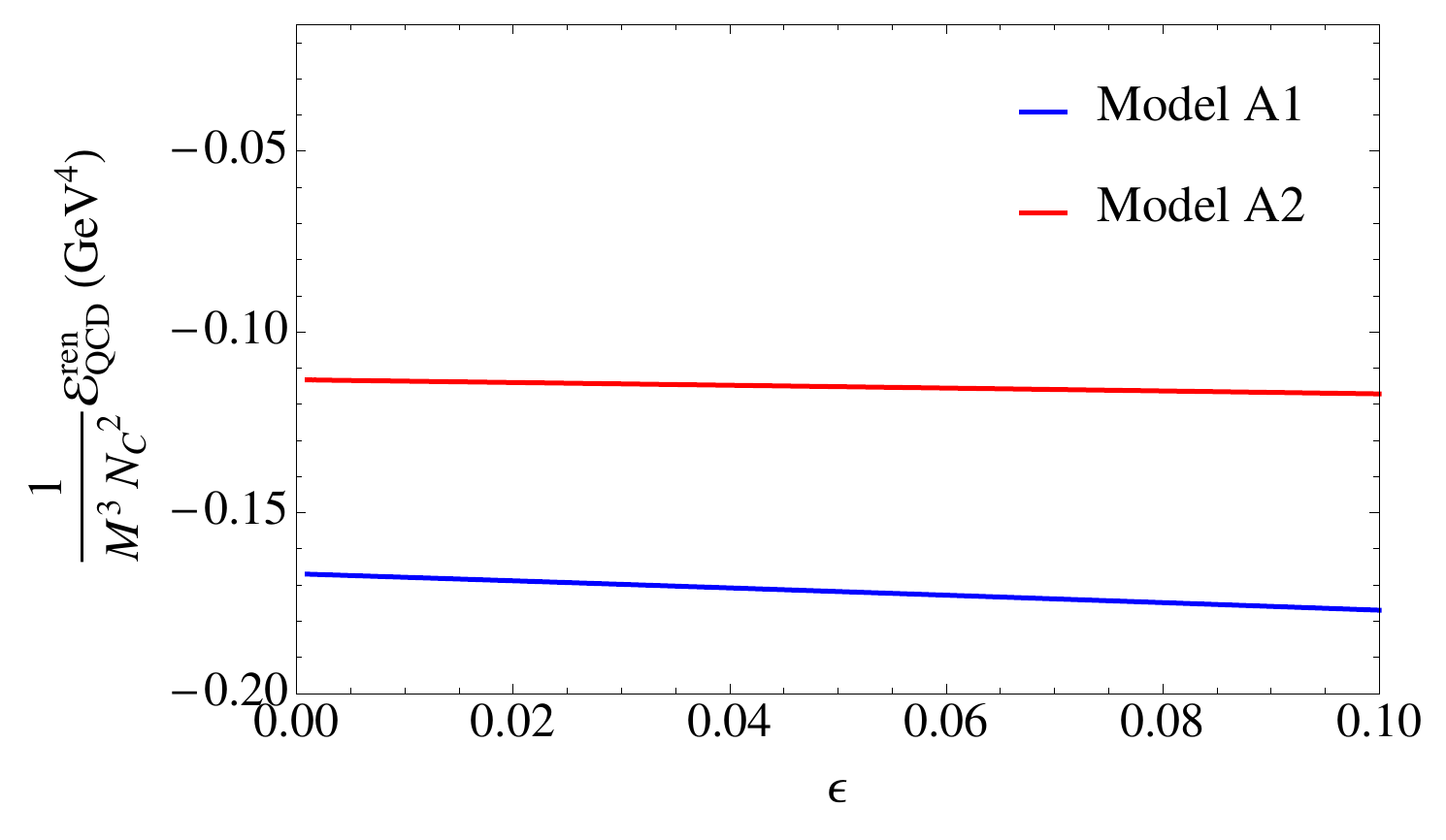} 
\caption{Evolution of the renormalized vacuum 
    energy density      $\mathcal{E}_{\text{QCD}}^{ren}$  with the conformal 
dimension $\epsilon$ for models A1 and A2. 
}
      \label{EQCDRunning}}

\section{The trace anomaly: from deformed CFTs to QCD}
\label{Sec:TraceAnomaly}

In this section we take advantage of the AdS/CFT dictionary and reproduce 
the universal result for the trace anomaly of deformed CFTs 
\cite{Skenderis:2002wp,Papadimitriou:2016yit} for the particular backgrounds 
considered in this work. This result in turn suggests a map between 
deformed CFTs and  large-$N_c$ QCD.  An important ingredient in this map is the 
reinterpretation of the CFT deformation  $\delta {\cal L} = \phi_{0}\, {\cal O}$ 
in terms of the large-$N_c$ Yang-Mills Lagrangian ${\cal L}_{YM}$. 

In Euclidean signature the large-$N_c$ Yang-Mills Lagrangian can be written as 
\cite{Lucini:2012gg}
\beqa
{\cal L}_{YM} = N_c \, \bar {\cal L}_{YM} \, , 
\eeqa 
with
\beqa
\bar {\cal L}_{YM} = \frac{1}{\lambda} \left ( \frac{1}{2}  {\rm Tr} \, F^2 
\right ) \,. \label{YMLag}
\eeqa
Here we can make use of the mass parameter $\Lambda$, defined previously in 
\eqref{DefParameters}, which has conformal dimension $1$. Multiplying and 
dividing equation \eqref{YMLag} by $\Lambda^{\epsilon}$, we get  
\beqa
\bar {\cal L}_{YM} = \frac{\Lambda^{\epsilon}}{\lambda}    \left ( \frac{1}{2}  
\frac{ {\rm Tr} \, F^2}{\Lambda^{\epsilon}}  \right ) \,. 
\eeqa
In this way we have dressed the inverse 't Hooft coupling $1/\lambda$ and the 
QCD operator $\frac{1}{2} {\rm Tr} \, F^2$ so that they acquire conformal 
dimensions $\epsilon$ and $4-\epsilon$, respectively. Matching the Lagrangian 
$\bar {\cal L}_{YM}$ to the CFT deformation $\delta {\cal L} = \phi_{0}\, 
{\cal O}$, we obtain the map 
\beqa
\phi_0 = \frac{\Lambda^{\epsilon}}{\lambda}\,, \qquad \quad 
{\cal O} = \frac{1}{2}  \frac{ {\rm Tr} \, F^2}{\Lambda^{\epsilon}} \,. 
\label{MapDefCFT}
\eeqa
Below we will use this map for calculating the gluon condensate and to find a 
connection between the trace anomaly of deformed CFTs and the QCD trace 
anomaly.

\subsection{The gluon condensate}

According to the AdS/CFT dictionary the VEV of an operator ${\cal O}$ is 
obtained from the variation of the on-shell action 
\cite{Gubser:1998bc,Witten:1998qj,Skenderis:2002wp,Papadimitriou:2016yit} 
\beqa
\delta S^{o-s} = \int d^4 x \, \delta \phi_0 \langle {\cal O} \rangle \,. 
\label{genvarSos}
\eeqa
where $\delta \phi_0$ is the variation of the source. In the case of holographic 
QCD backgrounds, the on-shell action was  obtained in subsection 
\ref{subsec:vacuumenergy} with the result 
\beqa
S^{{\rm o-s}} = 6M^3N_c^2 \int d^4 x \, e^{3A(z_0)} A'(z_0), 
\eeqa
with small $z_0$ and, as usual, the limit $z_0 \to 0$ will be taken 
at the very end of the calculation process. We then variate this on-shell 
action by considering the warp factor $A$ as a field whose dynamics is 
completely determined by the dilaton field $\Phi$, $A=A(\Phi)$. Therefore, we 
get 
\beqa
\delta S^{{\rm o-s}} &=& 6M^3N_c^2 \int d^4 x \, \delta \Phi \frac{d}{d \Phi}  
\left [ e^{3A(z_0)} A'(z_0) \right ] \cr
&=& 6M^3N_c^2 \int d^4 x \, \delta \phi_0 \, z_0^{\epsilon} \,  
\frac{1}{\Phi'(z_0)} \frac{d}{d z_0} \left [ e^{3A(z_0)} A'(z_0) \right ],
\label{varSosHQCD}
\eeqa
where we used the asymptotic form of the dilaton \eqref{dilaUV} to 
rewrite $\Phi$ in terms of the source $\phi_0$. From equations 
\eqref{genvarSos} and \eqref{varSosHQCD} we find the bare VEV of the operator 
${\cal O}$,
\beqa
\langle {\cal O} \rangle &=& 6M^{3}N_{c}^{2} z_{0}^{\epsilon} \, 
\frac{1}{\Phi'(z_0)}  e^{3A(z_0)}  \left [  A''(z_0) + 3 A'^{\, 2}(z_0) \right 
].
\eeqa
Using the field equations \eqref{IHQCDEqsv2} and the definition \eqref{xEq}, we 
may rewrite this VEV as  
\beqa
\langle {\cal O} \rangle &=& 24 M^{3}N_{c}^{2}z_{0}^{\epsilon} \, 
e^{3A(z_0)} \frac{A'(z_0) }{3 X(z_0)}   \left [  1 - X^2(z_0) \right ]. 
\label{finalbareVEV}
\eeqa
As it was done previously for the case of the bare energy density, we use the UV 
asymptotic form of $A$ and $X$  to identify the divergent and 
non-divergent terms in \eqref{finalbareVEV}. Again, we consider a MS scheme for renormalization. Eliminating the divergent terms and taking the $z_0 \to 0$ limit, we obtain the renormalized VEV 
\beqa
\label{renVEV}
\langle {\cal O} \rangle^{ren}=
\frac{16}{15} M^3N_c^2  (4 - \epsilon) G  \, .
\eeqa
Using the map \eqref{MapDefCFT} and the definitions in \eqref{DefParameters}, 
 we find the gluon condensate
\beqa
\langle {\rm Tr} \, F^2 \rangle^{ren}=
\frac{32}{15} M^3N_c^2  (4 - \epsilon) \Lambda^4  \, .
\label{gluoncondensate}
\eeqa
In figure~\ref{CondensateRunning} we plot our numerical results for the gluon 
condensate $\left\langle \frac{1}{4\pi^2} {\rm Tr} \, F^{2}\right\rangle^{ren}$ 
as a function of $\epsilon$ for the models A1 and A2. We remark that the gluon condensate in \eqref{gluoncondensate} was obtained in a particular renormalization scheme. Setting $M^3 N_c^2$ to 
unity and taking the limit $\epsilon \to 0$, the results for the gluon 
condensate $\left\langle \frac{1}{4\pi^2} {\rm Tr} \, F^{2}\right\rangle^{ren}$ 
are $0.063\,\text{GeV}^{4}$ and $0.047\,\text{GeV}^{4}$ for the models A1 and 
A2, respectively. For comparison, we present other results in the literature. 
The values obtained using SVZ sum rules are $0.013\,\text{GeV}^{4}$ 
\cite{Vainshtein:1978wd} and $0.012\,\text{GeV}^{4}$ \cite{Shifman:1978bx}. 
Previous results in holographic QCD include $0.043\,\text{GeV}^{4}$ 
\cite{Csaki:2006ji} and $0.01\,\text{GeV}^{4}$ \cite{Andreev:2007vn}. We also 
mention two different results in SU(3) lattice gauge theory: 
$0.10\,\text{GeV}^{4}$ \cite{Campostrini:1989uj} and $0.04\,\text{GeV}^{4}$ 
\cite{Rakow:2005yn}.  

\FIGURE{
\centering
\includegraphics[width=10cm,angle=0]{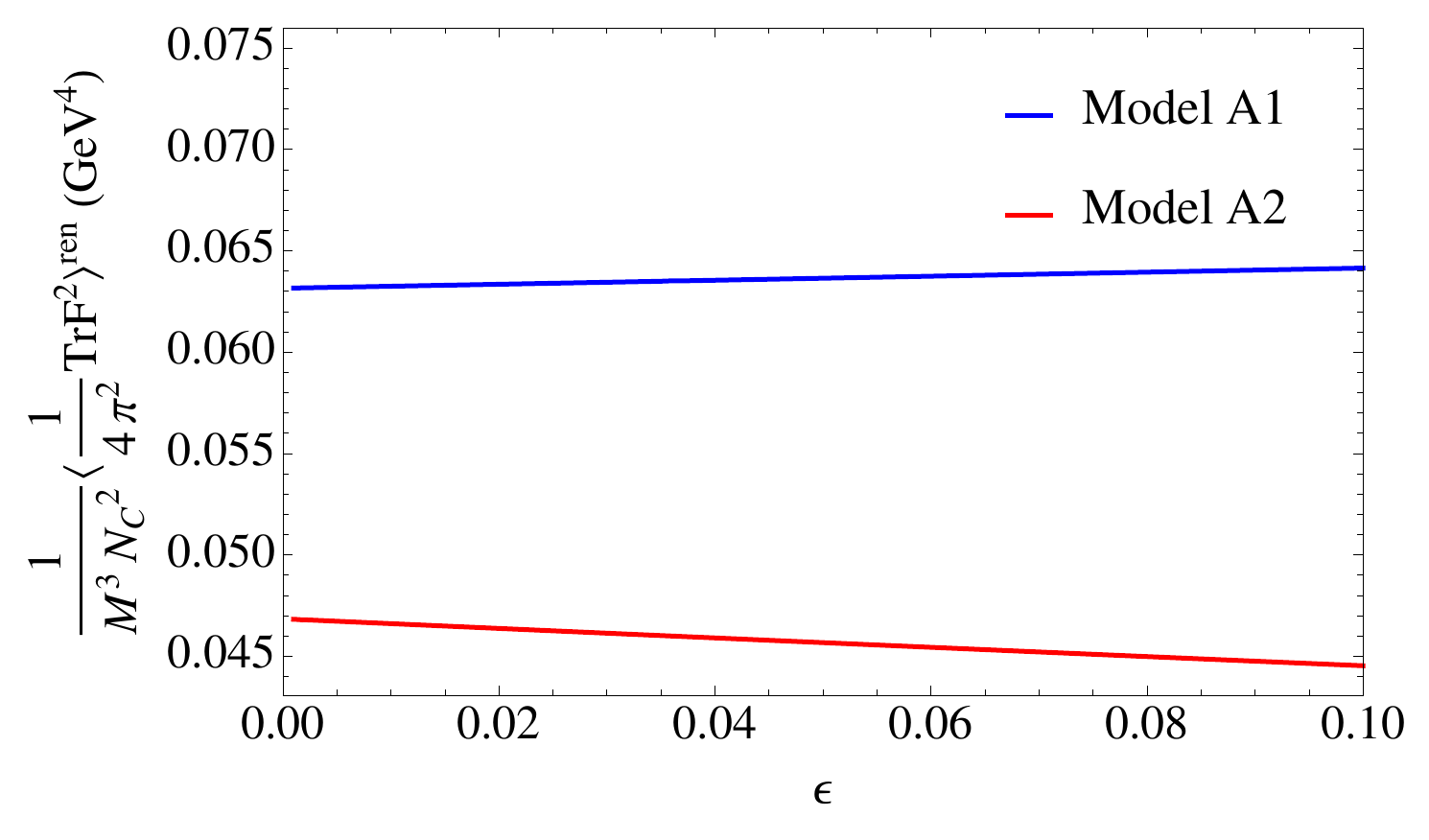} 
\caption{Evolution of the renormalized gluon condensate $\langle \frac{1}{4 
\pi^2} {\rm Tr} \, F^2 \rangle^{ren}$ with the conformal
    dimension $\epsilon$ for models A1 and A2.}
      \label{CondensateRunning}}

\subsection{The trace anomaly} 

The renormalized vacuum energy density $\mathcal{E}^{ren}_{\text{QCD}}$ 
for our model is given by equation \eqref{RNEnergyEq}.
At zero temperature, the pressure is just $- \mathcal{E}^{ren}_{\text{QCD}}$, 
so the trace of the energy momentum tensor is given by
\beqa
\label{TraceEMTensor}
\langle T^{\mu}_{\;\;\mu} \rangle^{ren} = 4 \, \mathcal{E}^{ren}_{\text{QCD}}=
-\frac{16}{15} M^3N_c^2\epsilon (4 - \epsilon) \phi_0 G  \, ,
\eeqa
From equations \eqref{renVEV} and \eqref{TraceEMTensor},  we find the relation
\beqa
\label{ConfAnomaly}
\langle T^{\mu}_{\;\; \mu} \rangle^{ren} = - \epsilon \, 
\phi_0 \langle {\cal O} \rangle^{ren} \,. 
\eeqa
This is the universal trace anomaly of 4-$d$ CFTs deformed by an operator 
${\cal O}$ with dimension $ \Delta = 4- \epsilon$ and coupling $\phi_0$ 
\cite{Skenderis:2002wp,Papadimitriou:2016yit}. The quantity $- \epsilon \, 
\phi_0$ is the classical $\beta$ function associated with the coupling $\phi_0$. We have reproduced this trace 
anomaly within the context of effective holographic QCD backgrounds, where the 
dilaton and warp factor depend solely on the radial coordinate $z$. This trace 
anomaly describes the explicit breaking of conformal symmetry and, as described 
in the previous section, a nontrivial consequence of this symmetry breaking is 
the discrete spectrum of scalar and tensor glueballs. It is interesting to note 
that the limit $\epsilon \to 0$, with $\phi_0$ fixed, corresponds to the case 
where conformal symmetry is spontaneously broken. In that case, as explained in 
appendix \ref{App:massless}, the first scalar glueball becomes a Nambu-Goldstone 
boson.  

The trace anomaly \eqref{ConfAnomaly} holds for more general backgrounds where 
the dilaton and metric are more involved. In any case the conformal dimension 
of ${\cal O}$ always maps to a mass term for the dilaton via the relation $M^2 
=\Delta (\Delta - 4)$. Although a general proof of \eqref{ConfAnomaly} was 
developed in \cite{Skenderis:2002wp,Papadimitriou:2016yit}, it is always 
illuminating to reproduce this trace anomaly case by case. In particular, it 
would be interesting to prove \eqref{ConfAnomaly} for the case of black hole 
solutions, such as those considered in  \cite{Gubser:2008ny,Gubser:2008yx}. 

We finish this section proposing a dictionary between the conformal trace anomaly 
\eqref{ConfAnomaly} and the QCD trace anomaly. Making use of the map 
\eqref{MapDefCFT} we can rewrite \eqref{ConfAnomaly} as 
\beqa
\label{ConfAnomalyV2}
\langle T^{\mu}_{\;\; \mu} \rangle^{ren} = - \frac{\epsilon}{2 \lambda} 
\langle {\rm Tr } F^2 \rangle^{ren} \,. 
\eeqa
This result looks very similar to the QCD trace anomaly 
\beqa
\label{QCDAnomaly}
\langle T^{\mu}_{\;\; \mu} \rangle^{ren} =  \frac{\beta}{2  \lambda^2}
\langle {\rm Tr } F^2 \rangle^{ren} \,, 
\eeqa
suggesting the identification $\epsilon = - \beta/\lambda$. We remind the reader 
that the CFT deformation takes place at some UV energy scale $E^{*}$ so the 
coupling $\lambda$ appearing in equation\eqref{ConfAnomalyV2} is actually 
evaluated at that scale.  

A few remarks are in order here. The dictionary proposed in this work differs 
from the original proposal \cite{Gubser:2008ny,Gubser:2008yx} because we map the 
conformal dimension $\epsilon$ to the $\beta$ function of the 4-$d$ theory, 
instead of the anomalous dimension of ${\rm Tr} \, F^2$. Our dictionary also 
differs significantly from \cite{Gursoy:2007cb,Gursoy:2007er} where the 4-$d$ 
$\beta$ function is mapped to the 5-$d$ field $X$. In 
\cite{Gursoy:2007cb,Gursoy:2007er} the 4-$d$ energy scale and coupling are 
mapped to the 5-$d$ warp factor and dilaton in a very natural way. However, 
evaluating correlation functions at any RG energy scale $E^{*}$ becomes a 
difficult task due to the necessity of introducing a geometric cutoff 
$z^{*}$.\footnote{See, however, the recent progress made in 
\cite{Kiritsis:2014kua}.} In the effective holographic approach considered here, 
a RG energy scale $E^{*}$ does not imply cutting the 5-$d$ geometry so that one 
can make full use of the AdS/CFT dictionary by embedding holographic QCD in the 
framework of holographic deformed CFTs, developed in 
\cite{Bianchi:2001de,Skenderis:2002wp,Papadimitriou:2016yit}. 
\FIGURE{
\centering
\includegraphics[width=10cm,angle=0]{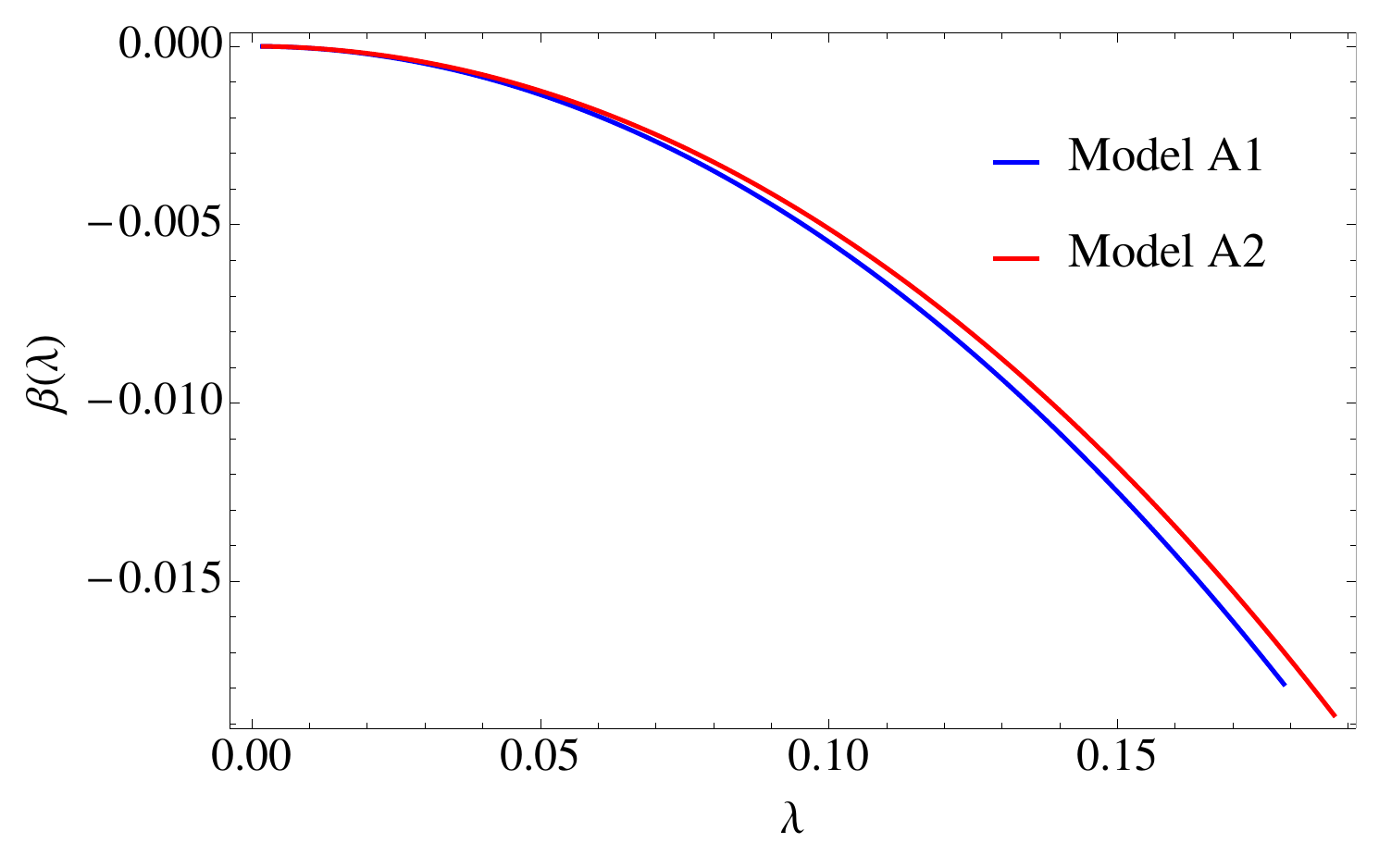} 
\caption{The 4-$d$ $\beta$ function in terms of $\lambda$, for models A1 and A2, 
obtained from the dictionary proposed in this section.}
      \label{BetaFunct}}

Figure~\ref{BetaFunct} shows the numerical results for the 4-$d$ $\beta$ 
function in terms of $\lambda$. According to the dictionary proposed in this 
work, these quantities are identified with $- \epsilon /{\hat \phi_0}$ and 
$1/{\hat \phi_0}$, respectively. We made a numerical fit for small $\lambda$  
and interestingly the fit takes the form $\beta= - b_0 \lambda^2 - b_1 
\lambda^3$. This is the same form arising in large-$N_c$ perturbative QCD at 
two-loops, with $b_0$ and $b_1$ being scheme independent. For the model A1 we 
find  $b_0 \approx 0.54$ and $b_1 \approx 0.48 \, b_0^2$, while for the model A2 
we obtain $b_0 \approx 0.49$ and $b_1 \approx 1.04 \, b_0^2$. For comparison, 
the coefficients of the large-$N_c$ perturbative QCD $\beta$ function are $b_0 
\approx 0.046$ and $b_1 \approx 0.42 \, b_0^2$.

We finish this section describing the relation between the anomalous dimension 
$\epsilon_{{\rm an}}$ of ${\rm Tr} \, F^2$ and the conformal dimension 
$\epsilon$. Following Ref. \cite{Gubser:2008yx}, the anomalous dimension 
$\epsilon_{{\rm an}}$ can be extracted from the QCD trace anomaly 
\eqref{QCDAnomaly} with the result 
\beqa
\epsilon_{{\rm an}} = - \lambda^2 \partial_{\lambda}  
\left [ \beta \lambda^{-2} \right ]. \label{AnDim}
\eeqa
At small $\lambda$ the anomalous dimension \eqref{AnDim} is approximated by 
$\epsilon_{{\rm an}} \approx b_1 \lambda^2$.
Using the identifications proposed in this work, namely $\hat \phi_0 =  
\frac{1}{\lambda}$ and $\epsilon = -\frac{\beta}{\lambda}\,$, this relation 
becomes 
\beqa
\epsilon_{{\rm an}} = \partial_{ {\hat \phi_0}} \left [- \epsilon \hat \phi_0 
\right ] = \partial_{ {\hat \phi_0}} \beta_{{\hat \phi_0}} \, , \label{AnDimv2}
\eeqa
where in the last equality we have introduced the beta function associated with 
$\hat \phi_0$, defined as the derivative of $\hat \phi_0$ with respect to $\log 
E^{*}$. Interestingly, the relation \eqref{AnDimv2} looks very similar to the 
anomalous dimensions that arise in holographic RG flows \cite{deBoer:1999tgo}.

\section{Conclusions}
 In this work we have investigated effective holographic models where QCD is 
described in terms of a 4-$d$ CFT deformation. The deformation is of the form 
$\delta {\cal L} = \phi_{0}\, {\cal O}$, where ${\cal O}$ is a relevant operator 
and $\phi_0$ the coupling, and takes place at a UV energy scale $E^{*}$. It is 
characterized by the conformal dimension of the relevant operator $\Delta = 
4-\epsilon$, which according to the AdS/CFT dictionary, maps to the 5-$d$  mass 
of the dual dilaton field. The IR dilaton asymptotics was constrained by the 
criteria of confinement and linear glueball spectrum; namely a dilaton quadratic 
in the radial coordinate $z$. We have proposed UV/IR semi-analytic 
interpolations that lead to a spectrum of scalar and tensor glueballs consistent 
with lattice QCD. A key ingredient in our description was the evolution of the 
model parameters with $\epsilon$. In particular, the evolution of the coupling 
$\phi_0$ with $\epsilon$ was essential to guarantee an explicit breaking of 
conformal symmetry consistent with the glueball spectrum.  
 
Making use of the AdS/CFT correspondence we have evaluated the renormalized 
vacuum energy density $\langle T^{00} \rangle$ and the VEV of the relevant 
operator $\langle {\cal O} \rangle$ in the 4-$d$ theory. Both quantities are 
different from zero as a consequence of the CFT deformation. We have mapped 
those quantities to the QCD vacuum energy and gluon condensate respectively. We 
have also reproduced the universal result for the trace anomaly in 4-$d$ 
deformed CFTs, namely $\langle T^{\mu}_{\;\;\mu} \rangle = - \epsilon \phi_0   
\langle {\cal O} \rangle$, and reinterpreted this result in terms of the QCD 
trace anomaly. This led us to suggest a map between the conformal dimension 
$\epsilon$ and the $\beta$ function of the QCD-like theory. The dictionary found 
in this work differs significantly  from the one proposed in 
\cite{Gursoy:2007cb,Gursoy:2007er}, but establish a novel connection between QCD 
and deformed CFTs. Moreover, from the evolution of $\phi_0$ with $\epsilon$ and 
the dictionary proposed in this work we found a 4-$d$ $\beta$ function that 
behaves qualitatively as the large-$N_c$ QCD $\beta$ function up to the 
perturbative regime. This nontrivial result indicates that the holographic 
description of QCD as a CFT deformation can be consistent with asymptotic 
freedom without the necessity of building a specific potential. 

There are some pieces of the dictionary that remain to be found, such as the 
relation between the 5-$d$ warp factor and the 4-$d$ energy. A map between the 
backgrounds developed in this work and the backgrounds developed in 
\cite{Gursoy:2007cb,Gursoy:2007er} would also be desirable. That map should be 
such that the metric and dilaton near a geometric cutoff $z^{*}$ in the 
background of \cite{Gursoy:2007cb,Gursoy:2007er} becomes an AdS metric slightly 
deformed by a massive dilaton. The holographic description of the 
Callan-Symanzik equations, following \cite{deBoer:1999tgo,Kiritsis:2014kua}, would also shed some 
light in the connection between holographic QCD and deformed CFTs.  

We finish this work mentioning some of its possible extensions. The 
description of mesons and chiral symmetry breaking in terms of gauge fields and 
a tachyonic field can be done, inspired by the progress made in 
\cite{Karch:2006pv,Gherghetta:2009ac,Jarvinen:2011qe}. Investigating black hole 
solutions at finite temperature will allow the description of a non-conformal 
plasma, complementing the progress made in \cite{Gubser:2008ny,Gubser:2008yx} 
and \cite{Gursoy:2008za}. In this context, a particular interesting phenomena is 
the so-called glueball melting \cite{Miranda:2009uw}. Finally, the study of 
higher spin glueballs and the pomeron could be pursued, along the lines of 
\cite{Brower:2006ea} and more recently 
\cite{Ballon-Bayona:2015wra,Capossoli:2016kcr,Nally:2017nsp,
Ballon-Bayona:2017vlm}.

\section*{Acknowledgments}

The authors would like to acknowledge conversations and correspondence with 
Nelson R. F. Braga, Renato Critelli, Stefano Finazzo, Elias Kiritsis, Carlisson Miller, Jorge 
Noronha and Robert C. Quevedo. This work was partially funded by Funda\c c\~ao 
de Amparo \`a Pesquisa do Estado de S\~ao Paulo (FAPESP), Brazil, Grants 
No.~2011/18729-1 (V. T. Z.),  No.~2013/17642-5 (L. A. H. M.) and 
No.~2015/17609-3 (A.B-B). V. T. Z. also thanks  Coordena\c{c}\~ao de 
Aperfei\c{c}oamento do Pessoal de N\'\i vel Superior (CAPES), Brazil, Grant 
No.~88881.064999/2014-01, and Conselho Nacional de Desenvolvimento Cient\'\i 
fico e Tecnol{\'o}gico  (CNPq), Brazil, Grant No.~308346/2015-7. A.B-B also 
acknowledges partial financial support from the  grant CERN/FIS-NUC/0045/2015. 
H. B. F. also acknowledges partial financial support from CNPq Grant No. 
307278/2015-8. 

\appendix

\section{The massless mode in the scalar sector}
\label{App:massless}
In this appendix we develop a semi-analytic approach to treat the massless mode. 
We find a general criteria to identify whether a massless 
mode appears or not. Such an approach follows the same idea of refs.
\cite{Kiritsis:2006ua,Gursoy:2007er}. 
The starting point is the Schr\"odinger like equation (\ref{schrodingerscaeq}), 
which can be written as \cite{Kiritsis:2006ua,Gursoy:2007er}
 \begin{equation}\label{eq1ap1}
 P^{\dag}P\psi_{s}(z)=\hat{m}_{s}^{2}\psi_{s}(z),
 \end{equation}
where $P=-\partial_{z}+B_{s}'(z)$ and  $P^{\dag}= \partial_{z}+B_{s}'(z)$.

Our goal here is to proof that there is a massless state with
normalizable wave function in the bulk. In order to do that, we need the 
asymptotic expansions of the dilaton in the UV and IR regimes, which are 
respectively of the form 
\begin{equation} \label{eq2ap1}
\begin{split}
\Phi(z)&= \phi_{0}\,z^{\Delta_{-}}+G\,z^{\Delta_{+}},
\quad z\rightarrow 0,\\ 
\Phi(z)&=C\,z^{2}, \qquad\qquad\qquad z\rightarrow \infty,
\end{split}
\end{equation}
where $\Delta_{+}=4-\Delta_{-}$.

Since we are looking for the wave function of the massless mode, we 
substitute $\hat{m}_{s}^{2}=0$ in equation (\ref{eq1ap1}) to get
\begin{equation}\label{eq3ap1}
\left[\partial_{z}+B_{s}'(z)\right] 
\left[-\partial_{z}+B_{s}'(z)\right]\psi_{s}(z)=0.
\end{equation}
The solutions of this equation may be written as
\cite{Kiritsis:2006ua}
\begin{equation}\label{eq4ap1}
\begin{split}
\psi_{s}^{(1)}(z)&=e^{B_{s}(z)}, \quad
\psi_{s}^{(2)}(z)=e^{B_{s}(z)}\int_{0}^{z}e^{-2B_{s}(z')}dz'.
\end{split}
\end{equation}
The next step is to verify the normalizability of such solutions, which means that 
the integral
\begin{equation}\label{eq4p6ap1}
I=\int_0^\infty dz\, \psi_{s}(z)\psi_{s}^{*}(z)
\end{equation}
must be finite. 
Following the procedure developed in section \ref{effectivepot}, 
we write down the leading terms of the asymptotic expansions for $B_{s}(z)$ in 
the UV and IR regimes:
\begin{equation}\label{eq4p5ap1}
\begin{split}
B_{s}(z)&= 
\log{\bigg[\frac{1}{3}\phi_{0}\,\Delta_{-}z^{(\Delta_{-}-3/2)}+
\frac{1}{3}G\,\Delta_{+}z^{(\Delta_{+}-3/2)}+\cdots\bigg]}, 
\qquad z\rightarrow 0;\\
B_{s}(z)&= -C\,z^{2}+\cdots, \qquad z\rightarrow \infty; 
\end{split}
\end{equation}
where $\phi_{0}$, $G$ and $ C$ are the constant parameters 
previously introduced, with $C$ being a positive real parameter.

We start by analyzing the first solution, $\psi_{s}^{(1)}(z)$,
in the IR regime. Using equations (\ref{eq4ap1}), (\ref{eq4p6ap1}) and (\ref{eq4p5ap1}),
we conclude that the first solution is normalizable. In fact, one has in this case 
\begin{equation}
\begin{split}
I&= \frac{\sqrt{\pi}}{2 \sqrt{C}}
 \text{Erf}[\sqrt{C} z]
\rightarrow 0 \qquad 
\text{as} \qquad z\rightarrow \infty,
\end{split}
\end{equation}
where $\text{Erf}[\sqrt{C}\, z]$ is the error function.
Therefore, the first solution is normalizable in the IR regime.

We now analyze the first solution in the UV regime. In such a limit, this solution can
be written as
\begin{equation}\label{eq5ap1}
\begin{split}
\psi_{s}^{(1)}(z)&= c_{1}\,z^{(-3/2+\Delta_{-})}+
c_{2}\,z^{(5/2-\Delta_{-})}, \quad \text{where}\\ \quad 
c_{1}&=\frac{1}{3}\Delta_{-}\phi_{0},\qquad
c_{2}=\frac{1}{3}\Delta_{+}G.
\end{split}
\end{equation}
We see that the coefficients $c_{1}$ and $c_{2}$ in 
equation (\ref{eq5ap1}) are 
related to the source $\phi_{0}$ and to the condensate $G$, 
respectively.

Considering the solutions of equation $\Delta(\Delta-4)=M^{2}$, 
and taking into account the Breitenlohner-Freedman (BF) bound \cite{Breitenlohner:1982bm},
$0\leq\Delta_{-}\leq2$ and 
$2\leq\Delta_{+}\leq 4$, we have the following four possibilities.

\begin{enumerate}[(i)]

\item $\Delta_{-}=0$ and $\Delta_{+}=4$. This represents the extremal case, for which $M=0$.
In this case the conformal dimension does not 
show in, but there are still the source and the condensate, the last effect 
being responsible for the spontaneous break of the conformal symmetry. The 
dilaton expansion in the UV becomes
\begin{equation}
\Phi(z)=\phi_{0}+G\, z^{4}, \qquad\quad z\rightarrow 0.
\end{equation}
By using the asymptotic expansion of $B_{s}=
\log{\left(\frac{4G}{3}\, z^{5/2}\right)}$, the integral (\ref{eq4p6ap1}) is 
evaluated to yield
\begin{equation}
\begin{split}
I&=
\frac{8}{27}G^{2}\,z^{6}\,, \qquad\quad \text{as} \quad 
z\rightarrow 0,
\end{split}
\end{equation}
from what follows that the solution is normalizable in the UV limit. Then, 
there exists a massless mode in the extremal case, as previously studied by
Csaki and Reece  \cite{Csaki:2006ji}.

\item $0<\Delta_{-}< 1$ and $3\leq\Delta_{+}<4$. In the case $0<\Delta_{-}< 
1$, the integral (\ref{eq4p6ap1}) gives
\begin{equation}\label{eq5p6ap1}
\begin{split}
I&=
\frac{1}{9}\left(\frac{G^{2}\Delta_{+}^{2}}{2(3-\Delta_{-})}\, 
z^{2(3-\Delta_{-})}
+G\,\phi_{0}\Delta_{-}\Delta_{+}\,z^{2}
+\frac{\phi_{0}^{2}\Delta_{-}^{2}}{2(\Delta_{-}-1)}z^{2(\Delta_{-}-1)}\right), 
\quad\text{as} \quad z\rightarrow 0.
\end{split}
\end{equation}
Since the exponent of the last term in (\ref{eq5p6ap1}) is negative, $\Delta_{-}-1<0$,
it diverges near the boundary unless its coefficient is zero.
This means that the source should be turned off, $\phi_{0}=0$, in order to have 
a resulting normalizable wave function,
\begin{equation}
I=
\frac{1}{9}\left(\frac{G^{2}\Delta_{+}^{2}}{2(3-\Delta_{-})}\, 
z^{2(3-\Delta_{-})}\right), 
\quad\qquad \text{as} \quad z\rightarrow 0.
\end{equation}
This means that we are breaking explicitly the conformal symmetry (because 
$\Delta_{+}=4-\Delta_{-}$), but the massless mode is present since the corresponding
wave function is normalizable from the UV to the IR limit in the bulk. However,
if the source is turned on, the wave function becomes non-normalizable in the 
UV ($I\rightarrow \infty$). The conclusion is that there is no massless mode in 
the presence of a source. In the case studied in this work, we have a small
deformation $\Delta_{-}=\epsilon$, that we are interpreting as conformal dimension,
and the source is turned on. This explains why there are no 
massless modes in the analysis of section \ref{Sec:Spectrum}. At this point it 
is worth to emphasize that we must have both, source and deformation, to 
eliminate the massless mode.

\item  Now let us look at the case $\Delta_{-}=1$ and $\Delta_{+}=3$. The integral 
(\ref{eq4p6ap1}) becomes
\begin{equation}
\begin{split}
I&=
\frac{1}{36}\left(G^{2}\Delta_{+}^{2}\,z^{4}
+4G\,\phi_{0}\Delta_{-}\Delta_{+}\,z
+4\phi_{0}^{2}\Delta_{-}^{2}\log{z}\right).
\end{split}
\end{equation}

The analytical model presented in the appendix G of ref.
\cite{Gursoy:2007er} is included here as a particular case with $\Delta_{-}=1$ and 
$\phi_{0}\neq 0$. As seen in the above analysis, there exists a massless mode if we turn 
off the source, and there is no massless modes when the source is turned on. An 
alternative analysis of this case, studying nearly marginal operators, was 
developed in \cite{Megias:2014iwa}.

\item $1<\Delta_{-}\leq2$ and $2\leq\Delta_{+}<3$. According 
to the exponent in the wave function (\ref{eq5ap1}) there are three
situations to be considered: \label{caseIII}

\begin{itemize}
\item $1<\Delta_{-}< 3/2$ and $5/2<\Delta_{+}<3$. In
this case the integral (\ref{eq4p6ap1}) is 
the same as the result (\ref{eq5p6ap1}) and it converges ($I\rightarrow 0$). 
Therefore, there is a massless mode for these ranges of $\Delta_{\pm}$ values.

\item $\Delta_{-}= 3/2$ and $\Delta_{+}=5/2$. For these specific values of $\Delta_{\pm}$,
the integral (\ref{eq4p6ap1}) becomes
\begin{equation}
\begin{split}
I&=
\frac{
\phi_{0}
^{3}}{20\,G}, \qquad\quad \text{as} \quad 
z\rightarrow 0.
\end{split}
\end{equation}
The last result depends on the source value (which is finite) and therefore the wave
function is normalizable. So, there is a massless mode also for these specific
values of $\Delta_{\pm}$.

\item $3/2<\Delta_{-}\leq 2$ and $2\leq\Delta_{+}<5/2$. In this case the 
integral (\ref{eq4p6ap1}) converge ($I\rightarrow 0$). 
Therefore, there exists a massless mode.

\end{itemize}

As the integral is finite in case \ref{caseIII}, it means 
that both terms of the wave function (\ref{eq5ap1}) are 
normalizable, and the source and the condensate 
can exchange their roles \cite{Bourdier:2013axa}.
\end{enumerate} 

Now let us look at the second solution, $\psi_{s}^{(2)}(z)$. In the IR regime
we have
\begin{equation}\label{eq7ap1}
\psi_{s}^{(2)}(z)=
e^{-Cz^{2}}\int_{0}^{z}e^{2Cz'^{2}}dz'
=\frac{1}{2}\sqrt{\frac{\pi}{2C}}e^{-Cz^{2}}\text{Erfi}[\sqrt{2C}z],
\end{equation}
where $\text{Erfi}[\sqrt{2C}z]$ is the imaginary error 
function. The last term in the above equation is divergent and then
the wave function is non-normalizable. On the other hand, in the UV regime we have
\begin{equation}
\psi_{s}^{(2)}(z)= 
e^{(\Delta_{-}-3/2)\log{z}}\int_{0}^{z}e^{-2(\Delta_{-}-3/2)\log{z'}}dz'.
\end{equation}
This solution might be or not normalizable in the UV. But as 
the solution in the IR is always non-normalizable, there is not a massless mode in this case. 

In conclusion, the normalizability of the first solution 
$\psi_{s}^{(1)}(z)$ depends on the values of $\Delta_{-}$ and $\phi_{0}$. 
As this solution is normalizable in the IR, by choosing appropriately 
these two parameters, it can be made normalizable also in the UV regime, 
and then a massless mode can be obtained in this case.
On the other hand, the second solution is non-normalizable in the IR. 
This result excludes the possibility to get a massless mode from the second solution.

\section{Models B: analytic form for the warp factor}
\label{App:ModelB}
Alternatively to what was presented in the main body of this work, we can choose 
to interpolate the warp factor between the UV and IR, instead of the dilaton. 
For future reference, as instance to explore the finite temperature effects, it 
is convenient to have an analytic warp factor. We call this approach of 
interpolating the warp factor as model B. Inspired by the interpolations done 
above for the dilaton field, we carefully seek the form of the interpolating 
functions so that they produce the asymptotic forms of equations 
(\ref{asymptotics2}) and (\ref{eqsuperpot3}), respectively. 

Our first choice, motivated by the model A1, is
\begin{equation}
A(z)=-\log{z}
-\frac{2\epsilon\,\phi_{0}^{2}}{9 (1+2\epsilon)}z^{2\epsilon}
-\frac{2\epsilon(4-\epsilon)\phi_{0}\,\Lambda^{4-\epsilon}}{45}\frac{z^{4}} 
{1+a_{1}(z)}
-\frac{\left(\Lambda z\right)^{8-2\epsilon}} 
{\frac{9(9-2\epsilon)}{2(4-\epsilon)}
+b_{1}(z)},
\end{equation}
where the functions $a_{1}(z)=7(\Lambda z)^{5}/50$ and $b_{1}(z)=5(\Lambda 
z)^{4}+3(\Lambda z)^{6-2\epsilon}/2$ were introduced to guarantee the 
compatibility of the results of the model with those of the lattice QCD 
\cite{Meyer:2004gx}. These functions also guarantee that the warp factor 
decreases monotonically from the UV to the IR, as required in Dilaton-Gravity 
models \cite{Kiritsis:2006ua}. We named this interpolation form as model B1.

The second model, called B2, that produces the asymptotic forms of  
equations (\ref{asymptotics2}) and (\ref{eqsuperpot3}), respectively, is
inspired by model A2. The chosen warp factor is
\begin{equation}
A(z)=-\log{z}
-\frac{2\epsilon\,\phi_{0}^{2}}{9 (1+2\epsilon)}z^{2\epsilon}
-\frac{2\epsilon(4-\epsilon)\phi_{0}\,\Lambda^{4-\epsilon}}{45} 
\frac{z^{4}}{1+a_{2}(z)}
-\frac{\left(\Lambda z\right)^{6}
\tanh^2{\left[\left(\Lambda z\right)^{1-\epsilon}\right]}}
{\frac{9(9-2\epsilon)}{2(4-\epsilon)}+b_{2}(z)}.
\end{equation}
As before, the functions $a_{2}(z)=7\left(\Lambda z\right)^{4}/50$ and 
$b_{2}(z)=733\, \left(\Lambda z\right)^{2}/100+3\, \left(\Lambda 
z\right)^{4}/2$ were adjusted to get compatible results with those of the 
lattice QCD \cite{Meyer:2004gx}, and to yield a monotonically decreasing warp 
factor.

To complete the analysis of these models, we show in figure 
\ref{ModelBPotentials}  the potentials of the Schr\"odinger like equations for 
both sectors, scalar and tensor, as indicated. 
The spectra calculated using these models are presented in table \ref{taba01}.

%
\FIGURE{
\begin{tabular}{*{2}{>{\centering\arraybackslash}p{.5\textwidth}}}
        \begin{center} \includegraphics[width=7.2cm,angle=0]{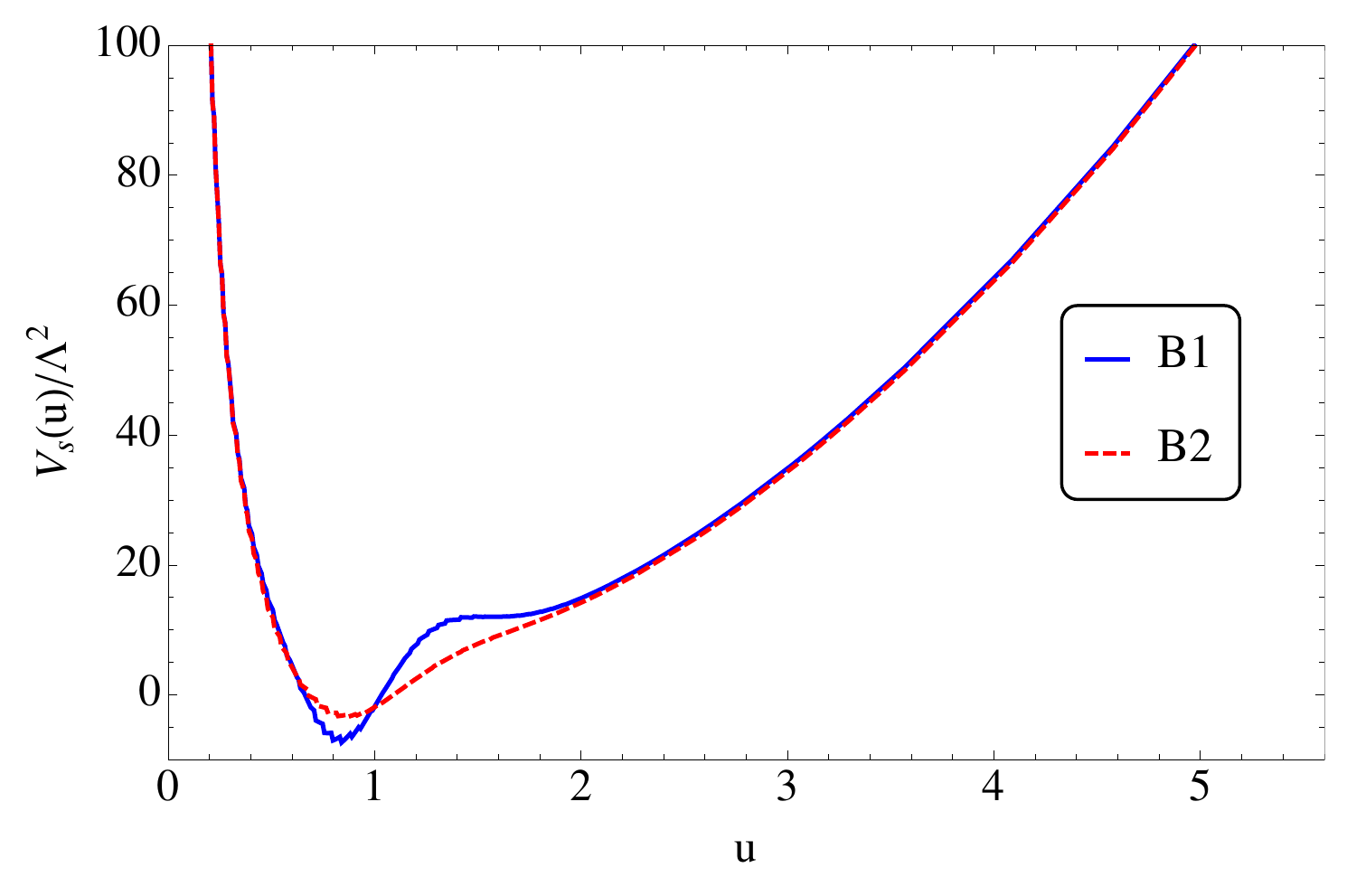} \end{center} & \begin{center}\hspace{-1.5cm}  \includegraphics[width=7.2cm,angle=0]{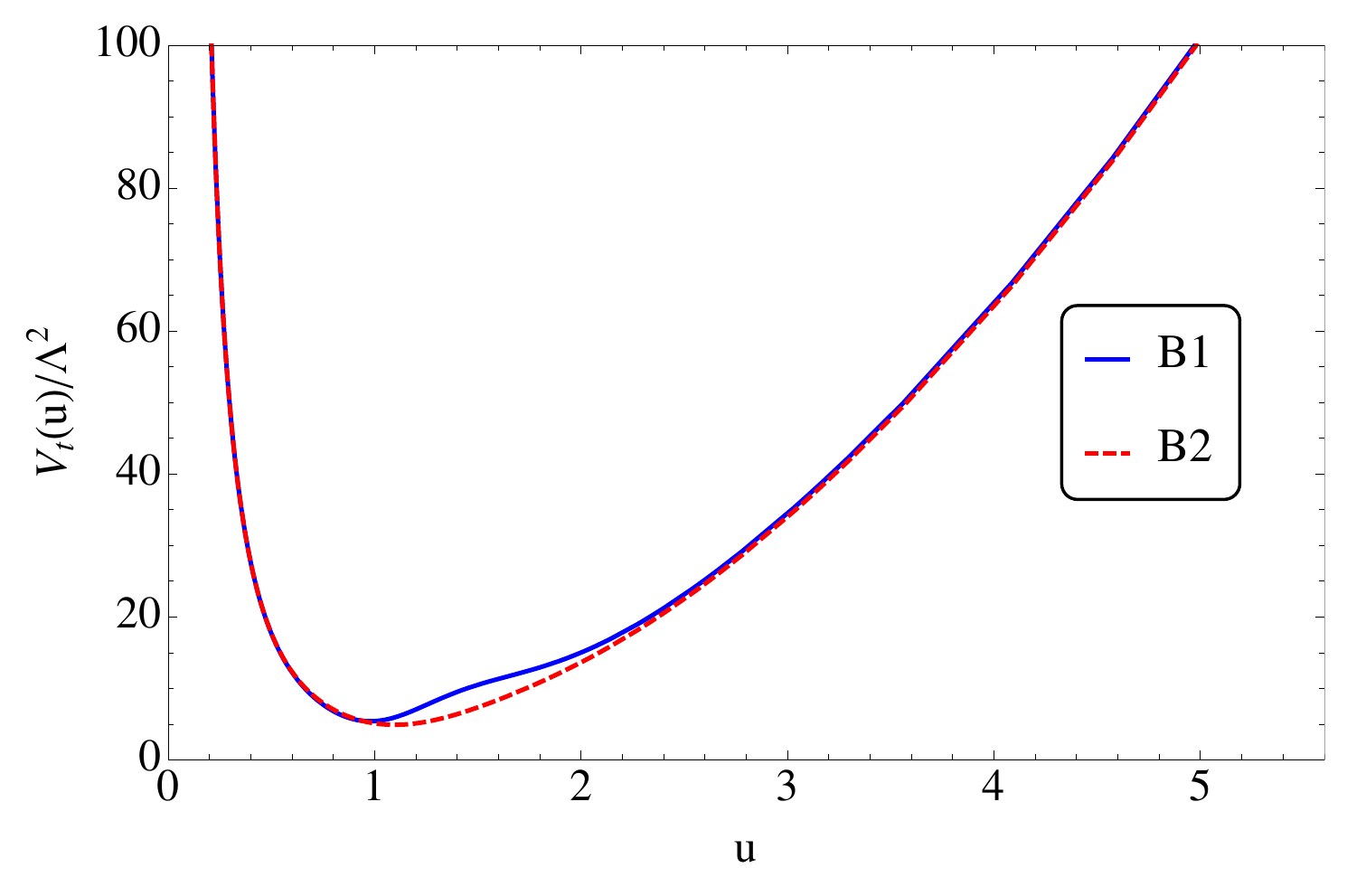} \end{center} \\
\end{tabular}
    \caption{The scalar ({\bf left panel}) and tensor ({\bf right panel}) normalized potentials for  $\epsilon=0.01$ and $\hat \phi_0=50$, in the models B1 and B2.} 
     \label{ModelBPotentials}}

\end{document}